\documentclass[default]{aa}
\usepackage[utf8]{inputenc}
\usepackage[english]{babel}
\usepackage{appendix}
\usepackage{graphicx,subcaption}
\usepackage{natbib}
\usepackage{listings}
\usepackage{colortbl} 
\usepackage{multirow}
\usepackage{xcolor}
\usepackage{soul}
\usepackage{orcidlink}

\usepackage{hyperref}
\hypersetup{
    colorlinks=true,
    linkcolor=cyan,
    urlcolor=blue,
    citecolor=blue
    }

\usepackage{stackengine}
\stackMath

\def\bfu {{\bf u}}
\def\bfv {{\bf v}}

\def\bfx {{\bf x}}

\def\bfB {{\bf B}}

\def\bfE {{\bf E}}
\def\bfF {{\bf F}}

\def\bfJ {{\bf J}}

\def\Dt {\Delta t}

\newcommand{\FIG}[1]{#1}

\title{\texttt{MPI-AMRVAC 3.0}: updates to an open-source simulation framework}
\author{R. Keppens\inst{1,}\orcidlink{0000-0003-3544-2733} \and B. Popescu Braileanu \inst{1} \and Y. Zhou\inst{1} \and W. Ruan\inst{1}  \and C. Xia \inst{2}\orcidlink{0000-0002-7153-4304} \and Y. Guo\inst{3} \and N. Claes\inst{1} \and F. Bacchini\inst{1,4}\orcidlink{0000-0002-7526-8154}}
\institute{Centre for mathematical Plasma Astrophysics, Department of Mathematics, KU Leuven, Celestijnenlaan 200B, B-3001 Leuven, Belgium \\\email{rony.keppens@kuleuven.be} \and
School of Physics and Astronomy, Yunnan University, Kunming 650500, China
\and
School of Astronomy and Space Science, Nanjing University, Nanjing 210023, China
\and
Royal Belgian Institute for Space Aeronomy, Solar-Terrestrial Centre of Excellence, Ringlaan 3, 1180 Uccle, Belgium}
\date{Received ???/ Accepted ???}

\abstract
{Computational astrophysics nowadays routinely combines grid-adaptive capabilities with modern shock-capturing, high resolution spatio-temporal integration schemes on challenging multi-dimensional hydro- and magnetohydrodynamic (MHD) simulations. A large, and still growing, number of community software efforts exist, and we here provide an update on recent developments within the open-source \texttt{MPI-AMRVAC} code.}
{Complete with online documentation, the \texttt{MPI-AMRVAC 3.0} release includes several recently added equation sets, and offers many options to explore and quantify the influence of implementation details. While showcasing this flexibility on a variety of hydro and MHD tests, we document new modules of direct interest for state-of-the-art solar applications.}
{Test cases address how higher order reconstruction strategies impact long term simulations of shear layers, with and without gas-dust coupling effects, how runaway radiative losses can transit to intricate multi-temperature, multi-phase dynamics, and how different flavors of spatio-temporal schemes and/or magnetic monopole control produce overall consistent MHD results in combination with adaptive meshes. We demonstrate the use of Super-Time-Stepping strategies for specific parabolic terms and give details on all the implemented Implicit-Explicit (IMEX) integrators. A new magnetofrictional module can be used for computing force-free magnetic field configurations or for data-driven time-dependent evolutions, while the Regularized-Biot-Savart-Law approach can insert fluxropes in 3D domains. Synthetic observations of 3D MHD simulations can now be rendered on-the-fly, or in post-processing, in many spectral wavebands.}
{A particle module as well as a generic fieldline tracing module, fully compatible with the hierarchical meshes, can be used to do anything from sampling information at prescribed locations, to follow dynamics of charged particles, or realize fully two-way coupled simulations between MHD setups and field-aligned non-thermal processes. We provide reproducible, fully demonstrated tests of all code functionalities.} {While highlighting the latest additions and various technical aspects (e.g.\ reading in datacubes for initial or boundary conditions), our open-source strategy welcomes any further code usage, contribution, or spin-off development.}

\begin{document}

\maketitle

\section{Introduction}
\subsection{AMR codes for astrophysics}
Adaptive Mesh Refinement (or AMR) is currently routinely available in many open-source, community-driven software efforts. The challenges associated with shock-dominated hydrodynamic simulations on hierarchically refined grids were already identified in the pioneering work by
\citet{1989Berger}, and have since been carried over to generic frameworks targeting the handling of multiple systems of partial differential equations. One such framework is the {\tt PARAMESH} \citep{2000MacNeice} package, offering support for parallelized AMR on logically Cartesian meshes. Codes that inherited the {\tt PARAMESH} AMR flexibility are the {\tt FLASH} code, that started as a pure hydro-AMR software for astrophysical applications \citep{2000Fryxell}. {\tt FLASH} has since been used in countless studies, and a recent example includes its favorable comparison with an independent simulation study~\citep{2022Orban} focusing to model the challenging radiative-hydro behavior of a laboratory, laser-produced jet. {\tt PARAMESH} has also been used in space-weather related simulations in 3D ideal MHD \citep{2012Feng}. For space weather applications, a similarly noteworthy forecasting framework employing AMR is discussed in \citet{2021Narechania}, where Sun-to-Earth solar wind simulations in ideal MHD are validated. Another AMR package in active development is the {\tt CHOMBO} library\footnote{
\url{https://commons.lbl.gov/display/chombo/}}, and this is how the {\tt PLUTO} code \citep{2012Mignone} inherits AMR functionality. Recent {\tt PLUTO} additions showcase how dust particles can be handled using a hybrid particle-gas treatment~\citep{2019Mignone}, or detail how novel non-local thermal equilibrium radiation hydro is performing~\citep{2019Colombo}. 

Various public-domain codes employ a native AMR implementation, such as the {\tt ENZO} code \citep{2014Bryan}, or the {\tt RAMSES} code which started as an AMR-cosmological hydrodynamics code \citep{2002Teyssier}. In {\tt Astrobear} \citep{2009Cunningham}, capable of using AMR on MHD simulations using constrained transport on the induction equation, the AMR functionality is known as the {\tt BEARCLAW} package. Radiative MHD functionality for {\tt Astrobear}, with a cooling function extending below 10000 K was demonstrated in \citet{2018Hansen}, when studying magnetized radiative shock waves. 
{\tt NIRVANA} has seen successive improvements in its AMR-MHD possibilities \citep{2005Ziegler,2008Ziegler}, and has more recently added a chemistry-cooling module described in \citet{2018Ziegler}. Another AMR-MHD code that pioneered the field was introduced as the {\tt BATS-R-US} code \citep{1999Powell}, which currently is the main solver engine used in the Space Weather Modeling Framework described in \citet{2012SWMF}. Their AMR functionality has been implemented in the Block-Adaptive-Tree Library {\tt BATL}, a {\tt Fortran}-AMR implementation. This shares various algorithmic details with the AMR implementation in {\tt MPI-AMRVAC}, described in \citet{2012Keppens}, whose {\tt 3.0} update forms the topic of this paper.

Meanwhile, various coding efforts anticipate the challenges posed by modern exascale High Performance Computing systems, such as that realized by task-based parallelism now available in the 
{\tt Athena++} \citep{2020Stone} effort. This code is in active use and development, with e.g.\ a recently added gas-dust module \citep{2022Huang}, similar to the gas-dust functionality available for {\tt MPI-AMRVAC} \citep{2014Porth}. This paper documents the latter code's novel options for implicit-explicit handling of various partial differential equation (PDE) systems, and shows its use for gas-dust coupling. {\tt GAMER-2} as presented in~\citet{2018Schive} is yet another community effort, offering AMR and many physics modules, where GPU acceleration in addition to hybrid {\tt OpenMP/MPI} allows effective resolutions of order $10000^3$. Even more visionary efforts in terms of adaptive simulations, where also multiple physics modules may want to be run concurrently on adapting grid hierarchies include the {\tt DISPATCH} \citep{2018Nordlund} and the {\tt PATCHWORK} \citep{2018Shiokawa} frameworks. This paper serves to provide an updated account of the {\tt MPI-AMRVAC} functionality, where future directions and potential links to such ongoing new developments are provided in our closing discussion.

\subsection{Open source strategy with \texttt{MPI-AMRVAC}}
With \texttt{MPI-AMRVAC}, we provide an open-source framework written in {\tt Fortran} where parallelization is achieved by a (possibly hybrid {\tt OpenMP-}){\tt MPI} implementation, where the block adaptive refinement strategy has evolved to the standard block-based quadtree-octree (2D-3D) organization. While originally used to evaluate efficiency gains affordable through AMR for multi-dimensional hydro (HD) and magnetohydrodynamics (MHD)~\citep{Keppens03}, later applications focused on special relativistic HD and MHD settings~\citep{2008vanderHolst,2012Keppens}. Currently, the {\tt github} source version\footnote{\url{https://github.com/amrvac}} is deliberately handling Newtonian dynamics throughout, and we refer to its \texttt{MPI-AMRVAC 1.0} version as documented in \citet{2014Porth}, while an update to \texttt{MPI-AMRVAC 2.0} is provided in \citet{2018Xia}. A more recent guideline on the code usability to solve generic PDE systems (including reaction-diffusion models) is found in \citet{AMRVAC2021}. Since \texttt{MPI-AMRVAC 2.0}, we have a modern library organization (using the code for 1D, 2D or 3D applications), have a growing number of automated regression tests in place, and provide a large number of tests or actual applications from published work under, e.g.\ the {\tt tests/hd} subfolder for all simulations using the hydro module {\tt src/hd}. This ensures full compliance with all modern requirements on data reproducibility and data sharing.

Our open-source strategy already led to various noteworthy off-spins, where e.g the AMR framework and its overall code organization got inherited to create completely new functionality: e.g.\ the {\tt Black Hole Accretion Code} or {\tt BHAC}\footnote{\url{http://bhac.science}} from
\citet{2017Porth} \citep[and its extensions, see, e.g.,][]{Bacchini2019,Olivares2019,Weih2020} realizes a modern general-relativistic MHD (GR-MHD) code, which was used in the GR-MHD code comparison project from \citet{2019Porth}. In \citet{2019Ripperda} the GR-MHD code {\tt BHAC} got extended to handle GR-resistive MHD (GR-RMHD) where implicit-explicit (IMEX) strategies handled stiff resistive source terms. We here document how various IMEX strategies can be used in Newtonian settings for \texttt{MPI-AMRVAC 3.0}. The hybrid {\tt OpenMP-MPI} parallelization strategy was optimized for {\tt BHAC} in 
\citet{2022Cielo}, and we inherited much of this functionality within \texttt{MPI-AMRVAC 3.0}. Other, completely independent GR-MHD software efforts that derived from earlier \texttt{MPI-AMRVAC} variants include {\tt GR-AMRVAC} by \citet{2016Meliani}, the {\tt Gmunu} code introduced in 
\citet{2021Cheong,2022Cheong}, or the {\tt NOVAs} effort presented in \citet{2022NOVA}.

The code is also used in the most recent update to the space weather modeling effort {\tt EUHFORIA}\footnote{\url{http://euhforia.com}}, introduced in \citet{2018Pomoell}. In the {\tt ICARUS}\footnote{\url{https://github.com/amrvac-icarus/icarus} or the {\tt tests/mhd/icarus} test case in the master branch.}
framework presented by \citet{2022Verbeke}, the most time-consuming aspect of the prediction pipeline is the 3D ideal MHD solver that uses extrapolated magnetogram data for solar coronal activity at 0.1 AU, to then advance the MHD equations till 2 AU, covering all $360^\circ$ longitudes, within a $\pm 60^\circ$ latitude band. This represents a typical use-case of \texttt{MPI-AMRVAC} functionality, where the user can choose a preferred flux scheme, the limiters, the many ways to automatically (de)refine on weighted, user-chosen (derived) plasma quantities, while adopting the radial grid stretching introduced in \citet{2018Xia} in spherical coordinates. In what follows, we provide an overview of current \texttt{MPI-AMRVAC 3.0} functionality that may be useful for future users, or for further spin-off developments.

\section{Available PDE systems}

The various PDE systems available in \texttt{MPI-AMRVAC 3.0} are listed in Table~\ref{tab:eqs}. These cover a fair variety of PDE types (elliptic, parabolic, but with an emphasis towards hyperbolic PDEs), and it is noteworthy that almost all modules can be exploited in 1D to 3D setups. They are all fully compatible with AMR and can be combined with modules that can meaningfully be shared between the many PDE systems. Examples of such shared modules are \begin{itemize} 
\item the particle module in {\tt src/particle}, which we briefly discuss in Section~\ref{sec:part},
\item the streamline/fieldline tracing module in {\tt src/modules/mod\_trace\_field.t} as demonstrated in Section~\ref{sec:trace},
\item additional physics in the form of source terms for the governing equations, such as {\tt src/physics/mod\_radiative\_cooling.t} to handle optically thin radiative cooling effects (see also Section~\ref{sec:otc}), or {\tt src/physics/mod\_thermal\_conduction.t} for thermal conduction effects, {\tt src/physics/mod\_viscosity.t} for viscous problems, \ldots
\end{itemize}
Table~\ref{tab:eqs} provides references related to module usage, while some general guidelines for adding new modules can be found in \cite{AMRVAC2021}. These modules share the code-parallelism, the grid-adaptive capacities and the various time-stepping strategies, e.g.\ the IMEX schemes mentioned below in Section~\ref{sec:IMEX}. In the next sections, we will highlight novel additions to the framework, with an emphasis on multi-dimensional (magneto)hydrodynamic settings. Adding a physics module to our open-source effort can follow the instructions in {\tt doc/addmodule.md} and the info in {\tt doc/contributing.md} to ensure that auto-testing is enforced. The code's documentation has two components: (1) the markup documents collected in the {\tt doc} folder, which appear as {\tt html} files on the code website \url{http://amrvac.org}; and (2) the inline source code documentation, which gets processed by {\tt Doxygen}\footnote{\url{http://doxygen.nl}} to deliver full dependency trees and documented online source code.

\begin{table*}[]
    \centering
    \caption{Equation sets available in \texttt{MPI-AMRVAC 3.0}.}
    \label{tab:eqs}
    \begin{tabular}{l l l}
        \hline
        \hline
         Module Name & Purpose & Equations or Reference \\[0.7ex]
        \hline
{\tt rho} & linear scalar advection & $\partial_t \rho+\mathbf{v} \cdot\nabla \rho =0$ \\
& constant velocity vector $\mathbf{v}$ & \\
\hline
{\tt rd} & reaction-diffusion systems & $\partial_t \mathbf{u} = {\mathrm{diag}}(D_i)\nabla^2 \mathbf{u} +\mathbf{f}(\mathbf{u})$ \\
 & 8 different PDE systems, 1 to 3 $\mathbf{u}$ components &   \cite{AMRVAC2021} \\
 & & {\tt doc/reaction\_diffusion.md} \\
       \hline
{\tt ard} & advection-reaction-diffusion systems & $\partial_t \mathbf{u} +(\mathbf{v}/p)\cdot \nabla \mathbf{u}^p= {\mathrm{diag}}(D_i)\nabla^2 \mathbf{u} +\mathbf{f}(\mathbf{u})$ \\
 & 8 different PDE systems, 1 to 3 $\mathbf{u}$ components &  \\
 & nonlinear advection for integer $p>1$ & {\tt doc/advection\_reaction\_diffusion.md}\\
       \hline
{\tt nonlinear} & scalar nonlinear advection & $\partial_t \rho +\nabla\cdot \mathbf{F}(\rho,\mathbf{x},t) = g$ \\
& inviscid Burgers or nonconvex equation, Korteweg-De Vries & \cite{2014KeppensPorth} \\
\hline
{\tt hd} & Euler to Navier-Stokes equations for gas dynamics & \cite{2014Porth} \\
 & with or without tracer quantities & \\
 & with or without added dust species &  see Section~\ref{sec:gasdust} and {\tt doc/dust.md} \\
 \hline
 {\tt mhd} & Ideal to (visco-)resistive (+Hall) MHD equations & \cite{2014Porth} \\
 & semirelativistic MHD equations & as in \cite{2002Gombosi} \\
 & with or without tracer quantities & \\
 & splitting strategies for $\mathbf{B}=\mathbf{B}_0+\mathbf{B}_1$ & \cite{2018Xia} \\
 & split-off magnetohydrostatic $-\nabla p_0+\rho_0 \mathbf{g}+\mathbf{J}_0\times\mathbf{B}_0 =\mathbf{0}$& \cite{nitin} \\
 \hline 
 {\tt rhd} & radiation hydrodynamics & \cite{2022Moens} \\
  & Flux-limited-diffusion approximation & \\
  \hline 
  {\tt mf}  & magnetofrictional module & Section~\ref{sec:mf} \\
   & 2D, 2.5D, and 3D magnetic field simulations & \\
  \hline 
  {\tt twofl} & plasma-neutral 2-fluid module & \cite{2022Braileanu} \\
  & chromospheric to coronal physics &  {\tt doc/twofluid.mf} \\
\hline 
    \end{tabular}
\end{table*}

\section{Schemes and limiters for HD and MHD}

Most \texttt{MPI-AMRVAC} applications employ a conservative finite volume type discretization, used in each substep of a multistage time-stepping scheme. This finite volume treatment, combined with suitable (e.g.\ doubly periodic or closed box) boundary conditions ensures conservation properties of mass, momentum and energy as demanded in pure hydrodynamic (HD) or ideal MHD runs. Available explicit time-stepping schemes include (1) a one-step forward Euler, (2) two-step variants such as predictor-corrector (midpoint) and trapezoidal (Heun) schemes, and (3) higher-order, multi-step schemes. Our default three-, four- and five-step time integration schemes fall into the strong stability preserving (SSP) Runge-Kutta schemes~\citep{2005Gottlieb}, indicated as SSPRK($s,p$) involving $s$ stages while reaching temporal order $p$. In that sense, the two-step Heun variant is SSPRK(2,2). In~\citet{2014Porth}, we provided all details of the three-step SSPRK(3,3), four-step SSPRK(4,3) and the five-step SSPRK(5,4) schemes, ensuring third, third, and fourth order temporal accuracy, respectively. Tables~\ref{tab:schemes_time} and \ref{tab:schemes_space} provide an overview of the choices in time integrators as well as the available shock-capturing spatial discretization schemes for the HD and MHD systems. The implicit-explicit IMEX schemes are further discussed in Sect.~\ref{sec:IMEX}. Note that \citet{2014Porth} emphasized that, instead of the standard finite volume approach, \texttt{MPI-AMRVAC} also allows for high order conservative finite difference strategies (in the \texttt{mod\_finite\_difference.t} module), but these will not be considered explicitly here. Having many choices for spatio-temporal discretization strategies allows one to select optimal combinations depending on available computation resources, or on robustness aspects when handling extreme differences in (magneto-)thermodynamical properties. The code allows to achieve higher than second order accuracy on smooth problems. In \citet{2014Porth}, where MPI-AMRVAC 1.0 was presented, we reported on setups that formally achieved up to fourth order accuracy in space and time. Figure 7 in that paper quantifies this for a 3D circularly polarized Alfv\'en wave test, while in the present paper, Fig.~\ref{fig:sts} shows third order accuracy on a 1.75D MHD problem involving ambipolar diffusion. The combined spatio-temporal order of accuracy reachable will very much depend on the problem at hand (discontinuity dominated or not), and on the chosen combination of flux schemes, reconstructions, and source term treatments.

\begin{table*}[]
    \centering
    \caption{The time integration methods in \texttt{MPI-AMRVAC 3.0}, as implemented in \texttt{mod\_advance.t}.}
    \label{tab:schemes_time}
    \begin{tabular}{l l l}
        \hline
        \hline
        Step & Explicit & IMEX \\[0.7ex]
        \hline
one-step & Forward Euler & IMEX-Euler or IMEX-SP\\
        \hline
        two-step & Predictor-Corrector (explicit midpoint) & IMEX-Midpoint\\
          &  SSPRK(2,2) (Heun's method) & IMEX-Trapezoidal  \\
          & RK2($\alpha$)  & IMEX222($\lambda$) \\
       \hline
       three-step & SSPRK(3,3) or SSP(3,2) & IMEX-ARK(2,3,2) or IMEX-SSP(2,3,2) \\
                  & RK3 (Butcher Table: Ralston3, RK-Wray3, Heun3, Nystrom3)  &  IMEX-ARS3 \\
                &  & IMEX-CB3a \\
       \hline
       four-step & SSPRK(4,3) or SSP(4,2) & -- \\
         & RK(4,4) & \\
       \hline
       five-step & SSPRK(5,4) (\citet{2005Gottlieb} or \citet{2002spiteri}) & -- \\
       \hline 
    \end{tabular}
\end{table*}

\begin{table*}[]
    \centering
    \caption{The choices for the numerical flux functions in \texttt{MPI-AMRVAC 3.0}, as implemented in \texttt{mod\_finite\_volume.t}. The full Roe-solver based schemes ($^a$) are discussed in~\citet{1996Toth}.}
    \label{tab:schemes_space}
    \begin{tabular}{l l}
    \hline\hline
    Module Name & Flux Scheme \\
    \hline
    {\tt hd}  & TVD-Lax-Friedrichs, HLL, HLLC, Roe (TVD/TVD-Muscl)$^a$ \\
    {\tt mhd}  & TVD-Lax-Friedrichs, HLL, HLLC, HLLD, Roe (TVD/TVD-Muscl)$^a$\\
    \hline 
    \end{tabular}
    \end{table*}

The finite-volume spatial discretization approach in each substep computes fluxes at cell volume interfaces, updating conservative variables stored as cell-centered quantities representing volume averages; however, when using constrained transport for MHD, we also have cell-face magnetic field variables. We list in Table~\ref{tab:schemes_space} the most common flux scheme choices for the HD and MHD systems. In the process where fluxes are evaluated at cell edges, a limited reconstruction strategy is used -- usually on the primitive variables -- where two sets of cell interface values are computed for each interface: one employing a reconstruction involving mostly left, and one involving mostly right cell neighbours. In what follows, we demonstrate some of the large variety of higher order reconstruction strategies that have meanwhile been implemented in \texttt{MPI-AMRVAC}. For explicit time integration schemes applied to hyperbolic conservation laws, temporal and spatial steps are intricately linked by the Courant-Friedrichs-Lewy (CFL) stability constraint. Therefore, combining high-order time-stepping and higher order spatial reconstructions is clearly of interest to resolve subtle details. Thereby, different flux scheme and reconstruction choices may be used on different AMR levels. Note that our AMR implementation is such that the maximum total number of cells that an AMR run can achieve is exactly equal to the maximum effective grid resolution, if the refinement criteria enforce the use of the finest level grid on the entire domain. Even when a transition to domain-filling turbulence occurs – where triggering finest level grids all over is likely to happen, a gain in using AMR versus a fixed resolution grid can be important, by cost-effectively computing a transient phase. In~\citet{Keppens03}, we quantified these gains for typical HD and MHD problems, and reported on efficiency gains by factors of 5 to 20, with limited overhead by AMR. Timings related to AMR overhead, boundary conditions, I/O and actual computing, are reported by \texttt{MPI-AMRVAC} in the standard output channel. For the tests discussed below, this efficiency aspect can hence be verified by rerunning the demo setups provided.

\subsection{Hydrodynamic tests and applications}

The three sections below contain a 1D Euler test case highlighting differences due to the employed reconstructions (Section~\ref{sec:1DHD}), a 2D hydro test without and with gas-dust coupling (Section~\ref{sec:gasdust}), and a 2D hydro test where optically thin radiative losses drive a runaway condensation and fragmentation (Section~\ref{sec:otc}). We note that the hydrodynamic {\tt hd} module of \texttt{MPI-AMRVAC} could also be used without solving explicitly for the total (i.e. internal plus kinetic) energy density evolution, in which case an isothermal or polytropic closure is assumed. Physical effects that can be activated easily include solving the equations in a rotating frame, adding viscosity, external gravity, thermal conduction and optically thin radiative losses. 

\subsubsection{TVD versus WENO reconstructions}\label{sec:1DHD}

\begin{figure}
    \centering
    \includegraphics[width=\columnwidth]{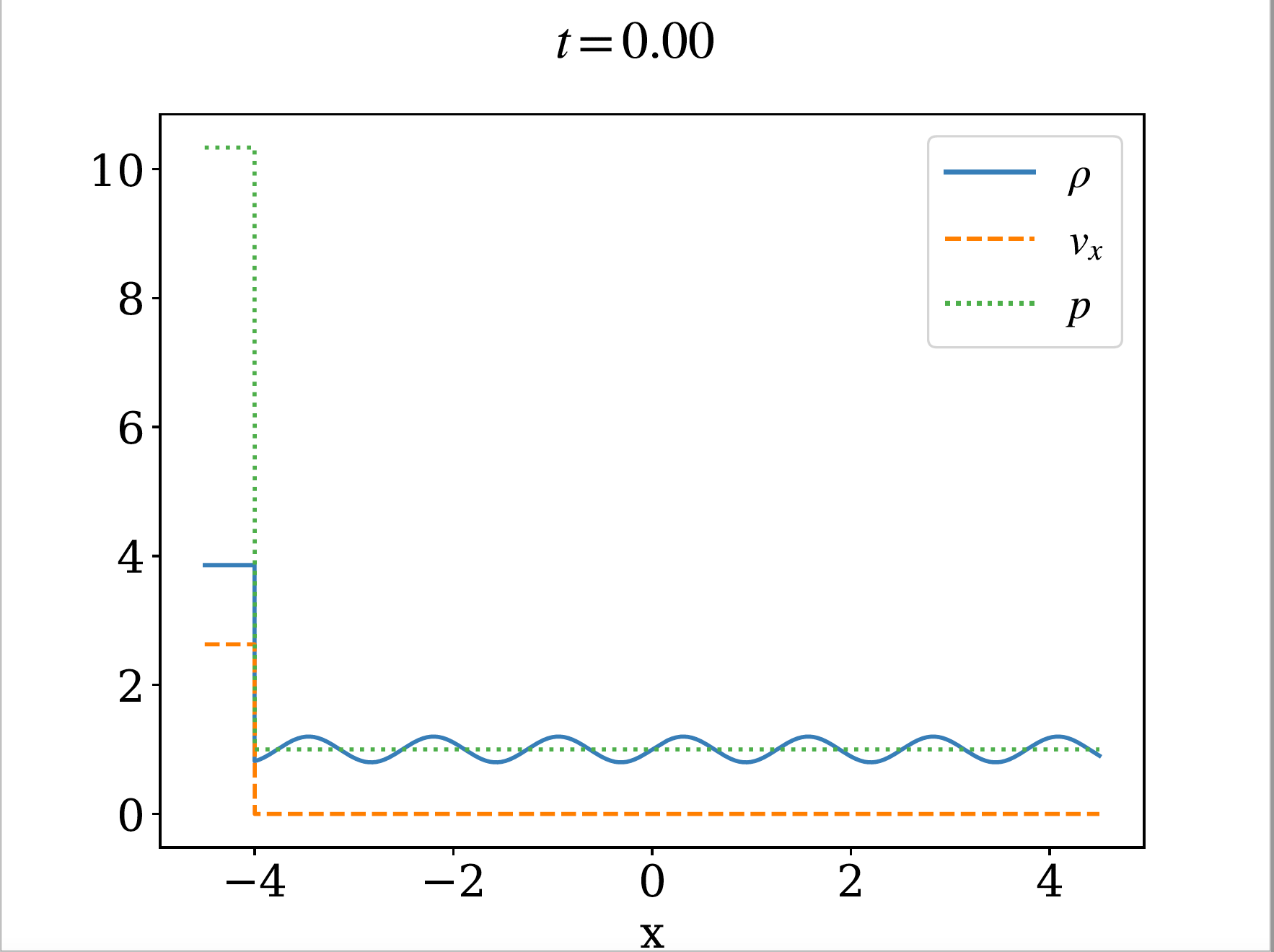} \includegraphics[width=\columnwidth]{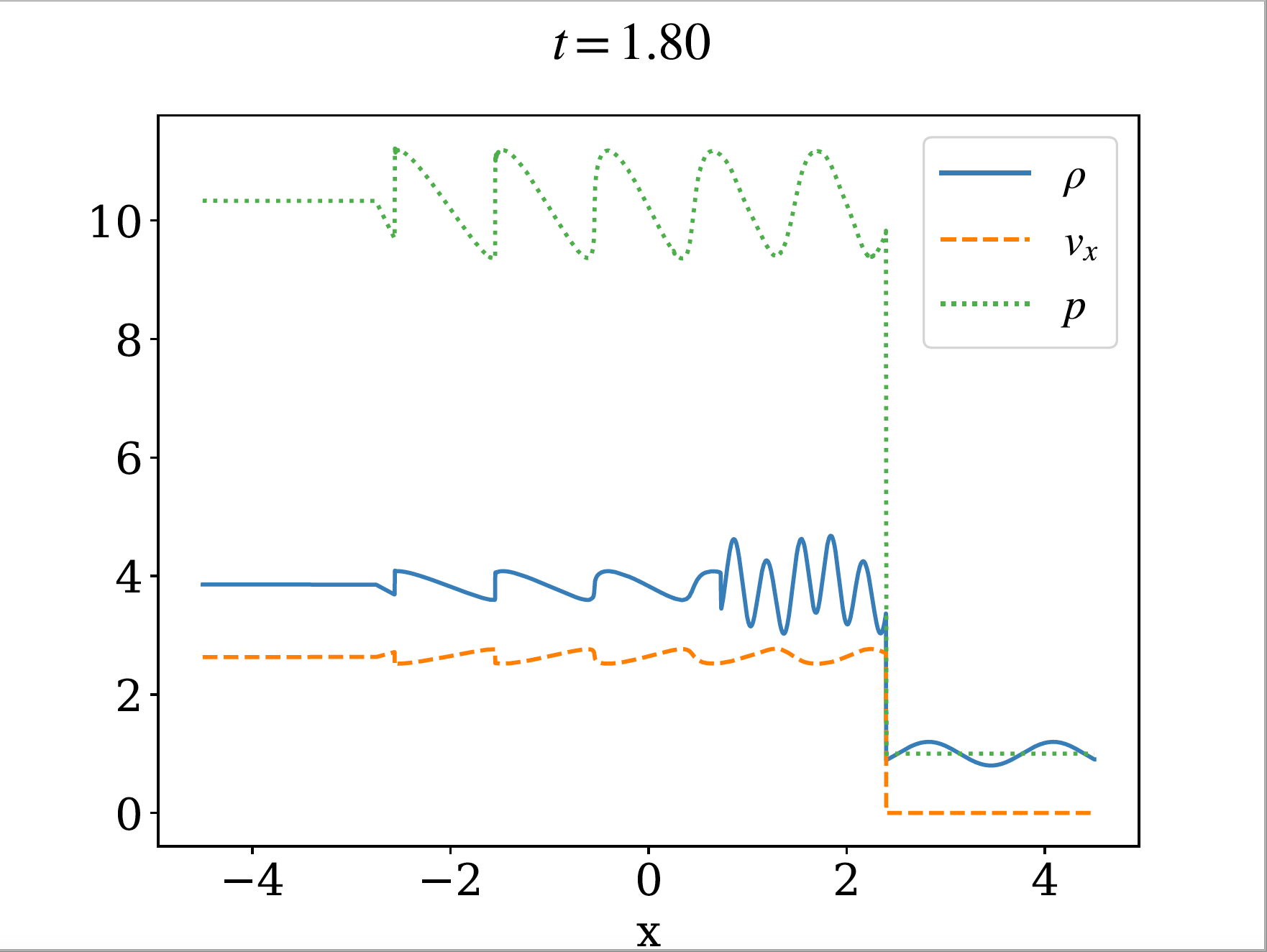} 
    \caption{1D Shu-Osher test. Density (blue solid line), velocity (orange dashed line) and pressure (green dotted line) for the initial time (top panel) and final time (bottom panel). This high resolution numerical solution was obtained using `wenozp5' limiter. An animation is provided.}
    \label{f:shu}
\end{figure}

\begin{figure}
    \centering
    \includegraphics[width=\columnwidth]{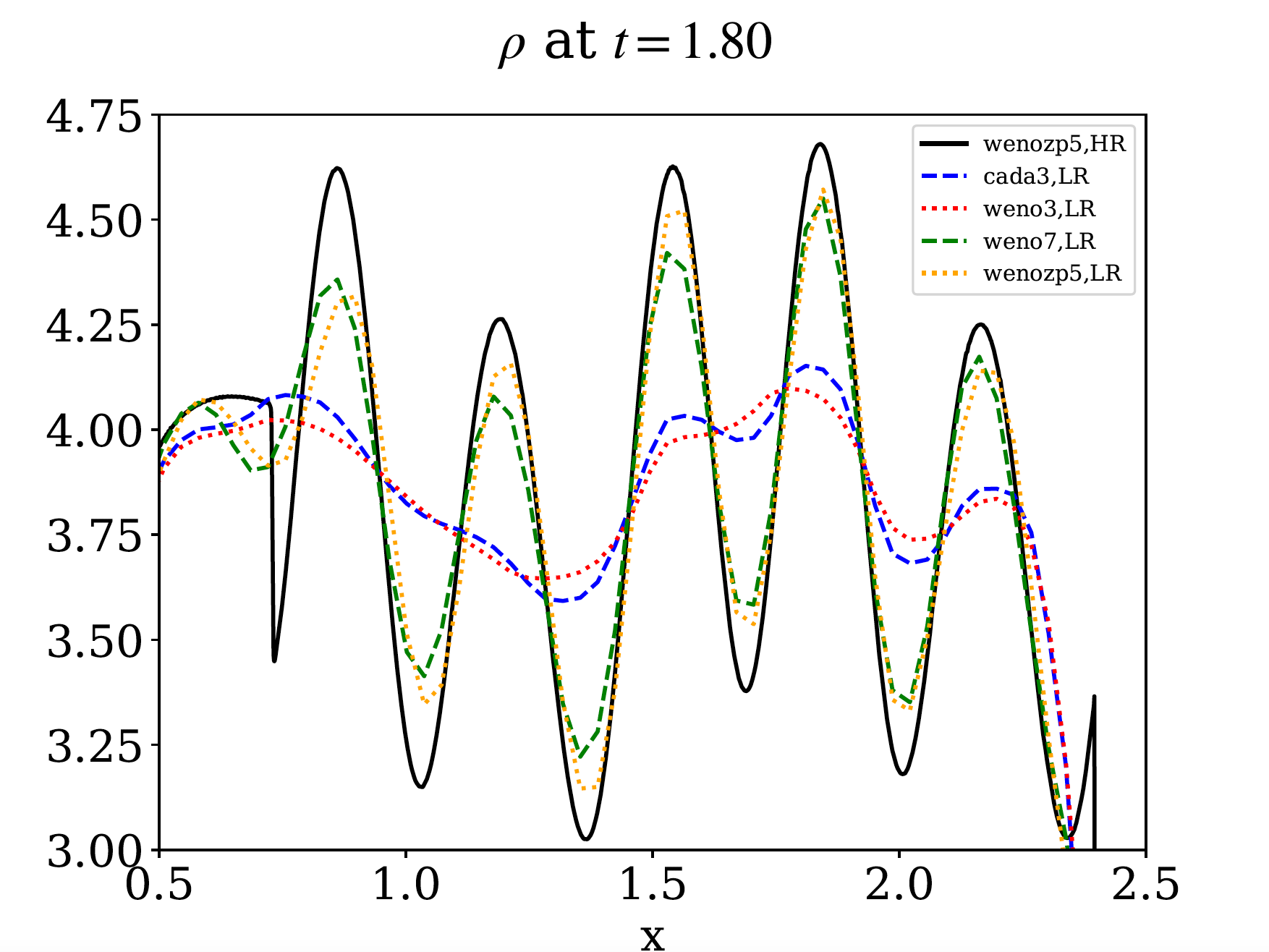} 
    \caption{1D Shu-Osher test. Comparison at final time $t=1.8$ between different type of limiters at low resolution (LR)
to the reference high resolution (HR) using
`wenozp5' limiter (black solid).
We zoom on the density variation for $x$-axis values between 0.5 and 2.5
and $\rho$-values between 3 and 4.75.}
    \label{f:shuzoom}
\end{figure}

Many of the implemented reconstruction and/or limiter choices in \texttt{MPI-AMRVAC} are briefly discussed in its online documentation\footnote{\url{http://amrvac.org/md_doc_limiter.html}, note that we use `limiter' and `reconstruction' in an interchangeable way.}. These are used when doing reconstructions on (usually primitive) variables from cell center to cell edge values, where their reconstructed values quantify local fluxes (on both sides of the cell face). They mostly differ in whether or not they ensure (1) the Total Variation Diminishing (TVD) property on scalar hyperbolic problems or rather build on the Essentially Non-Oscillatory (ENO) paradigm, (2) encode symmetry preservation, (3) achieve a certain theoretical order of accuracy (second or higher order possibilities). Various reconstructions/limiters are designed purely for uniform grids, others are compatible with non-uniform grid stretching. In the \texttt{mod\_limiter.t} module, one currently distinguishes many types as given in Table~\ref{t:limiters}. The choice of limiter impacts the stencil of the method, and hence the number of ghost cells used for each grid block in the AMR structure, as listed in Table~\ref{t:limiters}. In \texttt{MPI-AMRVAC}, the limiter (as well as the discretization scheme) can differ between AMR levels, where one may opt for a more diffusive (and usually more robust) variant at the highest AMR levels.  

\begin{table*}[]
    \centering
    \caption{Reconstruction/Limiter choices in \texttt{MPI-AMRVAC 3.0}, as typically used in the cell-centre-to-cell-face reconstructions. The formal order of accuracy (on smooth solutions), the needed number of ghost cells, and suitable references are indicated as well.}
    \label{t:limiters}
    \begin{tabular}{l l c c r}
        \hline
        \hline
        Limiter Type & Limiter & Order & Ghost Cells & Reference \\[0.7ex]
        \hline
        TVD limiter & `minmod' & 2 & 2 & e.g. \citet{1996Toth,2002leveque} \\
        & `superbee' & 2 & 2 & \citet{1985Roe,2002leveque} \\
        & `woodward' & 2 & 2 & \citet{1977VanLeer,1984Woodward} \\
        & `mcbeta' & 2 & 2 & \citet{1977VanLeer} \\
        & `vanleer' & 2 & 2 & \citet{1974VanLeer} \\
        & `albada' & 2 & 2 & \citet{1982Albada} \\
        & `koren' & 3 & 2 & \citet{1993Koren} \\
         & `ppm' & 3 & 3 & \citet{1984Colella,2005Mignone} \\
         & & 3 & 4 & \citet{2002MillerColella} \\
        \hline
        Beyond TVD & `cada' & 2 & 2 & \citet{2009Cada} \\
        & `cada3' & 3 & 2 & \citet{2009Cada} \\
        & `schmid1' & 3 & 2 & \citet{2016Schmidtmann} \\
        & `schmid2' & 3 & 2 & \citet{2016Schmidtmann}\\
        & `venk'  & 2 & 2  & \citet{1995Venk}\\
         & `mp5' & 5 & 3 & \citet{1997Suresh} \\
        \hline
        ENO-based & `weno3' & 3 & 2 & \citet{1996jiang} \\
                 & `wenoyc3' & 3 & 2 & \citet{2009Yamaleev,2014Arandiga} \\
         & `weno5(nm)' & 5 & 3 & \citet{1996jiang,2009Shu,2018Huang} \\
         & `wenoz5(nm)' & 5 & 3 & \citet{2008Borges,2018Huang} \\
         & `wenozp5(nm)' & 5 & 3 & \citet{2016Acker,2018Huang} \\
         & `weno5cu6' & 6 & 3 & \citet{2018HuangC} \\
         & `teno5ad' & 5 & 3 & \citet{2021Peng} \\
         & `weno7' & 7 & 4 & \citet{2000Balsara} \\
         & `mpweno7' & 7 & 4 & \citet{2000Balsara} \\
        \hline 
    \end{tabular}
\end{table*}

The option to use higher order Weighted ENO (WENO) reconstruction variants has been added recently, and here we show their higher-order advantage using the standard 1D hydrodynamic test from~\cite{1989ShuOsher}. This is run on a 1D domain comprised between $x=-4.5$ and $x=4.5$, and since it is 1D only, we compare uniform grid high resolution (25600 points), with low resolution (256 points) equivalents. This `low resolution' is inspired by actual full 3D setups, where it is typical to use several hundreds of grid cells per dimension. The initial condition in density, pressure and velocity is shown in Fig.~\ref{f:shu}, along with the final solution at $t=1.8$. A shock initially situated at $x=-4$ impacts a sinusoidally varying density field with left and right states as in
\begin{eqnarray*}
\left(\rho,v,p\right)_L & = &\left(3.86,2.63,10.33\right)\,,\\
\left(\rho,v,p\right)_R & = & \left(1.0+0.2\text{sin}(5 x),0,1.0\right)\,.
\end{eqnarray*}
We use an HLL solver \citep{1983Harten} in a three-step time integration, have zero gradient boundary conditions, and set the adiabatic index to $\gamma=1.4$.
In Fig.~\ref{f:shuzoom} we zoom in on the compressed density variation that trails the right-ward moving shock, where the fifth-order `wenozp5' limiter from \citet{2016Acker} is exploited in both high and low resolution. For comparison, low resolution third-order `cada3' \citep{2009Cada}, third-order `weno3', and seventh-order `weno7' \citep{2000Balsara} results show the expected behavior where higher order variants improve the numerical representation of the shock-compressed wave train. All files to reproduce this test are in the folder {\em \texttt{tests/demo/Shu\_Osher\_1D\_HD}}.

\subsubsection{2D Kelvin-Helmholtz: Gas and Gas-dust coupling}\label{sec:gasdust}

The Kelvin-Helmholtz (KH) instability is ubiquitous in fluids, gases and plasmas, and can cause intricate mixing. We here adopt a setup used in a recent study of KH-associated ion-neutral decouplings by~\citet{2019Hillier}, where a reference high resolution hydrodynamic run was introduced as well. We emphasize the effects of limiters in multi-dimensional hydro studies, by running the same setup twice, switching only the limiter exploited. We also demonstrate that \texttt{MPI-AMRVAC} can equally study the same processes in gas-dust mixtures, e.g.\ relevant in protoplanetary disk contexts. All files to reproduce these experiments are available at {\em \texttt{tests/demo/KelvinHelmholtz\_2D\_HD+dust}}.

\paragraph{2D KH and limiters.} The domain $(x,y) \in  [-1.5, 1.5] \times [-0.75,0.75]$, uses a base resolution of 128$\times$64 with 6 levels of refinement, hence we achieve 4096$\times$2048 effective resolution. This should be compared to the uniform grids used in~\citet{2019Hillier}, usually at 2048$\times$1024, but with one extreme run at 16384$\times$8192. Their Fig.~1 shows the density field at a very late time ($t=50$) in the evolution where multiple mergers and coalescence events between adjacent vortices led to large-scale vortices of half the box width, accompanied by clearly turbulent smaller-scale structures. The setup uses a sharp interface at $y=0$, with
\begin{eqnarray}
    y>0: & \,\,\,\,\, \rho_0=1.5, & \,\,\,\,\,v_{\rm x0}=\frac{1}{2.5} \Delta V \,,\\
 y\le0: & \,\,\,\,\, \rho_0=1, & \,\,\,\,\,v_{\rm x0}=-\frac{1.5}{2.5} \Delta V \,,
\end{eqnarray}
where $\Delta V = 0.2$, together with a
uniform gas pressure $p_0={1}/{\gamma}$ where $\gamma=5/3$. The vertical velocity is seeded by white noise with amplitude $10^{-3}$. However, the two runs discussed here use the exact same initial condition, i.e. the $t=0$ data is first generated using a noise realization, and used for both simulations. This demonstrates at the same time the code flexibility to restart from previously generated datafiles, needed to e.g.\ resume a run from a chosen snapshot, which can even be done on a different platform, using a different compiler. Note that the setup here uses a discontinuous interface at $t=0$, which is known to influence and preselect grid-scale fine-structure in the overall nonlinear simulations. \citet{Lecoanet2016} discussed how smooth initial variations can lead to reproducable KH behaviour (including viscosity), allowing to quantify convergence aspects. This is not possible with the current setup, but one can adjust this setup to the \citet{Lecoanet2016} configuration and activate viscosity source terms.

\begin{figure*}
  \FIG{
    \centering
    \includegraphics[width=0.92\textwidth]{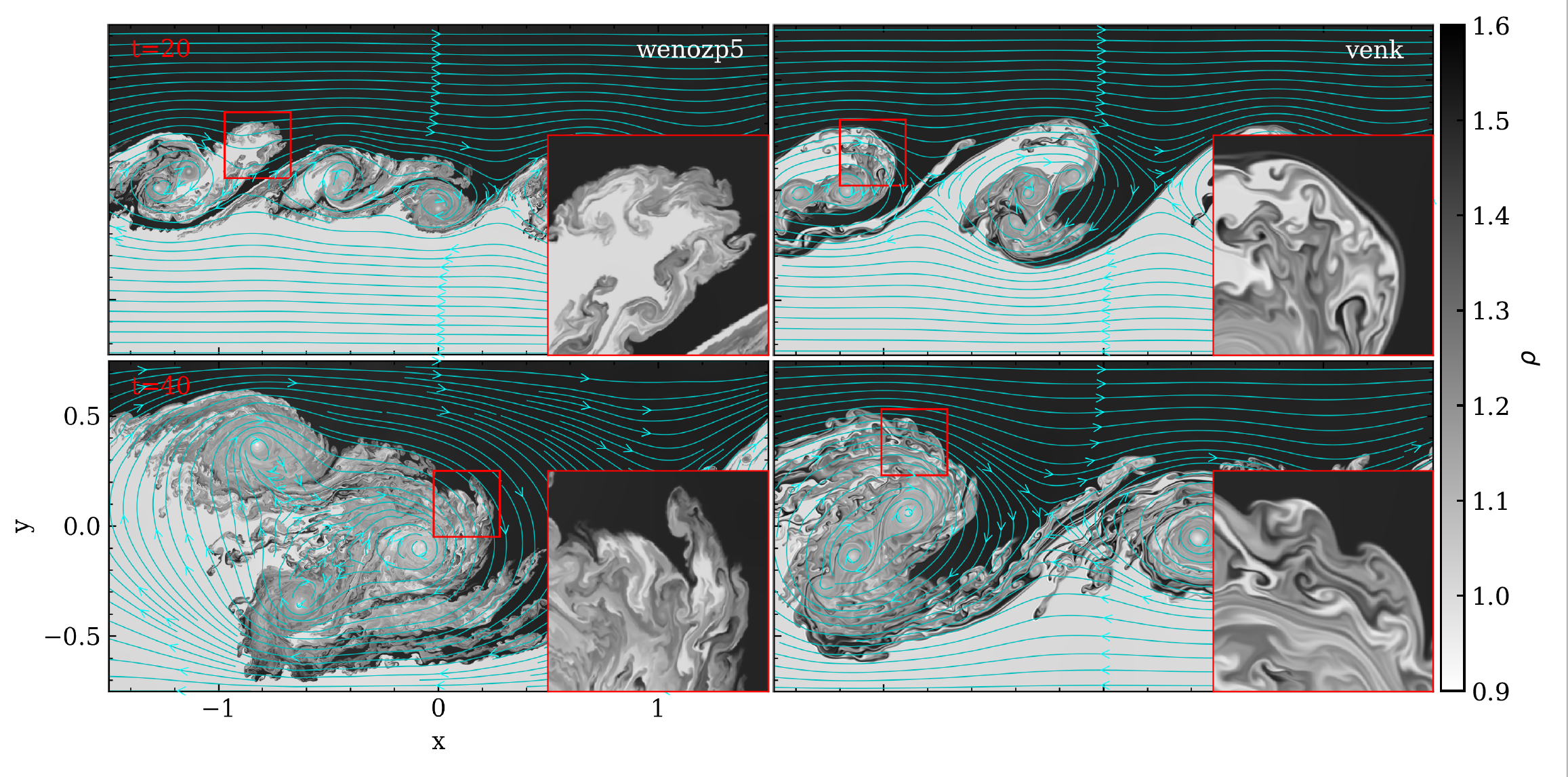}
    \caption{Purely hydrodynamic simulations of a 2D Kelvin-Helmholtz shear layer. The two runs start from the same initial condition, and only deviate due to the use of two different limiters in the center-to-face reconstructions: `wenozp5' (left column), `venk' (right column). We show density views at time t=20 (top row) and t=40 (bottom row). The flow streamlines plotted here are actually computed by {\tt MPI-AMRVAC} with its internal field line tracing functionality through the AMR hierarchy, as explained in Section~\ref{sec:trace}. Insets show zoomed details of the density variations in the red boxes as indicated. An animation is provided.
    }
    \label{fig:KH}
  }
\end{figure*}

We use a three-step time integrator, with periodic sides and closed up/down boundaries (the latter ensured by (a)symmetry conditions).
We use the HLLC scheme (see the review by~\citealt{2019Toro}),  known to improve the baseline HLL scheme \citep{1983Harten} in the numerical handling of density discontinuities. In Fig.~\ref{fig:KH}, we contrast two runs at times $t=20, 40$ that only differ in the limiter exploited, the left column again uses the `wenozp5' limiter \citep{2016Acker}, while at right the Venkatakrishnan \citep{1995Venk} limiter is used, which is a popular limiter on unstructured meshes. While both runs start from the same $t=0$ data, it is clear how the nonlinear processes at play in KH mixing ultimately lead to qualitatively similar, but quantitatively very different evolutions. The limiter is activated from the very beginning due to the sharp interface setup, and the simulation accumulates differences at each timestep. Note that the `wenozp5' run (left panels) clearly shows much more pronounced finer-scale structure than the `venk' run (right panels). Since the setup is using a discontinuous initial condition, some of the fine-structure is not necessarily physical \citep{Lecoanet2016}. If statistical properties specific to the turbulent substructures are of interest, one should exploit the higher order reconstructions, and perform multiple runs at varying effective resolution to fully appreciate physical versus numerical effects. Note that we did not (need to) include any hyperdiffusive terms or treatments here.

\paragraph{Gas-Dust KH evolutions.} The hydrodynamic module of \texttt{MPI-AMRVAC} provides the option to simulate drag-coupled gas-dust mixtures, introducing a user-chosen added number of dust species $n_d$ that differ in their `particle' size.  In fact, every dust species is treated as a pressureless fluid, adding its own continuity and momentum equation for density $\rho_{{\rm d}i}$ and momentum $\rho_{{\rm d}i}\mathbf{v}_{{\rm d}i}$, where interaction from dust species $i\in{1\dots n_d}$ is typically proportionate to the velocity difference $(\mathbf{v}-\mathbf{v}_{{\rm d}i})$, writing $\mathbf{v}$ for the gas velocity. This was demonstrated and used in various gas-dust applications~\citep{2011vanMarle,2012Meheut,2014Hendrix,2014Porth,2015Hendrix,2016Hendrix}. The governing equations as implemented are found in~\citet{2014Porth}, along with a suite of gas-dust testcases. Note that the dust species do not interact with each other, they only interact with the gas.

We here showcase a new algorithmic improvement specific to the gas-dust system: the possibility to handle the drag-collisional terms for the momentum equations through an implicit update. Thus far, all previous {\tt MPI-AMRVAC} gas-dust simulations used an explicit treatment for the coupling, implying that the (sometimes very stringent and erratic) explicit stopping time criterion could slow down a gas-dust simulation dramatically. For \texttt{Athena++}, \citet{2022Huang} recently demonstrated the advantage of implicit solution strategies allowing to handle extremely short stopping time cases. In \texttt{MPI-AMRVAC 3.0}, we now provide an implicit update option for the collisional terms in the momentum equations:
\begin{eqnarray}
\label{eqs:impl_dust}
 & &  (\rho_{{\rm d}i} \mathbf{v}_{{\rm d}i})^{n+1} = T_{{\rm d}i} + \Delta t\left( \alpha_i \rho \rho_{{\rm d}i} \left(\mathbf{v} - \mathbf{v}_{{\rm d}i}\right)\right)^{n+1}\,,\nonumber\\
 & & \qquad\forall i=1..n_d\,,\nonumber\\
 & & (\rho \mathbf{v})^{n+1} = T + \Delta t \left(\sum_{i=1}^{n_d}{\alpha_i \rho \rho_{{\rm d}i} \left(\mathbf{v}_{{\rm d}i}-\mathbf{v} \right)}\right)^{n+1}\,,
\end{eqnarray}
where we denote the end result of any previous (explicit) substage with $T, T_{{\rm d}i}$. Noting that when the collisional terms are linear, i.e. when we have the drag force $\mathbf{f}_{{\rm d}i}=\alpha_i \rho \rho_{{\rm d}i} \left(\mathbf{v}_{{\rm d}i}-\mathbf{v} \right)$ with a {\em constant} $\alpha_i$, one can do an analytic implicit update as follows
\begin{eqnarray}
 \label{eqs:impl_dust2}
 & & \left(\rho_{{\rm d}i} \mathbf{v}_{{\rm d}i}\right)^{n+1} - T_{{\rm d}i} = \frac{N_i}{D}\,,\forall i=1..n_d\,\nonumber\\
 & & \left(\rho \mathbf{v}\right)^{n+1} - T = \frac{N}{D}\,,
\end{eqnarray}
where 
\begin{eqnarray}
 & & D=1+ \sum_{k=1}^{n_d} {d_k (\Delta t)^k}, \nonumber \\
 & & {N_i}=\sum_{k=1}^{n_d} {n_{ik} (\Delta t)^k}\,,\forall i=1..n_d, \nonumber \\
 & & {N}=\sum_{k=1}^{n_d} {n_{k} (\Delta t)^k}\,.
\end{eqnarray}
Although the above is exact for any number of dust species $n_d$ when using proper expansions for $d_k$, $n_k$, and $n_{ik}$, in practice we 
implemented all terms up to second order in $\Delta t$, implying that the expressions used are exact for up to two species (and approximate for higher numbers), where we have
\begin{eqnarray}
 & & d_1 = \sum_{i=1}^{n_d}{\alpha_i (\rho + \rho_{{\rm d}i})}\,, \nonumber\\
 & & d_2 = \sum_{i=1}^{n_d} \sum_{j > i} {\alpha_i \alpha_j \rho(\rho + \rho_{{\rm d}i} + \rho_{{\rm d}j})}\,,\nonumber\\
 \end{eqnarray}
 where $\forall i=1..n_d$ we have
 \begin{eqnarray}
 & & \quad n_{i1} = \alpha_i (\rho_{{\rm d}i} T - \rho T_{{\rm d}i} ) \,;\,\nonumber\\
& & n_{i2} = \sum_{j\ne i}{\alpha_i \alpha_j \rho \left[ \rho_{{\rm d}i} (T_{{\rm d}j} + T) - (\rho + \rho_{{\rm d}j}) T_{{\rm d}i} \right]}\,,\nonumber\\
\end{eqnarray}
 while
 \begin{eqnarray}
 & & n_1 = \sum_{i=1}^{n_d}{\alpha_i (\rho T_{{\rm d}i} -\rho_{{\rm d}i} T )}\,,\nonumber\\
 & & n_2 = \sum_{i=1}^{n_d} \sum_{j > i} {\rho \alpha_i \alpha_j \left[\rho (T_{{\rm d}i} + T_{{\rm d}j}) - (\rho_{{\rm d}i} + \rho_{{\rm d}j}) T \right]}\,.
\end{eqnarray}
Eqs.~(\ref{eqs:impl_dust}) can be  written in a compact form, where the already explicitly updated variables $\vec{T}$ enter the implicit stage:
\begin{equation}
\vec{U}^{n+1} = \vec{T} + \Delta t \vec{P}(\vec{U}^{n+1})\,,
\end{equation}
where
\begin{equation}
\vec{U}=
\begin{pmatrix}
    \rho_{{\rm d}1} \mathbf{v}_{{\rm d}1}\\
        \vdots\\
    \rho_{{\rm d}n} \mathbf{v}_{{\rm d}n}\\
    \rho \mathbf{v}
  \end{pmatrix}\,, \quad   
\vec{P}(\vec{U})=
\begin{pmatrix}
\alpha_1 \rho \rho_{{\rm d}1} \left(\mathbf{v} - \mathbf{v}_{{\rm d}1}\right)\\
\vdots\\
\alpha_n \rho \rho_{{\rm d}n} \left(\mathbf{v} - \mathbf{v}_{{\rm d}n}\right)\\
\sum_{i=1}^{n_d}{\alpha_i \rho \rho_{{\rm d}i} \left(\mathbf{v}_{{\rm d}i}-\mathbf{v} \right)}
\end{pmatrix}\,.    
 \end{equation}
Following the point-implicit approach \citep[see, e.g.][]{toth2012}, $\vec{P}(\vec{U}^{n+1})$ is linearized in time after the explicit update, \begin{equation}
\vec{P}(\vec{U}^{n+1}) = \frac{\partial \vec{P}}{\partial \vec{U}}(\vec{T})\cdot \vec{U}^{n+1}\,.
\end{equation}
The elements of the Jacobian matrix $\partial \vec{P}/\partial \vec{U}$ contain in our case only elements of the form $\alpha_i\rho_{{\rm d}i}\rho$.
After the explicit update, the densities have already the final values at stage ${n+1}$.
Therefore, when $\alpha_i$ is constant, the linearization is actually exact, but when 
$\alpha_i$ also depends on the velocity, the implicit update might be less accurate.

The update of the gas energy density (being the sum of internal energy density $e_{\rm int}$ and kinetic energy density) due to the collisions is done in a similar way and  includes the frictional heating term,
\begin{equation}
   \left(e_{\rm int} + \frac{1}{2} \rho \mathbf{v}^2\right)^{n+1} = T + \Delta t\left( \frac{1}{2}\sum_{i=1}^{n_d}{\alpha_i \rho \rho_{{\rm d}i} \left(\mathbf{v}_{{\rm d}i}^2-\mathbf{v}^2 \right)}\right)^{n+1}\,.
\end{equation}
This is different from the previous implementation which only considered the work done by the momentum collisional terms \citep[see Eq.~(21) in ][]{2014Porth}, but this added
frictional heating term is generally needed for energy conservation \citep{1965Braginskii}.
The implicit update strategy can then be exploited in any multi-stage IMEX scheme, which are described in Section~\ref{sec:IMEX}. 

\begin{figure*}
  \FIG{
  \centering
  \includegraphics[width=\textwidth]{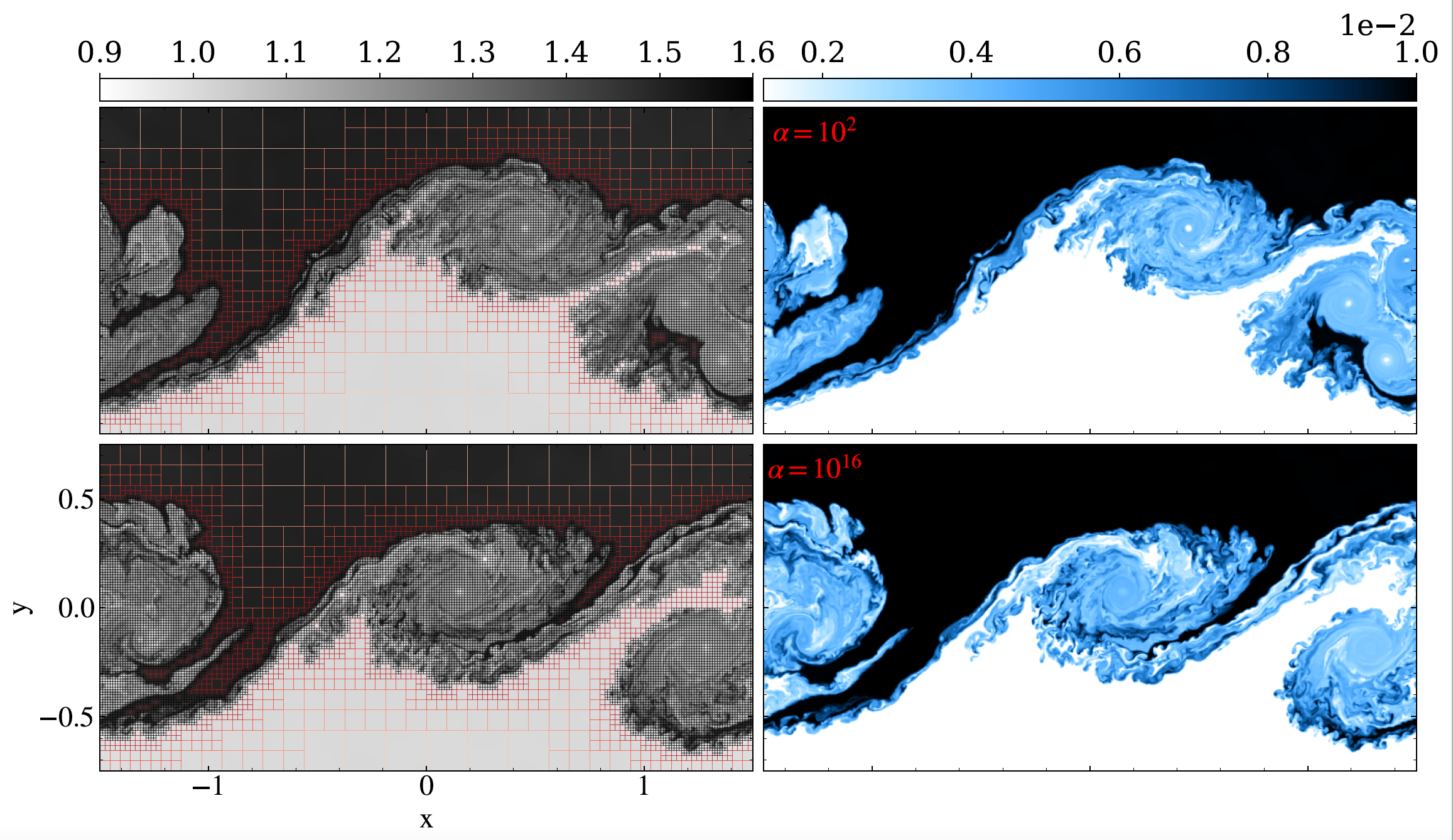}}
  \caption{As in Fig.~\ref{fig:KH}, but this time in a coupled gas-dust evolution, at time t=40, with one species of dust coupled by linear drag with $\alpha_{\rm drag}=10^{2}$ (Top Row). Left column: gas density, Right column: dust density. The limiter used was `wenozp5'. Bottom row: same for much stronger drag coupling $\alpha_{\rm drag}=10^{16}$. An animation is provided.
  }
  \label{fig:KHdust}
\end{figure*}

As a demonstration of its use, we now repeat the KH run from above with one added dust species, where the dust fluid represents a binned dust particle size of $[5 \left(b^{-1/2} - a^{-1/2}\right)/
\left(b^{-5/2} - a^{-5/2}\right)]^{1/2}$ where $a=5$~nm and $b=250$~nm.
We augment the initial condition for the gas with a dust velocity set identical to that of the gas by
${v_{\rm x0d}} = v_{\rm x0}$, but no velocity perturbation in the $y$-direction. The dust density is smaller than the gas density with a larger  density contrast below and above the interface, setting
$\rho_{0d}=\Delta \rho_d$ for $y>0$, $\rho_{0d}=0.1 \Delta \rho_d$ for $y\le0$
where
$\Delta \rho_d=0.01$. 
The time integrator used is a three step ARS3 IMEX scheme.

Results are shown in Fig.~\ref{fig:KHdust}, for two different coupling regimes, which differ in the adopted constant coupling constant $\alpha$, namely $100$ and $10^{16}$. The associated explicit stopping time would scale with $\alpha^{-1}$, so larger $\alpha$ would imply very costly explicit in time simulations. Shown in Fig.~\ref{fig:KHdust} are the AMR grid structure in combination with the gas density variation at left (note that we here used different noise realizations at $t=0$), as well as the single dust species density distribution at right, for $t=40$. The density field for the dust shows similarly intricate fine-structure within the large-scale vortices that have evolved from multiple mergers. We used the same `wenozp5' limiter  as in the left panels of Fig.~\ref{fig:KH}, and one may note how the gas dynamic vortex centers show clearly evacuated dust regions, consistent with the idealized KH gas-dust studies performed by \citet{2014Hendrix}. The top versus bottom panels from Fig.~\ref{fig:KHdust} show that the AMR properly traces the regions of interest, the  AMR criterion being based on density and temperature variables. The highly coupled case with $\alpha=10^{16}$  can be argued to show more fine structure, as the collisions might have an effect similar to the diffusion for the scales smaller than the collisional mean free path \citep{beatrice-rti}.
 Note that \citet{2022Huang} used corresponding $\alpha$ factors between $100-10^8$ on their 2D KH test case, and did not investigate the very far nonlinear KH evolution we address here.

\subsubsection{Thermally unstable evolutions}\label{sec:otc}

In many astrophysical contexts, one encounters complex multiphase physics, where cold and hot material coexist and interact. In solar physics, the million-degree hot corona is pervaded by cold (order 10000 K) condensations that appear as large-scale prominences or as more transient, smaller-scale coronal rain. Spontaneous in-situ condensations can derive from optically thin radiative losses, and  
\citet{2021Hermans} investigated how the precise radiative loss prescription can influence the thermal instability process and its further nonlinear evolution in 2D magnetized settings. In practice, optically thin radiative losses can be handled by the addition of a localized energy sink term, depending on density and temperature, and \texttt{MPI-AMRVAC} provides a choice among 20 implemented cooling tables, as documented in the appendix to \citet{2021Hermans}. The very same process of thermal instability, with its runaway condensation formation, is invoked for the so-called chaotic cold accretion \citep{2013Gaspari} scenario onto black holes, or for the multiphase nature of winds and outflows in Active Galactic Nuclei \citep{2021Waters}, or for some of the fine-structure found in stellar wind-wind interaction zones \citep{2012VanMarle}. Here, we introduce a new and reproducible test for handling thermal runaway in a 2D hydro setting. In \citet{2011MarleKeppens}, we intercompared explicit to (semi)implicit ways for handling the localized source term, and confirmed the exact integration method of \citet{2009Townsend} as a robust means to handle the extreme temperature-density variations that can be encountered. Using this method in combination with the {\tt SPEX\_DM} cooling curve $\Lambda(T)$ (from \citealt{2009Schure}, combined with the low-temperature behaviour as used by \citealt{1972DM}), we set up a double-periodic unit square domain, resolved by a $64\times 64$ base grid, and we allow for an additional 6 AMR levels. We use a five-step SSPRK(5,4) time integration, combined with the HLLC flux scheme, employing the `wenozp5' limiter. We simulate until time $t=7$, where the initial condition is a static (no flow) medium, of uniform pressure $p=1/\gamma$ throughout (with $\gamma=5/3$). The density is initially $\rho=1.1$ inside, and $\rho=1$ outside of a circle of radius $r=0.15$. To trigger this setup into a thermal runaway process, the energy equation not only has the optically thin $\propto \rho^2 \Lambda(T)$ sink term handled by the \citet{2009Townsend} method, but also adds a special energy source term that balances exactly these radiative losses corresponding to the exterior $\rho=1$, $p=1/\gamma$ settings.  A proper implementation where $\rho=1$ throughout would hence stay unaltered forever. Since the optically thin losses (and gains) require us to introduce dimensional factors (as $\Lambda(T)$ requires the temperature $T$ in Kelvin), we introduce units for length $L_u=10^9$ cm, for temperature $T_u=10^6$ K, and for number density $n_u=10^9 \,\mathrm{cm}^{-3}$. All other dimensional factors can be derived from these three.

\begin{figure*}
    \FIG{\centering
\includegraphics[width=\textwidth]{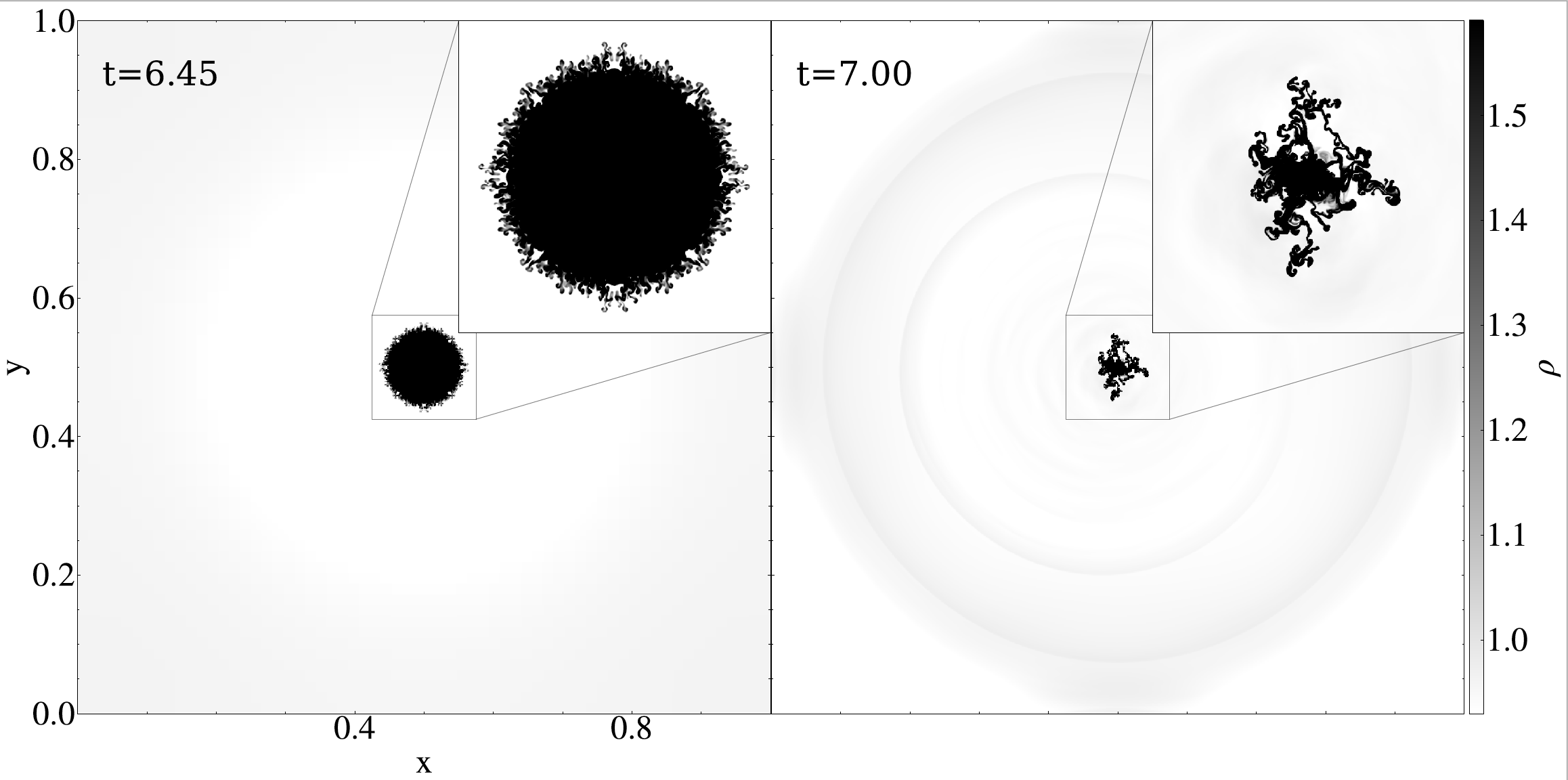} }
    \caption{Density distributions at times $t=6.45$ and $t=7$ for the runaway thermal condensation test. Insets zoom in for details. An animation of this 2D hydro test is provided.}
    \label{f:TI}
\end{figure*}

As losses overwhelm the constant heating term within the circle $r<0.15$, the setup naturally evolves to a largely spherically symmetric, constantly shrinking central density enhancement. This happens so rapidly that ultimately Rayleigh-Taylor driven substructures form on the `imploding' density. Time $t=6.45$ shown in Fig.~\ref{f:TI} typifies this stage of the evolution, where one notices the centrally shrunk density enhancement, and fine structure along its entire edge. Up to this time, our implementation never encountered any faulty negative pressure, so no artificial bootstrapping (briefly discussed in Section~\ref{sec:bootstrap}) was in effect. However, to get beyond this stage, we did activate an averaging procedure on density-pressure when an isolated grid cell did result in unphysical pressure values below $p<10^{-14}$. Doing so, the simulation can be continued up to the stage where a more erratically behaving, highly dynamical and filamentary condensation forms, shown in the right panel of Fig.~\ref{f:TI} at $t=7$ (see also the accompanying movie). A similar hydrodynamic transition - due to thermal instability and its radiative runaway - into a highly fragmented, rapidly evolving condensation is discussed in the appendix of \citet{2021Hermans}, in that case as thermal runaway happens after interacting sound waves damp away due to radiative losses. An ongoing debate \citep[e.g.,][]{2018McCourt,2020Gronke} on whether this process is best described as `shattering' versus `splattering', could perhaps benefit from this simple benchmark test to separate possible numerical from physical influences. The complete setup for this test is available under {\em \texttt{tests/demo/thermal\_instability\_HD}}.

\subsection{MHD tests and applications}

The following three sections illustrate differences due to the choice of the MHD flux scheme (see Table~\ref{tab:schemes_space}) in a 2D ideal MHD shock-cloud setup (Section~\ref{sec:alfven}), differences due to varying the magnetic monopole control in a 2D resistive MHD evolution (Section~\ref{sec:tilt}), as well as a 1D test showcasing ambipolar MHD effects on wave propagation through a stratified magnetized atmosphere (Section~\ref{sec:ambi}). We use the latter test to evaluate the behavior of the various super-time-stepping strategies available in \texttt{MPI-AMRVAC} for handling specific parabolic source additions. This test also employs the more generic splitting strategy usable in gravitationally stratified settings, also adopted recently in \citet{nitin}. We note that the {\tt mhd} module offers many more possibilities than showcased here: we can e.g.\ again drop the energy evolution equation in favor of an isothermal or polytropic closure, can ask to solve for internal energy density instead of the full (magnetic plus kinetic plus thermal) energy density, and have switches to activate anisotropic thermal conduction, optically thin radiative losses, viscosity, external gravity, as well as Hall and/or ambipolar effects. 

\subsubsection{Shock-cloud in MHD: Alfv\'en hits Alfv\'en}\label{sec:alfven}
\begin{figure}
    \FIG{
    \centering
    \includegraphics[width=\columnwidth]{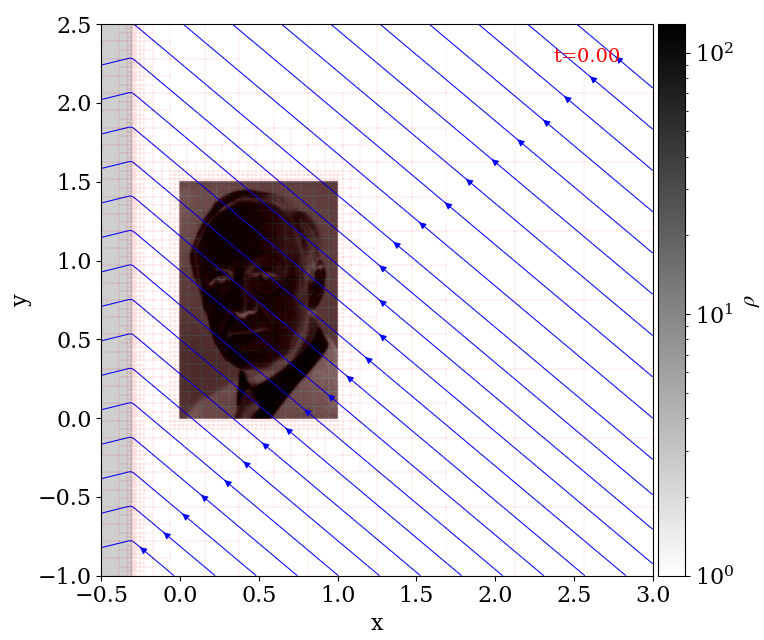} }
    \caption{The initial density variation for the 2D MHD Alfv\'en test: a planar Alfv\'en shock interacts with a density variation set from Alfv\'en's image. The AMR block structure and magnetic field lines are overlaid in red and blue, respectively.}
    \label{f:alfvenini}
\end{figure}
Shock-cloud interactions, where a shock front moves towards and interacts with a prescribed density variation, appear in many standard (M)HD code tests or in actual astrophysical applications. Here, we introduce an MHD shock-cloud interaction where an intermediate (also called Alfv\'en) shock impacts a cloud region that has a picture of Alfv\'en himself imprinted on it. This then simultaneously demonstrates how any multidimensional (2D or 3D) setup can initialize certain variables (in this case, the density at $t=0$) in a user-selected area of the domain by reading in a separate, structured data set: in this case a {\tt vtk}-file containing Alfv\'en's image as a lookup table (in {\em \texttt{tests/demo/AlfvenShock\_MHD2D}} stored as {\tt alfven.vtk}) on a rectangle $[0,1]\times[0,1.5]$. The 2D domain for the MHD setup takes $(x,y)\in [-0.5,3]\times[-1,2.5]$, and the pre-shock static medium is found where $x>-0.3$, setting $\rho=1$ and $p=1/\gamma$ ($\gamma=5/3$). The data read in from the image file is then used to change only the density in the subregion $[0,1]\times[0,1.5]$ to $\rho=1+f_sI(x,y)$ where a scale factor $f_s=0.5$ reduces the image $I(x,y)$ range (containing values between 0 and 256, as usual for image data). Note that the regularly spaced input image values will be properly interpolated to the hierarchical AMR grid, and that this AMR hierarchy auto-adjusts to resolve the image at the highest grid level in use. The square domain is covered by a base grid of size $128^2$, but with a total of 6 grid levels, we achieve a finest grid cell of size $0.0008545$ (to be compared to the $0.002$ spacing of the original image). 

\begin{figure*}
    \FIG{\centering
    \includegraphics[width=\textwidth]{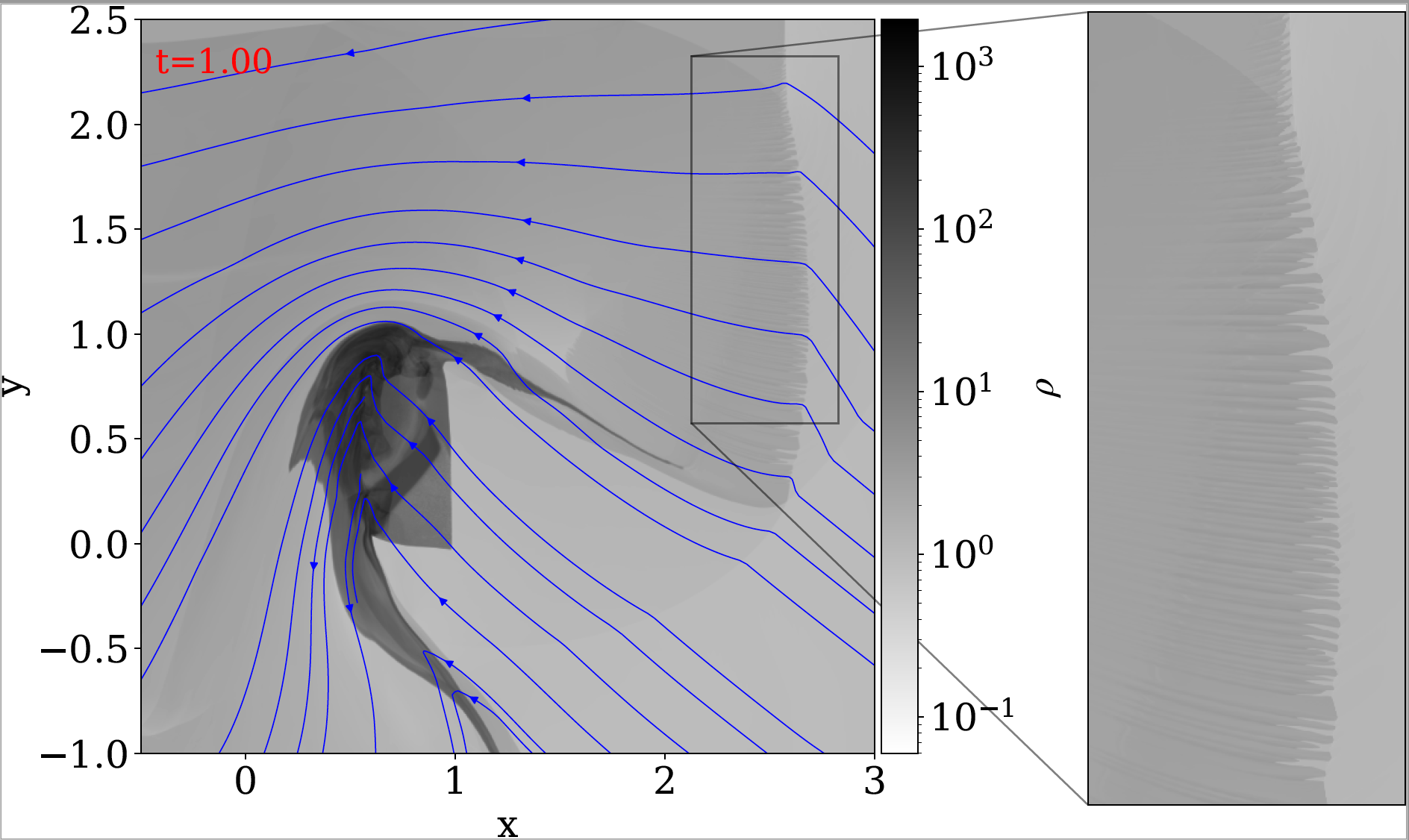} }
    \caption{A reference $t=1$ uniform grid result for the Alfv\'en test using HLLD and constrained transport. Uniform grid of 8192$\times$ 8192. We show density and magnetic field lines, zooming on the corrugated reflected shock at right.}
    \label{f:alfvenct}
\end{figure*}

To realize an Alfv\'en shock, i.e. a shock where the magnetic field lines flip over the shock normal (i.e. the $B_y$ component changes sign across $x=-0.3$), we solve for the intermediate speed solution of the shock adiabatic, parametrized by three input values: (1) the compression ratio $\delta$ (here quantifying the post-shock density); (2) the plasma beta of the pre-shock region; and (3) the angle between the shock normal and the pre-shock magnetic field. Ideal MHD theory constrains $\delta\in[1,(\gamma+1)/(\gamma-1)]$, and these three parameters suffice to then compute the three admissable roots of the shock adiabatic that correspond to slow, intermediate and fast shocks \citep[see, e.g.][]{2017Gurnett}. Selecting the intermediate root of the cubic equation then quantifies the upstream flow speed in the shock frame for a static intermediate shock. Shifting to the frame where the upstream medium is at rest then provides us with values for all post-shock quantities, fully consistent with the prevailing Rankine-Hugoniot conditions. In practice, we took $\delta=2.5$, an upstream plasma beta $2p/B^2= 0.1$, and set the upstream magnetic field using a $\theta=40^\circ$ angle in the pre-shock region, with $B_x=-B\cos(\theta)$ and $B_y=B\sin(\theta)$. This initial condition is illustrated in Fig.~\ref{f:alfvenini}, showing the density as well as magnetic field lines. The shock-cloud impact is then simulated in ideal MHD till $t=1$. Boundary conditions on all sides use continuous (zero gradient) extrapolation.

Since there is no actual (analytical) reference solution for this test, we run a uniform grid case at $8192^2$ resolution, i.e. above the effective $4096^2$ achieved by the AMR settings. Figure~\ref{f:alfvenct} shows the density and the magnetic field structure at $t=1$, where the HLLD scheme was combined with a constrained transport approach for handling magnetic monopole control. Our implementation of the HLLD solver follows \citet{2005Miyoshi} and \citet{2016Guo}, while divergence control strategies are discussed in the next section~\ref{sec:tilt}. A noteworthy detail of the setup involves the corrugated appearance of a rightward-moving shock front that relates to a reflected shock front that forms at first impact. It connects to the original rightward moving shock in a triple point still seen for $t=1$ at the top right $(x,y)\approx (2.6,2.2)$. This density variation, shown also in a zoomed view in Fig.~\ref{f:alfvenct}, results from a corrugation instability (that develops most notably beyond $t=0.7$). 

\begin{figure*}
    \FIG{\centering
    \includegraphics[width=\textwidth]{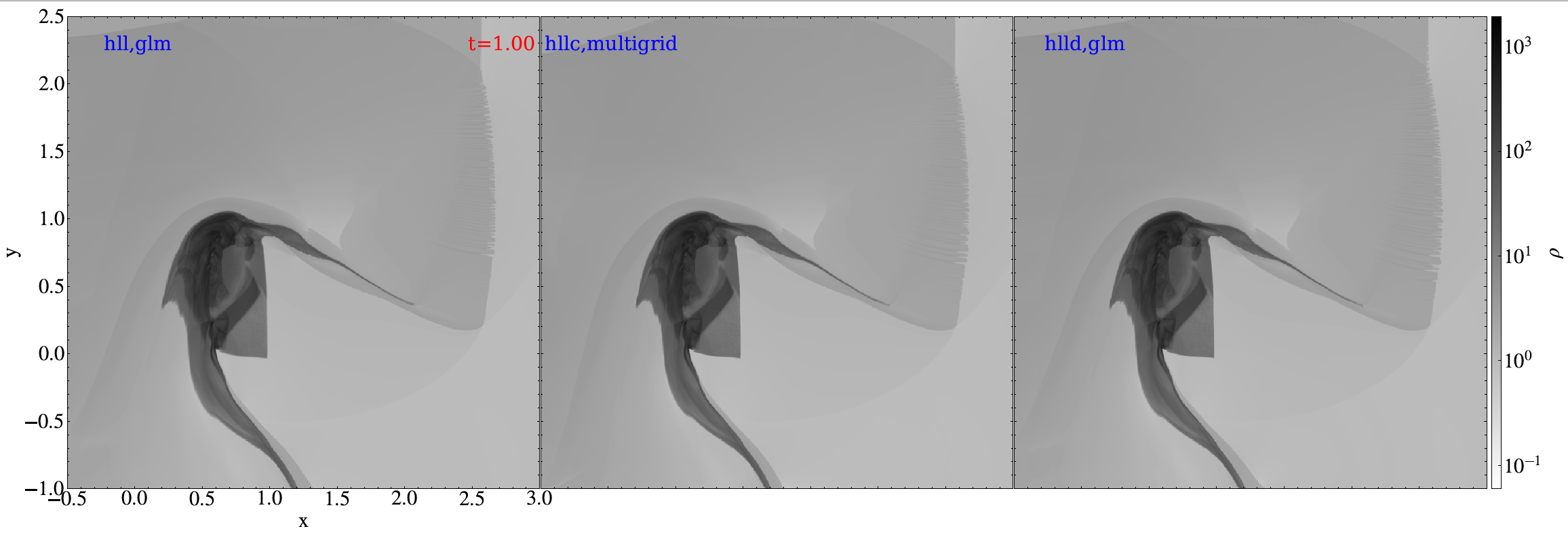} }
    \caption{The density view of the shock-cloud test, where an intermediate Alfv\'en shock impacts an `Alfv\'en' density field. Left: HLL and glm; Middle: HLLC and multigrid; Right: HLLD and glm. Compare to the reference run from Fig.~\ref{f:alfvenct}. An animation is provided.}
    \label{f:alfven4}
\end{figure*}

Fig.~\ref{f:alfven4} shows the final density distribution obtained with three different combinations of flux schemes, using AMR. We always employed a SSPRK(3,3) three-step explicit time marching with a `koren' limiter \citep{1993Koren}, but varied the flux scheme from HLL, over HHLC, to HLLD. The HLL and HLLD variants used the hyperbolic generalized lagrange multiplier (or `glm') idea from \citet{2002Dedner}, while the HLLC run exploited the recently added multigrid functionality for elliptic cleaning (see next section and \citealt{2019Teunissen}). The density views shown in Fig.~\ref{f:alfven4} are consistent with the reference result, and all combinations clearly demonstrate the corrugation of the reflected shock. 
We note that all the runs shown here did use a bootstrapping strategy (see Section~\ref{sec:bootstrap}) to recover automatically from local negative pressure occurrences (they occur far into the nonlinear evolution), where we used the averaging approach whenever one encounters a small pressure value below $10^{-7}$.

\subsubsection{Divergence control in MHD}\label{sec:tilt}

Here, we simulate a 2D resistive MHD evolution, that uses a uniform resistivity value $\eta=0.0001$. The simulation exploits a $(x,y)\in [-3,3]^2$ domain, with base resolution $128^2$ but effective resolution $1024^2$ (4 AMR levels). Always using a five-step SSPRK(5,4) time integration, the HLLC flux scheme, and a `mp5' limiter \citep{1997Suresh}, we simulate till $t=9$ from an initial condition where an ideal MHD equilibrium is unstable to the ideal tilt instability.
We use this test to show different strategies available for discrete magnetic monopole control, and how they lead to overall consistent results in a highly nonlinear, chaotic reconnection regime. This latter regime was used as a challenging test for different spatial discretizations (finite volume or finite differences) in \citet{2013Keppens}, and shown to appear already at ten-fold higher $\eta=0.001$ values. 

The initial density is uniform $\rho=1$, while the pressure and magnetic field derive from a vector potential $\bfB=\nabla\times A(r,\theta)\mathbf{e}_z$ where $(r,\theta)$ denote local polar coordinates. In particular,
\begin{equation}
A(r,\theta)= \left\{ \begin{array}{ccc} 
c \,J_1(r r_0) \cos(\theta) & \,\,\,\,\,\,\,\, & r\leq 1 \,,\\
 \left(r-\frac{1}{r}\right) \cos(\theta) & \,\,\,\,\,\,\,\,  &  r>1\,,
\end{array}\right.
\end{equation}
where $r_0=3.8317$ denotes the first root of the Bessel function of the first kind $J_1$. This makes the magnetic field potential exterior to the unit circle, but non force-free within. An exact equilibrium where pressure gradient is balanced by Lorentz forces can then take the pressure as the constant value $p_0=1/\gamma$ outside the unit circle, while choosing $p=p_0+0.5 [r_0 A(r,\theta)]^2$ within it. The constant was set to $c=2/(r_0 J_0(r_0))$. This setup produces two islands corresponding to anti-parallel current systems perpendicular to the simulated plane, which repel. This induces a rotation and separation of the islands whenever a small perturbation is applied: this is due to the ideal tilt instability (also studied in \citealt{2014Keppens}). A $t=0$ small perturbation is achieved by having an incompressible velocity field that follows $\bfv=\nabla\times \epsilon \exp(-r^2)\mathbf{e}_z$ with amplitude $\epsilon=10^{-4}$. 

This test case, fully provided under {\em \texttt{tests/demo/Tilt\_Instability\_MHD2D}} employs a special boundary treatment, where we extrapolate the primitive set of density, velocity components and pressure from the last interior grid cell, while both magnetic field components adopt a zero normal gradient extrapolation (i.e. a discrete formula $y_i=(-y_{i+2}+4y_{i+1})/3$ to fill ghost cells at a minimal edge, i.e. a left or bottom edge, and some analogous formula at maximal edges). This is done along all 4 domain edges (left, right, bottom, top). We note that ghost cells must ultimately contain correspondingly consistent conservative variables (density, momenta, total energy and magnetic field). 

\begin{figure*}
    \FIG{
    \includegraphics[width=0.98\textwidth]{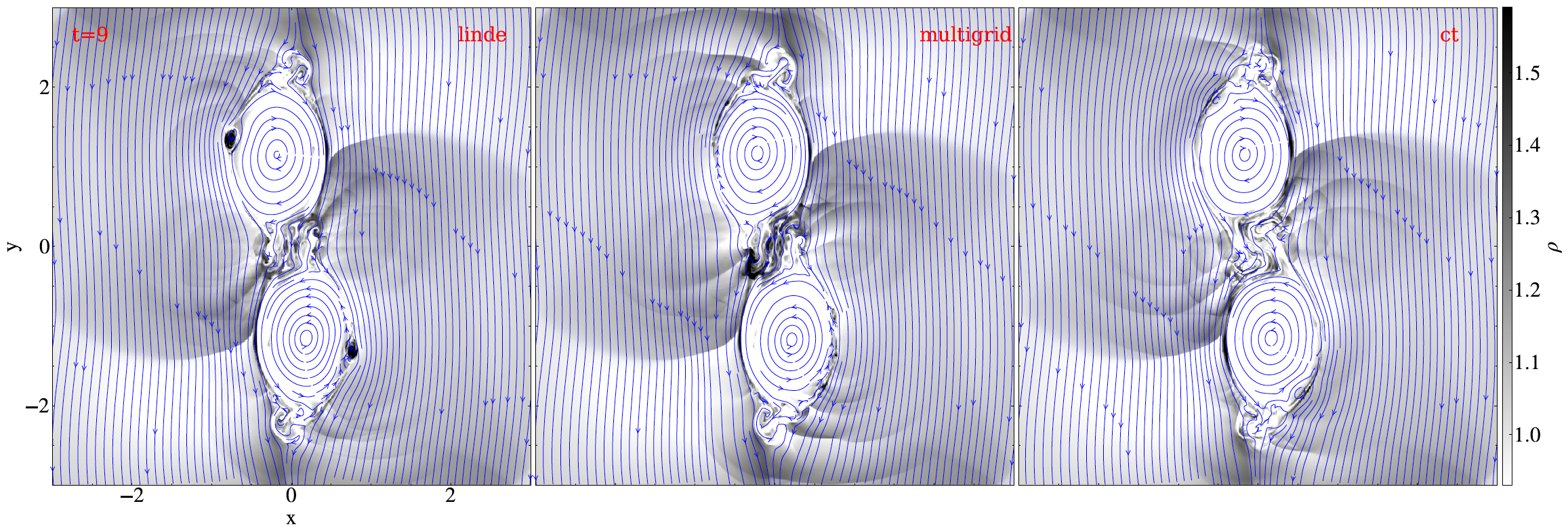} 
    \includegraphics[width=0.98\textwidth]{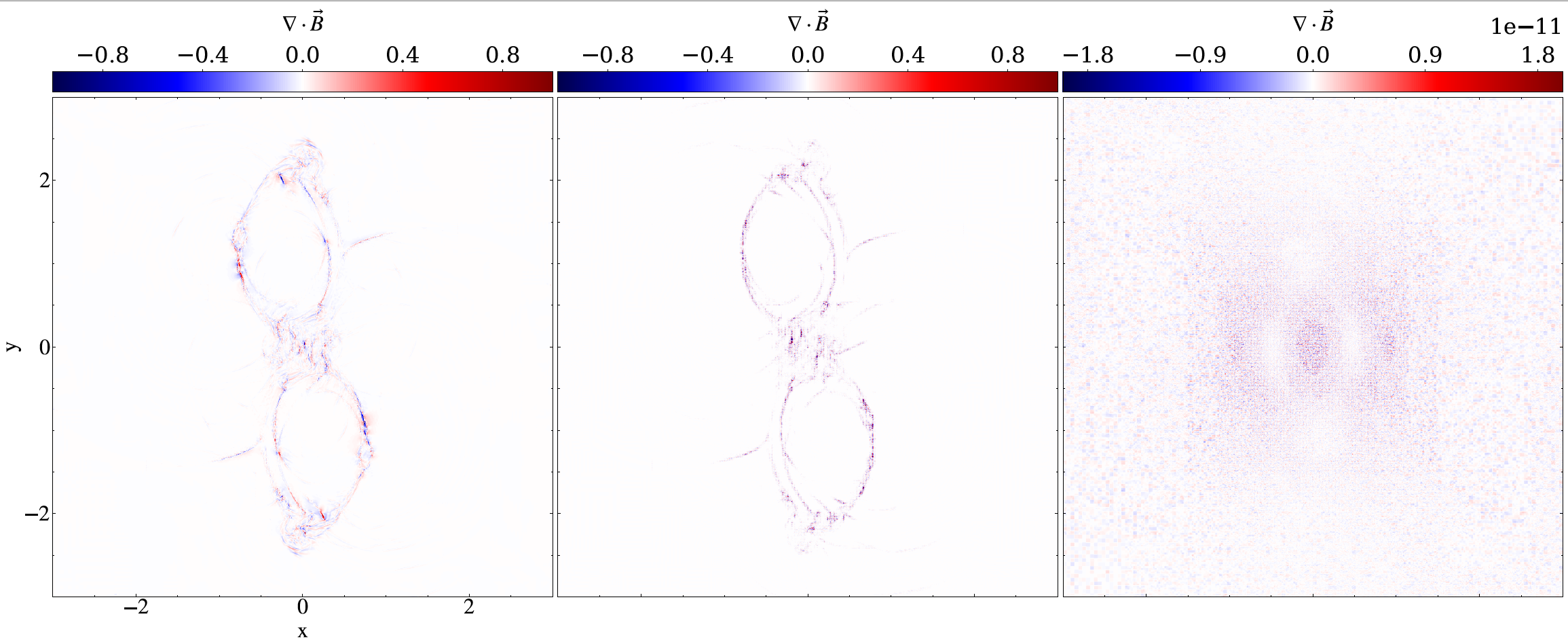} 
    }
    \caption{Snapshots at time $t=9$ for the resistive tilt evolution, using different magnetic field divergence cleaning methods: `linde', `multigrid' and `ct'. First row: density. The magnetic field lines are overplotted with blue lines, and as in Fig.~\ref{fig:KH}, these are computed by \texttt{MPI-AMRVAC} by field line tracing (see Section \ref{sec:trace}). Second row: Divergence of magnetic field. An animation is provided.}
    \label{f:tilt}
\end{figure*}

We use this test to highlight differences due to the magnetic monopole control strategy, for which \texttt{MPI-AMRVAC 3.0} offers a choice between 10 different options. These are listed in Table~\ref{t:divb}, along with relevant references. Note that we provide options to mix strategies (e.g.\ `lindeglm' both diffuses monopole errors in a parabolic fashion and uses an added hyperbolic variable to advect monopoles).  There is a vast amount of literature related to handling monopole errors in combination with shock capturing schemes, e.g.\ the seminal contribution by \citet{2000Toth} discusses this at length for a series of stringent ideal MHD problems. Here, we demonstrate the effect of three different treatments on a resistive MHD evolution where in the far nonlinear regime of the ideal tilt process, secondary tearing events can occur along the edges of the displaced magnetic islands. These edges correspond to extremely thin current concentrations, and the $\eta=0.0001$ value ensures we can get chaotic island formation. We run the setup as explained above with three different strategies, namely `linde', `multigrid' and `ct'. The `linde' strategy was already compared on ideal MHD settings (a standard 2D MHD rotor and Orszag-Tang problem) in \citet{Keppens03}, while the constrained transport strategy is adopted in analogy to its implementation in the related GR-RMHD {\tt BHAC} code \citet{2019Olivares} with an additional option of using the contact-mode upwind constrained transport method by \citet{2005Gardiner} . Note that the use of `ct' requires us to handle the initial condition, as well as the treatment of the special boundary extrapolations, in a staggered-field tailored fashion, to ensure no discrete monopoles are present from initialization or boundary conditions. The `multigrid' method realizes the elliptic cleaning strategy as mentioned originally in \citet{1980BB} on our hierarchical AMR grid. This uses a geometric multigrid solver to handle Poisson's equation $\nabla ^2 \phi =\nabla \cdot \bfB_{\mathrm{before}}$, followed by an update $\bfB_{\mathrm{after}} \leftarrow \bfB_{\mathrm{before}} -\nabla \phi$, as described in \citet{2019Teunissen}.

\begin{table}[]
    \centering
    \caption{Options for $\nabla\cdot\bfB$ control in \texttt{MPI-AMRVAC 3.0}. Some of these come along with different options in terms of their control parameters or their detailed algorithmic implementation.}
    \label{t:divb}
    \begin{tabular}{l l}
        \hline
        \hline
        Monopole Control & Reference  \\[0.7ex]
        \hline
  none  &  - \\
  powel &  \citet{1999Powell}\\
  janhunen & \citet{2000Janhunen} \\
  glm & \citet{2002Dedner} \\
  linde & \citet{Keppens03} \\
  lindejanhunen &  - \\
  lindepowel &  - \\
  lindeglm &  - \\
  ct &  as in {\tt BHAC}, \citet{2019Olivares} \\
  multigrid & \citet{2019Teunissen} \\
        \hline 
    \end{tabular}
\end{table}

The evolution of the two spontaneously separating islands occurs identical for all three treatments, and it is noteworthy that all runs require no activation of a bootstrap strategy at all, i.e. always produce positive pressure and density values. We carry out all three simulations up to $t=9$, and our endtime is shown in Fig.~\ref{f:tilt}. The top panels show the density variations (the density was uniform initially), and one can see many shock fronts associated with the small-scale magnetic islands that appear. Differences between the three runs manifest themselves in where the first secondary islands appear, and how they evolve with time. This indicates how the $1024^2$ effective resolution, combined with the SSPRK(5,4)-HLLD-`mp5' strategy still is influenced by numerical discretization errors (numerical `resistivity'), although $\eta=0.0001$. Relevant length scales of interest are the cross-sectional size of the plasmoids obtained, which should be resolved by at least several tens of grid cells. A related study of 2D merging flux tubes showing plasmoid formation \citep{2019RipperdaB} in resistive, relativistic MHD setting, noted that effective resolutions beyond $8000^2$ were needed to confidently obtain strict convergence at high Lundquist numbers. 

We also plot the discrete divergence of the magnetic field in the bottom panels. Obviously, the rightmost `ct' variant realizes negligible (average absolute values at $10^{-12}-10^{-11}$ throughout the entire evolution) divergence in its pre-chosen discrete monopole evaluation. Because of a slow accumulation of roundoff errors due to the divergence-preserving nature of the constrained transport method, this divergence can become larger than machine-precision zero, but remains very low. However, in any other discrete evaluation for the divergence, also the `ct' run displays monopole errors of similar magnitude and distribution as seen in both leftmost bottom panels of Fig.~\ref{f:tilt}. Indeed, truncation-related monopole errors may approach unity in the thinning current sheets, at island edges, or at shock fronts. Note that the fieldlines as shown in the top panels have been computed by the code's fieldtracing module discussed in section~\ref{sec:trace}. 

\subsubsection{Super-time-stepping and stratification splitting}\label{sec:ambi}

\begin{figure*}
  \FIG{
  \centering
  \includegraphics[width=0.9\textwidth]{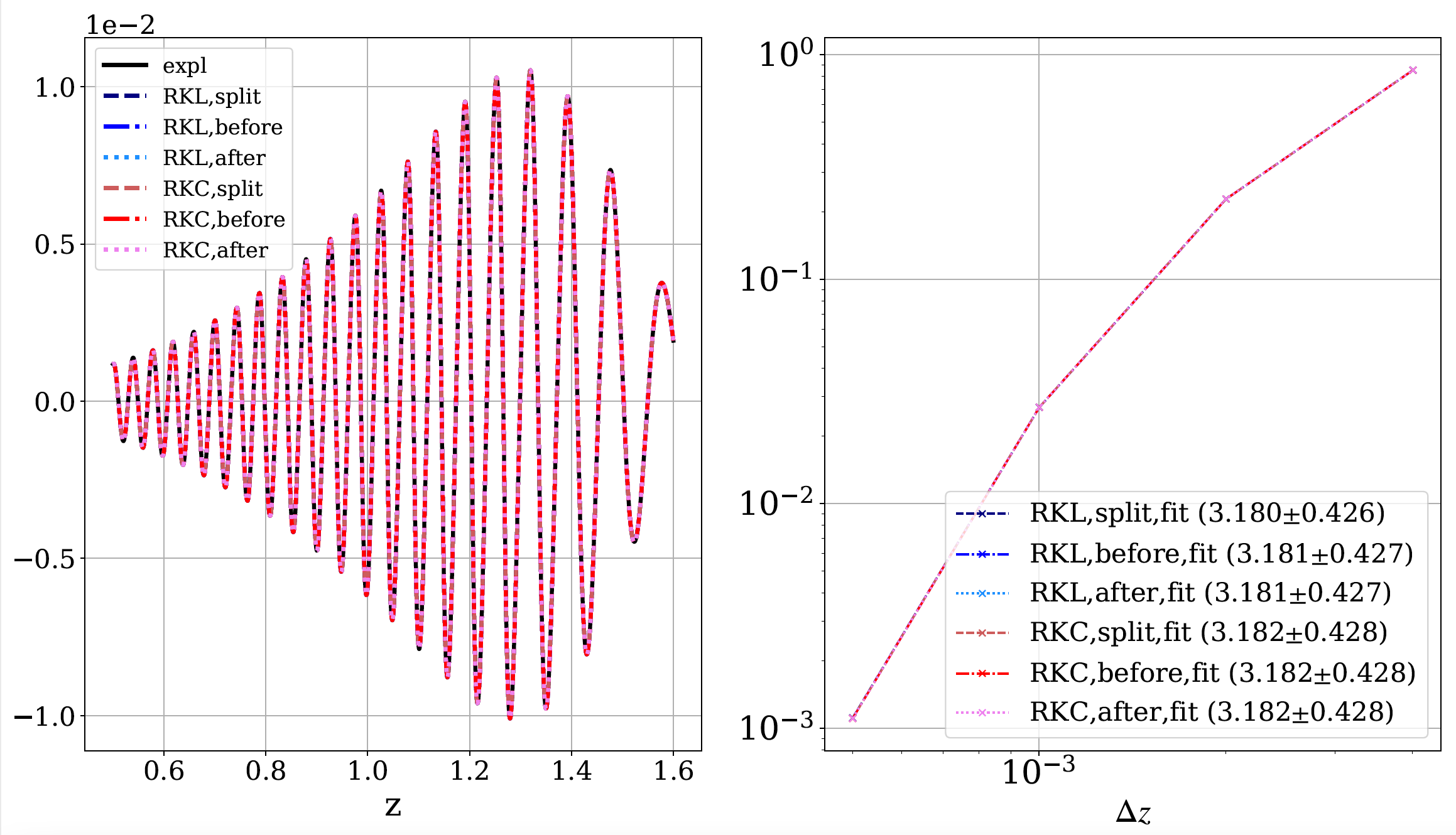}
  }
  \caption{Left: Vertical velocity profile of the 1.75D ambipolar MHD wave test, for the two STS and three different splitting approaches (and an explicit reference run). Right: Normalized error $\mathcal{E}$ from Eq.~(\ref{eq:error}) as a function of the cell size, comparing 
  the numerical solution obtained using STS with a reference numerical solution obtained in an explicit implementation. All variants produce nearly identical results, such that all curves seem overlapping.}
  \label{fig:sts}
\end{figure*}

In a system of PDEs, parabolic terms may impose a very small timestep for an explicit time advance strategy, as 
$\Delta t\propto \Delta x^2$, according to the CFL condition. In combination with AMR, this can easily become too restrictive. 
This issue can be overcome in practice by the super-time-stepping (STS) technique which allows the use of a relatively large (beyond the $\Delta x^2$ restriction) explicit super-timestep $\Delta t_{s}$ for the parabolic terms, by subdividing $\Delta t_{s}$ into carefully chosen smaller substeps. This $\Delta t_{s}$ can e.g.\ follow from the hyperbolic terms in the PDE alone, when parabolic and hyperbolic updates are handled in a split fashion. Super-time-stepping across $\Delta t_{s}$ involves an $s$-stage Runge-Kutta scheme, and its number of stages $s$ and the coefficients used in each stage get adjusted to ensure stability and accuracy. With the $s$-stage Runge-Kutta in a two-term recursive formulation, one can determine
the substep length
by writing the amplification factor for each substep as one involving an orthogonal family
of polynomials that follow a similar two-term recursion. The free parameters involved can be fixed by matching the Taylor expansion of the solution to the desired accuracy.
The use of either Chebyshev or Legendre polynomials
gives rise to two STS techniques described in the literature: RKC 
 \citep{1996Vasilios} and RKL \citep{2014Meyer}. The latter second-order accurate RKL2 variant was demonstrated on stringent anisotropic thermal conduction in multi-dimensional MHD settings by \citet{2014Meyer}, and in \texttt{MPI-AMRVAC}, the same strategy was first used in a 3D prominence formation study \citep{2016XiaRK} and a 3D coronal rain setup \citep{2017Xiaetal}. We detailed in \citet{2018Xia} how the discretized parabolic term for anisotropic conduction best uses the slope-limited symmetric scheme introduced by \citet{2007Sharma}, to preserve monotonicity. RKL1 and RKL2 variants are also implemented in {\tt Athena++} \citep{2020Stone}. RKC variants were demonstrated on Hall MHD and ambipolar effects by \citet{2006Osullivan} and \citet{2007Osullivan}, and used for handling ambipolar diffusion in MHD settings in the codes {\tt MANCHA3D} \citep{2018Gonzalez} and {\tt Bifrost} \citep{2020Nobrega}.

The STS method eliminates the timestep restriction of explicit schemes
and it is faster than standard subcycling.
As pointed out in \cite{2014Meyer}, compared to the RKC methods, the RKL variant ensures stability during every substep (instead of ensuring stability at the end of the super-time-step); have a larger stability region; do not require adjusting the final timestep (roundoff errors) and are more efficient (smaller number of subcycles). However, RKL methods require four times more storage compared to the RKC methods.

Meanwhile, both STS methods have been implemented in {\tt MPI-AMRVAC 3.0} and could be used for any parabolic source term. We specifically use STS for (anisotropic) thermal conductivity and ambipolar effects in MHD. The strategy can also be used for isotropic HD conduction or in the plasma component of a plasma-neutral setup. 
There are three splitting strategies to add the parabolic source term in {\tt MPI-AMRVAC}: before the divergence of fluxes are added (referred to as ``before''), after (``after'') or in a split (``split'') manner, meaning that the source is added for half a timestep before and half a timestep after the fluxes.

As a demonstration of the now available STS-usage for ambipolar effects, we perform a 1D MHD test of a fast wave traveling upwards in a gravitationally stratified atmosphere where partial ionization effects are included through the ambipolar term. Due to this term, such a wave can get damped as it travels up. In the following, we tested both RKC and RKL methods combined with the three strategies ``before'', ``after'', and ``split'' for adding the source. 

The setup is similar to that employed for a previous study of the ambipolar effect on MHD waves in a 2D setup in \cite{2021Popescu}.
The MHD equations solved are for (up to nonlinear) perturbations only, where the variables distinguish between equilibrium (a hydrostatically stratified atmosphere with fixed pressure $p_0(z)$ and density $\rho_0(z)$) and perturbed variables, as described in 
\cite{nitin} (see Eqs.~(4)-(9)). 
The geometry adopted is 1.75~D (i.e. all three vector components are included, but only 1D $z$-variation is allowed). The background magnetic field is horizontal, with a small gradient in the magnetic pressure that balances the gravitational equilibrium. It is important to note that the ambipolar diffusion terms are essential in this test, in order to get wave damping, since a pure MHD variant would see the fast wave amplitude increase, in accord with the background stratification. Ambipolar damping gets more important at higher layers, as the adopted ambipolar diffusion coefficient varies inversely with density-squared. \cite{2021Popescu} studied cases with varying magnetic field orientation, and made comparisons between simulated wave transformation behavior and approximate local dispersion relations. Here, we use a purely horizontal field and retrieve pure fast mode damping due to ambipolar diffusion.

We performed spatio-temporal convergence tests where we ran the simulation using an explicit implementation and 3200 grid points, having this as a reference solution. 
Left panel in Fig.~\ref{fig:sts} shows the reference numerical solution in its vertical velocity profile $v_z(z)$ at $t=0.7$. This panel also overplots the numerical solution for 3200 points for the six STS cases and
we see how the seven solutions overlap.
The right panel of Fig.~\ref{fig:sts} shows the normalized error 
\begin{equation}
\mathcal{E}= \sqrt{\sum_{i=1}^N{(u\left[ i\right]-r\left[ i\right])^2}/\sum_{j=i}^N{r\left[ i\right]^2}} \,, \label{eq:error}
\end{equation}
as a function of the cell size $\Delta z=
\{5 \times 10^{-4}, 2.5 \times 10^{-4}, 1.25 \times 10^{-4}, 6.25 \times 10^{-5}\}$, where
$u$ is the numerical solution obtained using STS and $r$ is the reference numerical solution. Then we ran simulations using all six STS combinations using 3200, 1600, 800 and 400 points. We can observe that in all six cases the error curve is the same, and shows an order of convergence larger than 3. We used HLL flux scheme with a `cada3' limiter \citep{2009Cada}. The temporal scheme was a  SSPRK(3,3) three-step explicit time.

\begin{table*}
  \centering
  \caption{Comparison of computational time between explicit, RKL and RKC methods (always exploiting 8 cores).}
  \begin{tabular}{llll}
  \hline
  \hline  
  Method & Timestep & Number of iterations & Computational time\\
  \hline  
  Explicit & 3.38  $\times$  10$^{-6}$ & 295880 & 233933 s\\
RKL(split) &  4.24 $\times$ 10$^{-5}$ & 23589 & 17196 s\\ 
  RKC (split) &  4.24 $\times$ 10$^{-5}$ & 23589 & 14367 s\\ 
  \hline
  \end{tabular}
  \label{table:comp_ambi}
\end{table*}

Table~\ref{table:comp_ambi} shows the computational cost of this simulation, run with the same number of cores using an explicit implementation and the two variants of the STS technique.
We can observe that when the STS technique is employed, the computational time drops by a factor $>10$, being slightly smaller for RKC. The test can be found in {\tt tests/demo/AmbipolarMHD\_fastwave\_1D}.

\section{IMEX variants}\label{sec:IMEX}

The generic idea of Implicit-Explicit time integrators is to separate off all stiff parts for implicit evaluations, while handling all non-stiff parts using standard explicit time advancement. If we adopt the common (Method-Of-Lines or MOL) approach where the spatial discretization is handled independently from the time dimension, we must time-advance equations of the form
\begin{equation}
  \label{eq:splitting}
  \partial_t \bfu = \bfF(\bfu) = \bfF_{\mathrm{im}}(\bfu) + \bfF_{\mathrm{ex}}(\bfu)\,.
\end{equation}

\subsection{Multi-step IMEX choices}
\paragraph{One-step IMEX schemes.} When we combine a first order, single step forward Euler (FE) scheme for the explicit part, with a first order backward Euler (BE) scheme for the implicit part we arrive at an overall first order accurate scheme, known as the IMEX Euler scheme. We can write the general strategy of this scheme as
\begin{equation}
\bfu^{n+1} = \bfu^{n}+\Dt \left[ \bfF_{\mathrm{ex}}(\bfu^{n})+ \bfF_{\mathrm{im}}(\bfu^{n+1})\right]  \,, \label{q-imexeuler}
\end{equation}
and we can denote it by a combination of two Butcher tableau's, as follows:
\begin{equation}
\begin{array}{c}
\mathrm{IMEX\,\,\,Euler} \\
\begin{array}{ccc}
\begin{array}{c}
\mathrm{FE} \\
\begin{array}
{c|cc}
0\\
1 & 1\\
\hline
& 1& 0
\end{array}
\end{array}
& \hspace*{1cm} & 
\begin{array}{c}
\mathrm{BE} \\
\begin{array}
{c|cc}
0\\
1 & 0 & 1\\
\hline
& 0& 1
\end{array} 
\end{array}
\end{array} 
\end{array}
\end{equation}
Instead, the IMEX SP combination (with SP denoting a {\it sp}litting approach), operates as follows: first do an implicit BE step, then perform an explicit FE step, as in
\begin{eqnarray}
\bfu^{(1)} & = & \bfu^n + \Dt \bfF_{\mathrm{im}}(\bfu^{(1)}) \,, \nonumber \\
\bfu^{n+1} & = & \bfu^n + \Dt \left[ \bfF_{\mathrm{ex}}(\bfu^{n})+ \bfF_{\mathrm{im}}(\bfu^{(1)})\right] \,. \label{q-imexsp}
\end{eqnarray}
\begin{equation}
\begin{array}{c}
\mathrm{IMEX\,\,\,SP} \\
\begin{array}{ccc}
\begin{array}{c}
\mathrm{FE} \\
\begin{array}
{c|c}
0\\
\hline
& 1
\end{array}
\end{array}
& \hspace*{1cm} & 
\begin{array}{c}
\mathrm{BE} \\
\begin{array}
{c|c}
1 &  1\\
\hline
& 1
\end{array} 
\end{array}
\end{array}
\end{array}
\label{q-imexeul}
\end{equation}
The above two schemes fall under the one-step strategy in \texttt{MPI-AMRVAC}, since only a single explicit advance is needed in each of them.

\paragraph{Two-step IMEX variants.}
A higher order accurate IMEX scheme, given in~\cite{2003hundsdorfer} (Eq.~(4.12) of their chapter IV), is a combination of the
implicit trapezoidal (or Crank-Nicholson) scheme and the explicit trapezoidal (or Heun) scheme, and writes as:
\begin{align}
  \bfu^{(n+1)*} &= \bfu^n + \Delta t \bfF_{\mathrm{ex}}(\bfu^n) + \tfrac{1}{2} \Delta t \left [\bfF_{\mathrm{im}}(\bfu^n) + \bfF_{\mathrm{im}}(\bfu^{(n+1)*})\right] \,,\nonumber \\
  \bfu^{n+1} &= \bfu^n + \tfrac{1}{2} \Delta t \left[\bfF(\bfu^n) + \bfF(\bfu^{(n+1)*})\right]\,. \label{q-imextrap}
\end{align}
\begin{equation}
\begin{array}{c}
\mathrm{IMEX\,\,\,trapezoidal} \\
\begin{array}{ccc}
\begin{array}{c}
\mathrm{Heun} \\
\begin{array}
{c|cc}
0\\
1 & 1 \\
\hline
& \scriptstyle{1/2} & \scriptstyle{1/2}
\end{array}
\end{array}
& \hspace*{1cm} & 
\begin{array}{c}
\mathrm{CN} \\
\begin{array}
{c|cc}
0\\
1 & \scriptstyle{1/2} & \scriptstyle{1/2}\\
\hline
& \scriptstyle{1/2}& \scriptstyle{1/2}
\end{array} 
\end{array}
\end{array} 
\end{array}
\label{q-imex-trap}
\end{equation}
This scheme is known as the IMEX trapezoidal scheme (or sometimes denoted as IMEX CN, as it uses an implicit Crank-Nicholson step). Since it involves one implicit stage, and two explicit stages, while achieving second order accuracy, the IMEX trapezoidal scheme is denoted as an IMEX(1,2,2) scheme. The IMEX Euler and IMEX SP are both IMEX(1,1,1). 

The three IMEX($s_{\mathrm{im}},s_{\mathrm{ex}},p$) schemes given by Eqs.~\eqref{q-imexeuler}-\eqref{q-imexsp}-\eqref{q-imextrap} differ in the number of stages used for the implicit ($s_{\mathrm{im}}$) versus explicit ($s_{\mathrm{ex}}$) parts, and in the overall order of accuracy $p$. Both IMEX(1,1,1) first order schemes from Eq.~\eqref{q-imexeuler}-\eqref{q-imexsp} require one explicit stage, and one implicit one. The IMEX(1,2,2)  trapezoidal scheme from Eq.~\eqref{q-imextrap} has one implicit stage, and two explicit ones. We can design another IMEX(1,2,2) scheme by combining the implicit midpoint scheme with a twostep explicit midpoint or Predictor-Corrector scheme. This yields the following double Butcher tableau:
\begin{equation}
\begin{array}{c}
\mathrm{IMEX\,\,\,midpoint} \\
\begin{array}{ccc}
\begin{array}{c}
\mathrm{PC} \\
\begin{array}
{c|cc}
0\\
\scriptstyle{1/2} & \scriptstyle{1/2}\\
\hline
& 0& 1
\end{array}
\end{array}
& \hspace*{1cm} & 
\begin{array}{c}
\mathrm{IM} \\
\begin{array}
{c|cc}
0\\
\scriptstyle{1/2} & 0 & \scriptstyle{1/2}\\
\hline
& 0& 1
\end{array} 
\end{array}
\end{array} 
\end{array}
\label{q-imex-mp}
\end{equation}
and corresponds to the second order IMEX midpoint scheme
\begin{align}
  \bfu^{(n+1)*} &= \bfu^n + \tfrac{1}{2}\Delta t \bfF_{\mathrm{ex}}(\bfu^n) + \tfrac{1}{2} \Delta t \bfF_{\mathrm{im}}(\bfu^{(n+1)*}) \,,\nonumber \\
  \bfu^{n+1} &= \bfu^n + \Delta t \bfF(\bfu^{(n+1)*})\,. \label{q-imexmp}
\end{align}

Another variant of a two-step IMEX scheme available in \texttt{MPI-AMRVAC} is known as the IMEX222($\lambda$) scheme from \cite{pareschi2005}, where a $\lambda$ parameter can be varied, but the default value $\lambda=1-1/\sqrt{2}$ ensures that the scheme is SSP and L-stable \citep{Izzo2017}. It has implicit evaluations at fractional steps $\lambda$ and $(1-\lambda)$. Its double Butcher table reads 
\begin{equation}
\begin{array}{c}
\mathrm{IMEX222(\lambda)} \\
\begin{array}{ccc}
\begin{array}{c}
\mathrm{Heun} \\
\begin{array}
{c|cc}
0\\
1 & 1\\
\hline
& \scriptstyle{1/2} & \scriptstyle{1/2}
\end{array}
\end{array}
& \hspace*{1cm} & 
\begin{array}{c}
\mathrm{IM} \\
\begin{array}
{c|cc}
\lambda & \lambda \\
1-\lambda  & 1-2\lambda & \lambda \\
\hline
& \scriptstyle{1/2} & \scriptstyle{1/2}
\end{array} 
\end{array}
\end{array} 
\end{array}
\label{q-imex-222l}
\end{equation}

\paragraph{Three-step IMEX variants.}
Since we thus far almost exclusively handled Butcher tableau's with everywhere positive entries, we may prefer the IMEX-ARK(2,3,2) scheme~\citep{giraldo2013} which has also two implicit stages, three explicit stages, at overall second order. It writes as
\begin{equation}
\begin{array}{c}
\mathrm{IMEX-ARK(2,3,2)} \\
\begin{array}{ccc}
\begin{array}{c}
\mathrm{explicit} \\
\begin{array}
{c|ccc}
0\\
2\delta & 2\delta \\
1 & 1-\nu & \nu  \\
\hline
& \scriptstyle{\frac{1}{2\sqrt{2}}} & \scriptstyle{\frac{1}{2\sqrt{2}}} & \delta
\end{array}
\end{array}
&  & 
\begin{array}{c}
\mathrm{implicit} \\
\begin{array}
{c|ccc}
0\\
2\delta & \delta &\delta \\
1  &  \scriptstyle{\frac{1}{2\sqrt{2}}} & \scriptstyle{\frac{1}{2\sqrt{2}}} & \delta \\
\hline
 & \scriptstyle{\frac{1}{2\sqrt{2}}}& \scriptstyle{\frac{1}{2\sqrt{2}}} & \delta
\end{array} 
\end{array}
\end{array}
\label{q-imex-ark}
\end{array}
\end{equation}
where we use the fixed values $\delta =1-{1}/{\sqrt{2}}$ while $\nu= (3+2\sqrt{2})/6$.

Thus far, in terms of the double Butcher tableau's, we have mostly been combining schemes that have the same left column entries (i.e.\ substep time evaluations) for the implicit and the explicit stages. A possible exception was the IMEX222($\lambda$) scheme. The implicit part was always in diagonally implicit Runge-Kutta type (or DIRK). Since one in practice implements the implicit stages separately from the explicit ones, one can relax the condition for implicit and explicit stages to be at the same time. In~\cite{rokhzadi2018} an IMEX-SSP(2,3,2) scheme with 2 implicit and 3 explicit stages was introduced which indeed relaxes this, and it writes as
\begin{equation}
\begin{array}{c}
\mathrm{IMEX-SSP(2,3,2)} \\
\begin{array}{ccc}
\begin{array}{c}
\mathrm{explicit} \\
\begin{array}
{c|ccc}
0\\
\scriptstyle{0.712} & \scriptstyle{0.712} \\
\scriptstyle{0.994} & \scriptstyle{0.077}&   \scriptstyle{0.917} \\
\hline
& \scriptstyle{0.399} & \scriptstyle{0.346} &  \scriptstyle{0.255}
\end{array}
\end{array}
& \hspace*{-0.5cm} & 
\begin{array}{c}
\mathrm{implicit} \\
\begin{array}
{c|ccc}
0\\
\scriptstyle{0.708} & \scriptstyle{0.354} &  \scriptstyle{0.354} \\
1&   \scriptstyle{0.399} & \scriptstyle{0.346} & \scriptstyle{0.255} \\
\hline
& \scriptstyle{0.399} & \scriptstyle{0.346}  & \scriptstyle{0.255} 
\end{array} 
\end{array}
\end{array}
\end{array}
\label{q-imex-ssp}
\end{equation}

If we allow for tableau's with also negative entries, we may even get third order IMEX schemes, e.g.\ the ARS(2,3,3) scheme by~\cite{ars1997} (also denoted as IMEX-ARS3) where 
\begin{equation}
\begin{array}{c}
\mathrm{IMEX-ARS3 \,\, or \,\, ARS(2,3,3)} \\
\begin{array}{cccc}
\begin{array}{c}
\mathrm{explicit} \\
\begin{array}
{c|ccc}
0\\
\gamma & \gamma \\
1 -\gamma & \gamma-1 & 2(1-\gamma) \\
\hline
& 0 & \scriptstyle{1/2}& \scriptstyle{1/2}
\end{array}
\end{array}
& \hspace*{-0.7cm} & 
\begin{array}{c}
\mathrm{implicit} \\
\begin{array}
{c|ccc}
0\\
\gamma & 0 & \gamma\\
1 -\gamma & 0 & 1-2\gamma & \gamma \\
\hline
 & 0&  \scriptstyle{1/2}& \scriptstyle{1/2}
\end{array} 
\end{array}
\end{array}
\end{array}
\label{q-imex-ars3}
\end{equation}
which uses the fixed value $\gamma ={(3+\sqrt{3})}/{6} $. This has the advantage of having only 2 implicit stages, which are usually more costly to compute than explicit stages. This ARS3~\citep{ars1997} scheme has been shown to achieve better than second order accuracy on some tests~\citep{koto2008}, while needing three explicit and two implicit stages. 

Finally, the IMEX-CB3a scheme, denoted as IMEXRKCB3a in \citet{Cavaglieri2015} uses three explicit steps, in combination with two implicit stages to arrive at overall third order, so it is an IMEX(2,3,3) variant:
\begin{equation}
\begin{array}{c}
\mathrm{IMEX-CB3a} \\
\begin{array}{cccc}
\begin{array}{c}
\mathrm{explicit} \\
\begin{array}
{c|ccc}
0\\
c_2 & c_2 \\
c_3 & 0 & c_3 \\
\hline
& 0 & b_2 & b_3
\end{array}
\end{array}
&  & 
\begin{array}{c}
\mathrm{implicit} \\
\begin{array}
{c|ccc}
0\\
c_2 & 0 & c_2\\
c_3 & 0 & c_3-a_{33} & a_{33} \\
\hline
 & 0&  b_2 & b_3
\end{array} 
\end{array}
\end{array}
\end{array}
\label{q-imex-cb3a}
\end{equation}
The following relations fix all the values in its Butcher representation
\begin{eqnarray}
c_2 & = & \frac{1}{54}\left(27+\sqrt[3]{2187-1458\sqrt{2}}+9\sqrt[3]{3+2\sqrt{2}}\right) \,, \nonumber\\
 &  \approx & 0.89255 \,, \nonumber\\
c_3 & = & \frac{c_2}{6{c_2^2}-3c_2+1} \,, \nonumber\\
b_2 & = & \frac{3c_2-1}{6{c_2^2}} \,, \nonumber \\
b_3 & = & 1- b_2 \,, \nonumber\\
a_{33} & = & \frac{\frac{1}{6}-b_2 {c_2^2}-b_3 c_2 c_3}{b_3(c_3-c_2)} \,.
\end{eqnarray}
This scheme has the advantage that a low storage implementation (using 4 registers) is possible.

\subsection{IMEX implementation and usage in \texttt{MPI-AMRVAC}}
\paragraph{IMEX implementation.} The various IMEX schemes are shared between all equation/physics modules, and a generic implementation strategy uses the following pseudo-code ingredients for its efficient implementation. First, we introduced a subroutine
\begin{lstlisting}
global_implicit_update($\alpha\Dt$,$t_n+\beta\Dt$,$\bfu^a$,$\bfu^b$)
\end{lstlisting}
which solves the (usually global) problem on the instantaneous AMR grid hierarchy given by
\begin{equation}
\bfu^a = \bfu^b + \alpha \Dt \bfF_{\rm im}(\bfu^a) \,. \label{eq:implicit}
\end{equation}
This call leaves $\bfu^b$ unchanged, and returns $\bfu^a$ as the solution of this implicit problem. On entry, both states are available at time $t_n+\beta\Dt$. On exit, state $\bfu^a$ is advanced by $\alpha\Dt$ and has its boundary updated. Second,
\begin{lstlisting}
evaluate_implicit($t$,$\bfu^a$)
\end{lstlisting}
just replaces the $\bfu^a$ state with its evaluation (at time $t$) in the implicit part, i.e. $\bfu^a\rightarrow \bfF_{\rm im}(t,\bfu^a)$. Finally, any explicit substep is handled by a subroutine
\begin{lstlisting}
advect1($\alpha\Dt$,$t_a$,$\bfu^a$,$t_b$,$\bfu^b$)
\end{lstlisting}
which advances the $\bfu^b$ state explicitly according to
\begin{equation}
\bfu^b(t_b+\alpha\Dt) = \bfu^b(t_b) + \alpha \Dt \bfF_{\rm ex}(\bfu^a(t_a)) \,,
\end{equation}
along with boundary conditions on $\bfu^b(t_b+\alpha\Dt)$.

\paragraph{IMEX usage in \texttt{MPI-AMRVAC}.} Currently, the IMEX schemes are used for handling (1) stiff diffusion terms, such as encountered in pure reaction-diffusion (or advection-reaction-diffusion) problems, or in the radiation-hydro module using flux-limited-diffusion \citep{2022Moens}; (2) stiff coupling terms, such as in the gas-dust treatment as explained in Section~\ref{sec:gasdust}, or in the ion-neutral couplings in the two-fluid module \citep{2022Braileanu}.

\begin{figure*}[ht]
    \centering
    \includegraphics[width=\textwidth]{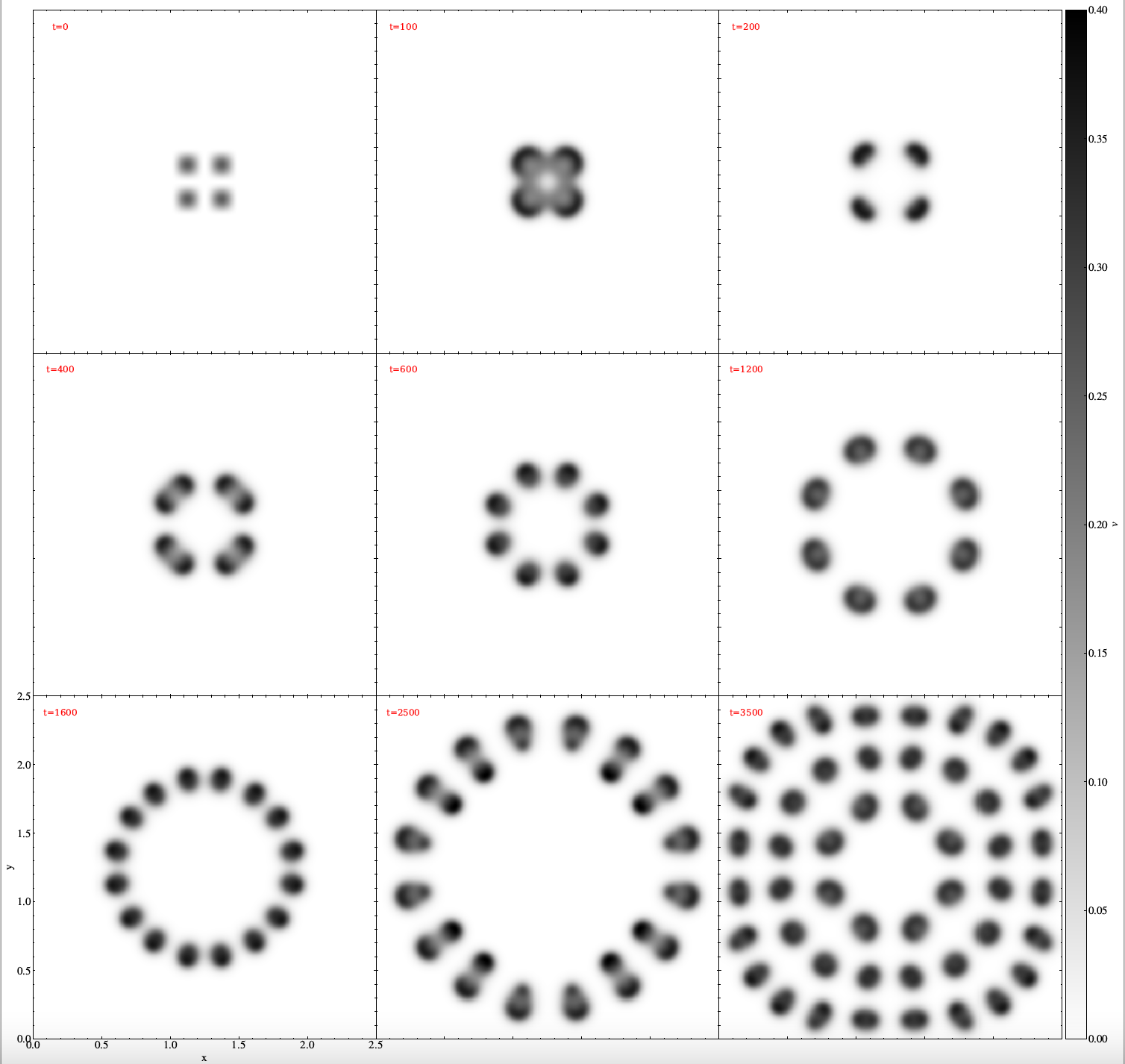}
    \caption{Temporal evolution of $v(x, y, t)$ in a pure reaction-diffusion 2D Gray-Scott spot replication simulation. An animation is provided.}
    \label{fig: GS2-evolution}
\end{figure*}

As an example of the first (handling stiff diffusion) IMEX use-case, we solve the pure reaction-diffusion 2D Gray-Scott spot replication simulation from~\cite{2003hundsdorfer} (which also appears in~\citealt{pearson1993complex}).
The Gray-Scott two-component PDE system for $\bfu(\bfx,t)=\left(u(\bfx,t),v(\bfx,t)\right)$,
has the following form
\begin{align}
  \label{eq:gray-scott}
  \partial_t u &= D_u \nabla^2 u - u v^2 + F(1 - u)\,,\\
  \partial_t v &= D_v \nabla^2 v + u v^2 - (F + k) v\nonumber\,,
\end{align}
where $F$ and $k$ are positive constants; $D_u$ and $D_v$ are constant diffusion coefficients. Note that the feeding term $F(1 - u)$ drives the concentration of $u$ to one, whereas the term $-(F + k) v$ removes $v$ from the system. A wide range of patterns can be generated depending on the values of $F$ and $k$ \citep{pearson1993complex}, here we take $F = 0.024$ and $k = 0.06$. The diffusion coefficients have values $D_1 = 8 \times 10^{-5}$ and $D_2 = 4 \times 10^{-5}$. The initial conditions consist of a sinusoidal pattern in the center of the domain $[0, 2.5]^2$:
\begin{equation} \label{eq: GS_initial_conditions}
    \begin{split}
        u(x,y,0) &= 1 - 2 v(x,y,0), \\      
        v(x,y,0) &=
        \begin{cases}
            \frac{1}{4} \sin(4\pi x)^2 \sin(4\pi y)^2 & \text{if } x,y \in [1,1.5] \\
            0 & \text{elsewhere}
        \end{cases}.
    \end{split}
\end{equation}
Fig.~\ref{fig: GS2-evolution} shows the temporal evolution of $v(x, y, t)$ for a high resolution AMR simulation with a base resolution of $256^2$ cells. The five levels of refinement allow for a maximal effective resolution of $4096^2$ cells. This long-term, high-resolution run then shows how the AMR quickly adjusts to the self-replicating, more volume-filling pattern that forms: while at $t=100$ the coarsest grid occupies a large fraction of 0.859 of the total area, while the finest level covers only the central 0.066 area, this evolves to 0.031 (level 1) and 0.269 (level 5) at time $t=3500$, the last time shown in Fig.~\ref{fig: GS2-evolution}. Note that on a modern 20-CPU desktop [using Intel Xeon Silver 4210 CPU at 2.20GHz], this entire run takes only 1053 seconds, of which less than 10 percent is spent on generating 36 complete file dumps in both the native \texttt{.dat} format, and the on-the-fly converted \texttt{.vtu} format (suitable for visualization packages such as {\tt ParaView} or {\tt VisIt}, see Section~\ref{sec:vtu}). 

\begin{figure}[h]
    \centering
    \includegraphics[width=\columnwidth]{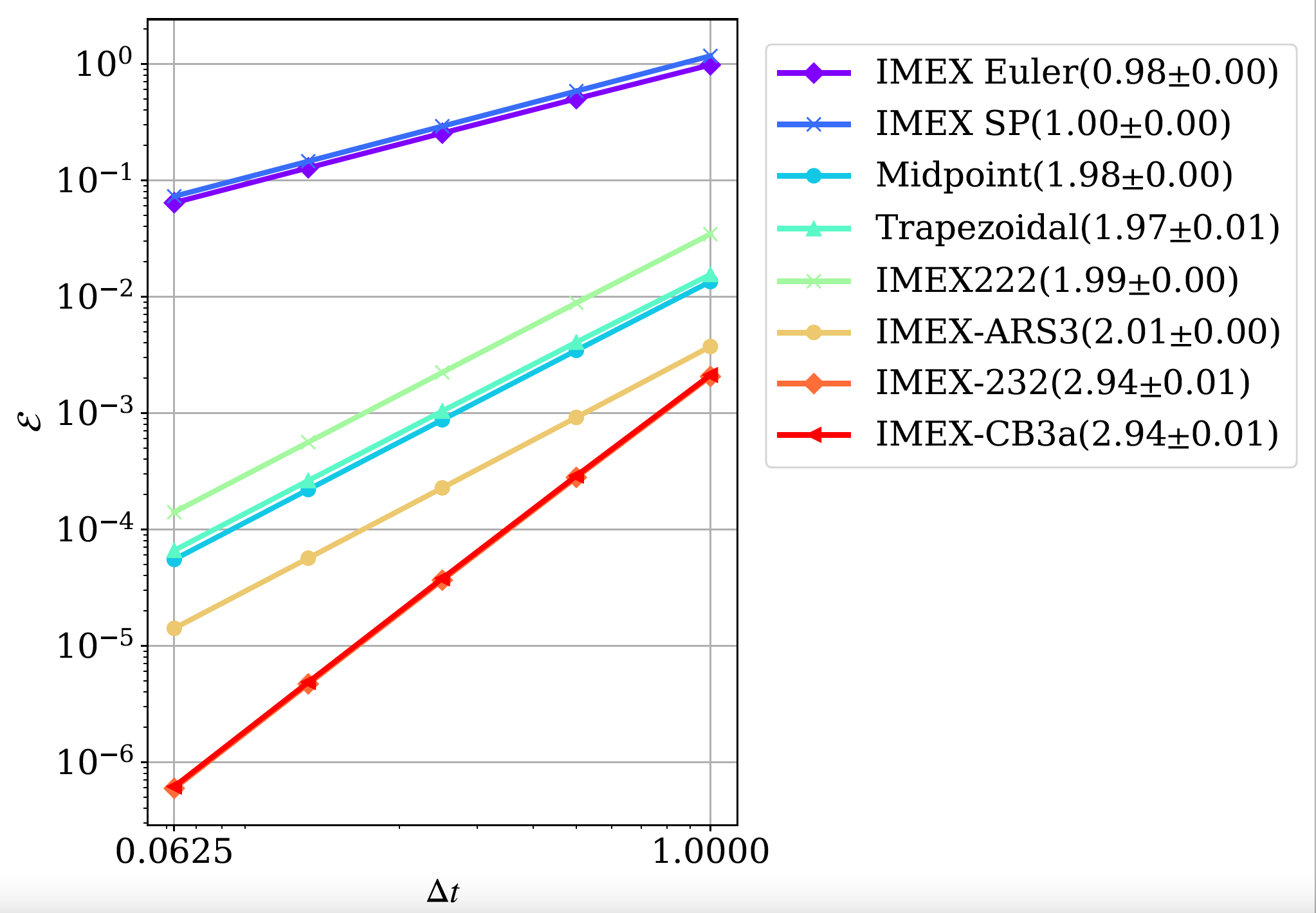}
    \caption{Temporal convergence of IMEX-RK schemes in MPI-AMRVAC.  The error computed as $[\sum_{i=1}^{N}{(u[i]-u_{\rm ref}[i])^2 + (v[i]-v_{\rm ref}[i])^2}]^{1/2}$, where $N$ is the total number of grid points, is plotted as a function of the timestep used to obtain the numerical solutions $u$ and $v$ using IMEX schemes. $u_{\rm ref}$ and $v_{\rm ref}$ are the reference numerical solutions obtained using an explicit scheme with a  much smaller timestep.
    }
    \label{fig: GS_temporal_convergence}
\end{figure}

In order to perform a convergence study in time we take the same setup as Fig.~\ref{fig: GS2-evolution}, but for a uniform grid of $256^2$ cells (corresponding to a cell size of $\approx 0.01^2$) and a final time $t_\text{end} = 100$ (corresponding to the second panel of Fig.~\ref{fig: GS2-evolution}). We compare every simulation to a reference solution obtained with a classical fourth-order explicit Runge-Kutta scheme for $\Delta t = 10^{-3}$. When explicitly solving the reaction-diffusion system corresponding to the initial conditions, the explicit timesteps are $\Delta t_\text{d, expl} = 0.149$ and $\Delta t_\text{r, expl} = 5.803$  associated with diffusion and the reaction terms, respectively. Hence,  the use of the IMEX schemes reduces the computational cost by a factor of 
$\Delta t_\text{r, expl}/\Delta t_\text{d, expl} \approx  39$ for this particular problem. For the convergence tests themselves the value of the largest timestep is fixed to unity, followed by four successive timesteps smaller by a factor two. The resulting convergence graph for $\Delta t \in \{0.0625, 0.125, 0.25, 0.5, 1\}$ is shown in Fig.~\ref{fig: GS_temporal_convergence}, showing good correspondence between the theoretical and observed convergence rates.

Note that in this Gray-Scott problem, the implicit update from Eq.~\eqref{eq:implicit} is actually a problem that can be recast to the following form
\begin{equation}
\nabla^2 u^a - \frac{1}{\alpha\Dt D_u} u^a = -\frac{1}{\alpha\Dt D_u}u^b \,,
\end{equation}
and similarly for $v$. For solving such generic elliptic problems on our AMR grid, we exploit the efficient algebraic multigrid solver as introduced in \citet{2019Teunissen}. 
This test is available at {\em \texttt{tests/demo/Gray\_Scott\_2D}}.

\section{Special (solar) physics modules}

In recent years, \texttt{MPI-AMRVAC} has been actively applied to solar physics, where 3D MHD simulations are standard, although they may meet very particular challenges. Even when restricting attention to the solar atmosphere (photosphere to corona), handling the extreme variations in thermodynamic quantities (density, pressure and temperature) in combination with strong magnetic field concentrations, already implies large differences in plasma beta. Moreover, a proper handling of the chromospheric layers, along with the rapid temperature rise in a narrow transition region, really forces one to use advanced radiative-MHD treatments (accounting for frequency-dependent, non-local couplings between radiation and matter, true non-local-thermal-equilibrium physics affecting spectral line emission/absorption, \ldots). Thus far, all these aspects are only handled approximately, with e.g.\ the recently added plasma-neutral {\tt src/twofl} module~\citep{2022Braileanu} as an example where the intrinsically varying degree of ionization throughout the atmosphere can already be incorporated. To deal with large variations in plasma beta, we provided options to split off a time-independent (not necessarily potential) magnetic field $\bfB_0$ in up to resistive MHD settings \citep{2018Xia}, meanwhile generalized \citep{nitin} to split off entire 3D magnetostatic force-balanced states $-\nabla p_0+\rho_0\mathbf{g}+\mathbf{J}_0\times\bfB_0=\mathbf{0}$. For MHD and two-fluid modules, we add an option to solve internal energy equation instead of total energy equation to avoid negative pressure  when plasma beta is extremely small. 

A specific development relates to numerically handling energy and mass fluxes across the sharp transition region variation, which under typical solar conditions and traditional Spitzer-type thermal conductivities can never be resolved accurately in multidimensional settings. Suitably modifying the thermal conduction and radiative loss prescriptions can preserve physically correct total radiative losses and heating aspects~\citep{2020Johnston}. This led to the Transition-Region-Adaptive-Conduction or {\tt TRAC} approaches \citep{2020Johnston,2021Iijima,2021Johnston,2021Zhou}, with e.g.\ 
\citet{2021Zhou} introducing various flavors where the field line tracing functionality (from Section~\ref{sec:trace}) was used to extend the originally 1D hydro incarnations to multi-dimensional MHD settings. Meanwhile, truly local variants~\citep{2021Iijima,2021Johnston} emerged, and \texttt{MPI-AMRVAC 3.0} provides multiple options\footnote{In fact, the original 1D hydro `infinite-field' limit where MHD along a geometrically prescribed, fixed-shape field line is believed to follow the hydrodynamic laws with field-line projected gravity, can also activate the {\tt TRAC} treatment in the {\tt src/hd} module, in combination with a user-set area variation along the `fieldline', which was e.g.\ shown to be important in \citet{2013Mikic}.} collected in {\tt src/mhd/mod\_trac.t}. In practice, up to 7 variants of the {\tt TRAC} method can be distinguished in higher dimensional ($>$ 1D) setups, including the use of a globally fixed cut-off temperature, the (masked and unmasked) multi-dimensional {\tt TRACL} and {\tt TRACB} methods introduced in \citet{2021Zhou}, or the local fix according to~\citet{2021Iijima}.

Various modules are available which implement frequently recurring ingredients in solar applications. These are, e.g., a Potential-Field-Source-Surface ({\tt PFSS}) solution on a 3D spherical grid that extrapolates magnetic fields from a given bottom magnetogram (see {\tt src/physics/mod\_pfss.t} as evaluated in \citealt{2014Porth}),
a method to extrapolate a magnetogram into a linear force-free field in a 3D Cartesian box (see {\tt src/physics/mod\_lfff.t}), or a modular implementation of the frequently employed 3D Titov-D\'emoulin~\citep{1999TD} analytic flux rope model (see {\tt src/physics/mod\_tdfluxrope.t}), or the functionality to perform non-linear force-free field extrapolations from vector magnetograms (see {\tt src/physics/mod\_magnetofriction.t} as evaluated in \citealt{2016Guo2,2016Guo1}).

In what follows, we demonstrate more recently added solar-relevant functionality, namely the addition of a time-dependent magnetofrictional module in Section~\ref{sec:mf}, the possibility to insert flux ropes using the regularized Biot-Savart laws (RBSL) from~\citet{2018Titov} in Section~\ref{sec:rbsl}, and the way to synthesize 3D MHD data to actual EUV images in a simple on-the-fly fashion in Section~\ref{sec:euv}.

\subsection{Magneto-frictional module}\label{sec:mf}

The magneto-frictional (MF) method is closely related to the MHD relaxation process \citep[e.g.,][]{1981Chodura}. It is proposed by \citet{1986Yang} and considers both the momentum equation and the magnetic induction equation:
\begin{eqnarray}
  \label{eq:momentum}
  & & \rho \left(\frac{\partial \bfv}{\partial t} + \bfv \cdot \nabla \bfv\right) = \bfJ \times \bfB - \nabla p + \rho \boldsymbol{g} -\nu \bfv ,  \\
  \label{eq:induction}
  & & \frac{\partial \bfB}{\partial t} = \nabla \times (\bfv \times \bfB) ,
\end{eqnarray}
where $\rho$ is the density, $\bfv$ the velocity, $\bfJ=\nabla \times \bfB/\mu_0$ the electric current density, $\bfB$ the magnetic field, $p$ the pressure, $\boldsymbol{g}$ the gravitational field, $\nu$ the friction coefficient, and $\mu_0$ the vacuum permeability. To construct a steady-state force-free magnetic field configuration, the inertial, pressure-gradient, and gravitational forces are omitted in Eq.~(\ref{eq:momentum}) and one only uses the simplified momentum equation to give the MF velocity:
\begin{eqnarray}
  \label{eq:momentum2}
  & & \bfv = \frac{1}{\nu} \bfJ \times \bfB . 
\end{eqnarray}
Eqs.~(\ref{eq:induction}) and (\ref{eq:momentum2}) are then combined together to relax an initially finite-force magnetic field to a force-free state where $\bfJ\times \bfB=\mathbf{0}$ with appropriate boundary conditions.

The MF method has been adopted to derive force-free magnetic fields in 3D domains \citep[e.g.,][]{1996Roumeliotis,2005Valori,2016Guo2,2016Guo1}. It is commonly regarded as an iteration process to relax an initial magnetic field that does not need to have an obvious physical meaning. For example, if the initial state is provided by an extrapolated potential field together with an observed vector magnetic field at the bottom boundary, the horizontal field components can jump discontinuously there initially, and locally are probably not in a divergence-free condition. The MF method can still relax this unphysical initial state to an almost force-free and divergence-free state (the degree of force-freeness and its solenoidal character can be quantified during the iterates and monitored). On the other hand, the MF method could also be used to actually mimic a time-dependent process \citep[e.g.,][]{2008Yeates, 2012Cheung}, although there are caveats about using the MF method in this way \citep{2013Low}. The advantage of such time-dependent MF method is that it consumes much less computational resources than a full MHD simulation to cover a long-term quasi-static evolution of nearly force-free magnetic fields. This allows us to simulate the global solar coronal magnetic field over a very long period, for instance, several months or even years. 

We implemented a new MF module ({\tt src/mf}), parallel to the existing physics modules like MHD, in \texttt{MPI-AMRVAC}. This module can be used in 2D and 3D, and in different geometries, fully compatible with (possibly stretched) block-AMR. We set the friction coefficient $\nu=\nu_0 B^2 $, where $\nu_0=10^{-15}$ s cm$^{-2}$ is the default value. The magnitude of the MF velocity is smoothly truncated to an upper limit $v_{max}=30$ km s$^{-1}$ by default to avoid extremely large numerical speed near magnetic null points \citep{Pomoell2019}.  $\nu_0$ and $v_{max}$ are input parameters {\tt mf\_nu} and {\tt mf\_vmax} with dimensions. We allow a smooth decay of the MF velocity towards the physical boundaries to match line-tied quasi-static boundaries \citep{2012Cheung}. In contrast to the previous MF module (still available as {\tt src/physics/mod\_magnetofriction.t}) used in \citet{2016Guo1}, this new MF module in {\tt src/mf} includes the time-dependent MF method and now fully utilizes the framework for data I/O with many more options of numerical schemes. Especially, the constrained transport scheme \citep{Balsara1999}, compatibly implemented with the staggered AMR mesh \citep{Olivares2019}, to solve the induction equation (\ref{eq:induction}) is recommended when using this new MF module. This then can enforce the divergence of magnetic field to near machine precision zero.

\begin{figure}
    \centering
    \includegraphics[width=\columnwidth]{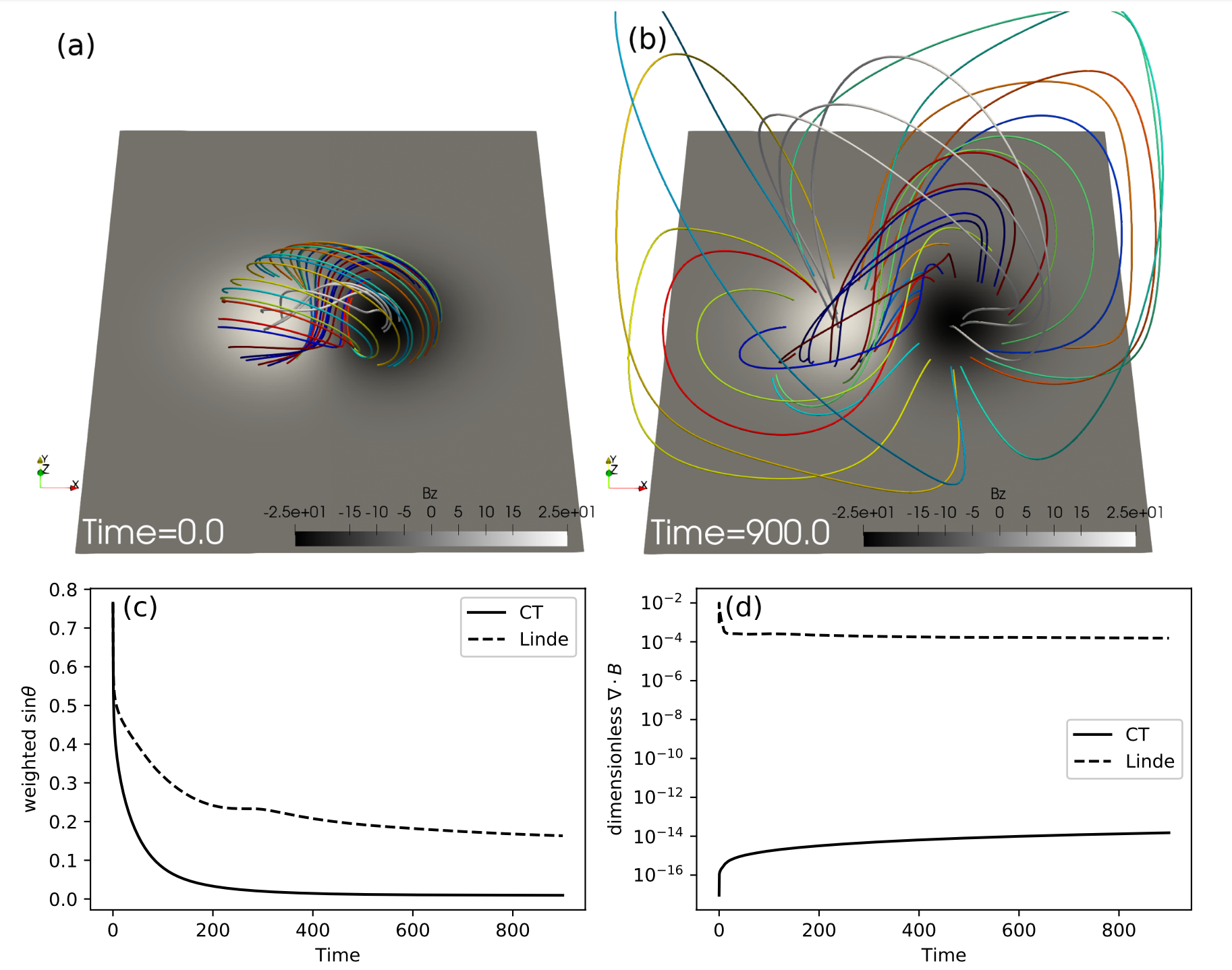}
    \caption{MF relaxation of a non-force-free twisted magnetic flux tube. (a) magnetic field lines of the flux tube in different colors at time 0, (b) magnetic field lines starting from the same footpoints at time 900 of the `ct' run, (c) degree of force-free as the weighted average of the sine of the angle between the magnetic field and the current density for the `linde'  and `ct' run, (d) degree of divergence-free as the average dimensionless divergence of the magnetic field for both runs.}
    \label{fig:fbm}
\end{figure}

To validate the new MF module, we set up a test, fully provided in {\tt tests/demo/Mackay\_bipole\_Cartesian\_3D}, starting from a non-force-free bipolar twisted magnetic field \citep{Mackay2001} to verify that the MF module can efficiently relax it to a force-free state. The magnetic field is given by
\begin{eqnarray}
  \label{eq:Bx}
  & & B_x=B_0 e^{0.5}\left(\frac{z}{L_0}e^{-\xi}+4\beta\frac{xy}{L_0^2}e^{-2\xi}\right) ,  \\
  \label{eq:By}
  & & B_y=2 \beta B_0 e^{0.5}\left(1-\frac{x^2+z^2}{L_0^2}\right)e^{-2\xi}, \\
  \label{eq:Bz}
  & & B_z=B_0 e^{0.5}\left(-\frac{x}{L_0}e^{-\xi}+4\beta\frac{yz}{L_0^2}e^{-2\xi}\right) , \\
  & & \xi=\frac{0.5(x^2+z^2)+y^2}{L_0^2},
\end{eqnarray}
where $B_0=50$ G is the peak flux density, $\beta=1$ is a dimensionless parameter to control the twist of the magnetic field, and $L_0=10$ Mm is the half distance between the peaks of two polarities on the $z=0$ boundary. The  test is performed in a 3D Cartesian box bounded by $-40\leq x\leq 40$ Mm, $-40\leq y \leq 40$ Mm, and $0\leq z \leq 80$ Mm with 4-level AMR grid of effective $256\times256\times256$ resolution. The (magnetofrictional) velocities in the ghost cells are set to zero. The induction equation is solved with a finite-volume scheme combining the HLL Riemann flux \citep{Harten83} with \v{C}ada's third-order limiter \citep{2009Cada} for reconstruction and a three-step Runge--Kutta time integration. The magnetic field is extrapolated on the side and top boundaries assuming zero normal gradient. We compare two divergence control strategies, namely `linde' versus `ct' (from Table~\ref{t:divb}). In the run using `linde', we use Courant number 0.3 for stability and a diffusive term in the induction equation  diminishes the divergence of the magnetic field \citep{Keppens03}. To keep the bottom magnetic flux distribution, the magnetic field vectors are fixed at the initial values in the first-layer (next to the physical domain) ghost cells of the bottom boundary and extrapolated to deeper layers of ghost cells with divergence-free condition. In the run `ct', we use Courant number 0.8 and the constrained transport method with the initial face-centered magnetic field integrated from the edge-centered vector potential to start with zero numerical divergence of the magnetic field. The tangential electric field on the bottom surface is fixed to zero to preserve magnetic flux distribution. Magnetic structures at the initial time 0 and the final time 900 are presented by magnetic field lines in Fig.~\ref{fig:fbm}a and \ref{fig:fbm}b, respectively. The initial torus-like flux tube expands and relaxes to a twisted flux rope wrapped by sheared and potential arcades. In Fig.~\ref{fig:fbm}c and \ref{fig:fbm}d, we present the force-free degree by the weighted average of the sine of the angle between the magnetic field and the current density as Eq.~(13) of \citet{2016Guo1} and the divergence-free degree by the average dimensionless $\nabla\cdot\bfB$ as the Eq.~(11) of \citet{2016Guo1}, respectively. In the `ct' run, the force-free degree rapidly decreases from 0.76 to lower than 0.1 within time 84, and converges to 0.0095. The divergence-free degree levels off to $1.5\times10^{-14}$. In the run using `linde', the force-free degree decreases similarly until 0.6 and slowly converges to a worse value of 0.16. The divergence-free degree quickly reaches a peak value of 0.098 and decreases to $1.5\times 10^{-4}$. Further investigation locates the large $\nabla\cdot\bfB$ errors at the two main polarities in the first-layer cells above the bottom boundary in the `linde' run.

\begin{figure*}
  \FIG{
    \centering
    \includegraphics[width=0.8\textwidth]{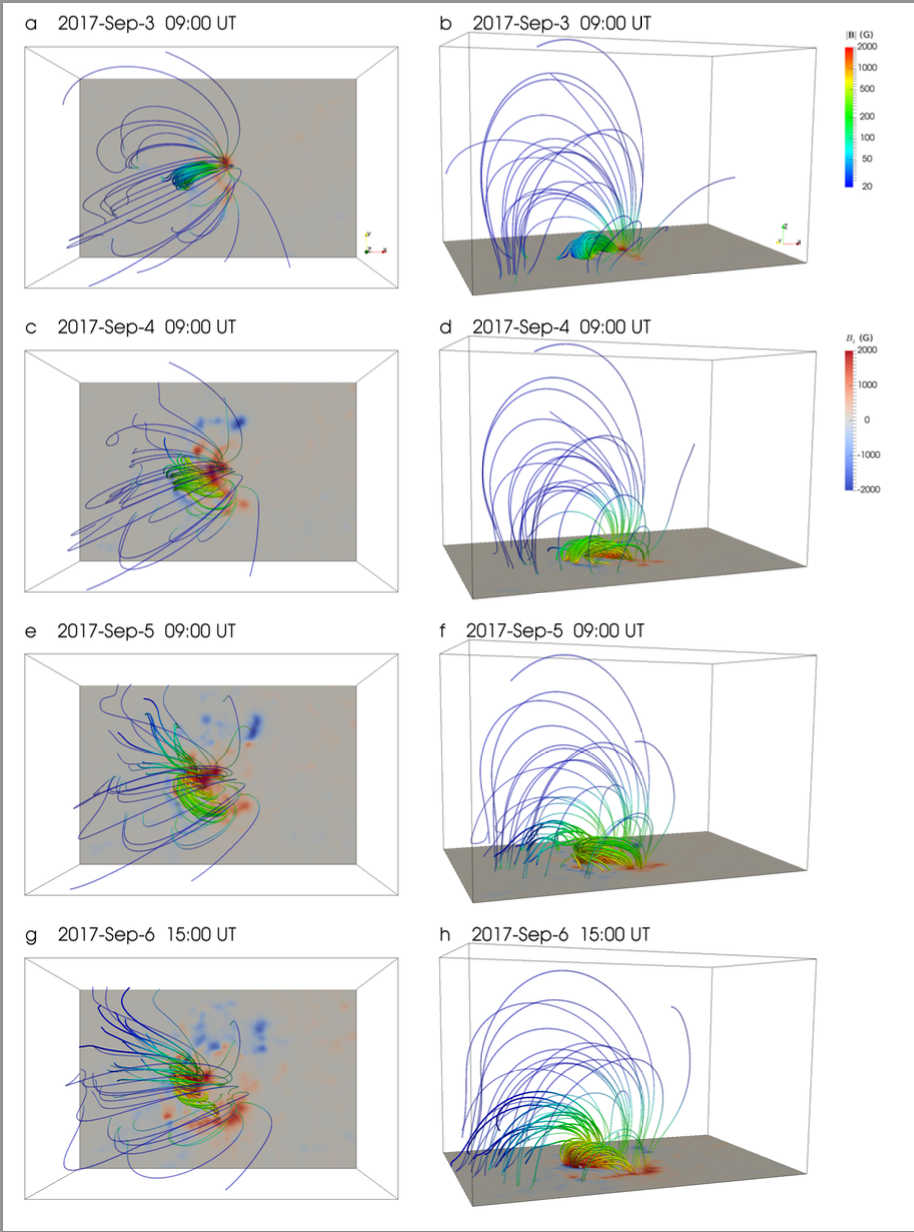}
    \caption{Evolution of magnetic field lines in the data-driven time-dependent magneto-frictional simulation. The left column shows a top view and the right column shows a side view. The slice on the bottom displays normal magnetic field, $B_z$. The magnetic field lines are colored by the field strength.}
    \label{fig:tmf}
  }
\end{figure*}

We also apply the new MF module to observations (provided in {\tt tests/mf/data\_driven\_tmf}) for time-dependent evolution. Fig.~\ref{fig:tmf} shows an example of the application of the MF module in the solar active region 12673. We select the time range between 09:00 UT on 2017 September 3 and 15:00 UT on 2017 September 6 to do the simulation, which includes the buildup period for two X-class flares peaking at 09:10 UT (X2.2) and 12:02 UT (X9.3) on 2017 September 6, respectively. SDO/HMI observes a temporal sequence of vector magnetic field on the photosphere. We use the SDO/HMI Active Region Patch (SHARP) data with a cadence of 12 minutes \citep{2014Hoeksema}. The series name of the data is ``hmi.sharp\_cea\_720s.7115''. The vector velocity field is also derived by the inversion of the magnetic field using the Differential Affine Velocity Estimator for Vector Magnetograms (DAVE4VM; \citealt{2008Schuck}). Then, both the temporal sequence of the vector magnetic field and the velocity field are used to fill the first-layer ghost cells at the bottom boundary to drive the evolution of the coronal magnetic field. The initial condition is a potential field at 09:00 UT on 2017 September 3 as shown in Fig.~\ref{fig:tmf}a. On a uniform (domain-decomposed) grid of $340\times220\times220$ cells, the induction equation is solved with the same numerical schemes as in the `linde' run of the bipolar test. The magnetic field line evolution in Fig.~\ref{fig:tmf} indicates that a twisted magnetic flux rope is formed along the polarity inversion line towards the explosion of the major flares. The resulting 3D magnetic field evolution can be compared to actual observations (in terms of their emission morphology), or can be used to start full 3D follow-up MHD simulations that incorporate actual plasma dynamics.

Note that we here did exploit `linde' divergence control, since the `ct' method requires a strict divergence-free boundary condition of magnetic field for numerical stability. For static cases (as in the much more idealized test from Fig.~\ref{fig:fbm}), this can be realized easily. However, for data driven boundary conditions in which magnetic fields are directly given by actual observations, such strict divergence-free condition can not always be ensured. With the `linde' method, spurious divergence induced by a data driven boundary can still be diffused and reduced, and the numerical scheme is stable even though locally the discrete divergence of magnetic field can be relatively large (as stated above for the test from Fig.~\ref{fig:fbm}, typically in the first grid cell layer). When we tried to apply the `ct' method to actual SDO data, code stability was compromised due to residual magnetic field divergence. Future work should focus on more robust, fully AMR-compatible means for performing data-driven runs using actual observational vector magnetic and flow data.

\subsection{Inserting flux ropes using regularized Biot--Savart laws}\label{sec:rbsl}

Solar eruptive activities, such as flares and CMEs, are believed to be driven by the eruption of magnetic flux ropes. Many efforts have been devoted to model the magnetic field of such a configuration, such as the analytical Gibson--Low model \citep{1998Gibson}, Titov--D\'emoulin model \citep{1999Titov}, and Titov--D\'emoulin modified model \citep{2014Titov}. Alternatively, nonlinear force-free field extrapolations are also applied to model flux ropes numerically \citep[e.g.,][]{2009Canou,2010Guo}. They use the vector magnetic field observed on the photosphere as the boundary condition, solve the force-free equation $\nabla \times \bfB = \alpha \bfB$, and derive the 3D coronal magnetic field. There are some drawbacks in these analytical and numerical methods. Most analytical solutions assume some geometric symmetries, such as a toroidal arc in the Titov--D\'emoulin model. On the other hand, many numerical techniques cannot derive flux rope structures in weak magnetic field region or when they detach from the photosphere, such as for intermediate or quiescent prominences. One way to alleviate this problem is to adopt the flux rope insertion method \citep{2004vanBallegooijen}. However, this method uses an initial state far from equilibrium, which asks for many numerical iterations to relax and the final configuration is difficult to control.

\citet{2018Titov} proposed the RBSL method to overcome the aforementioned drawbacks. A force-free magnetic flux rope with arbitrary axis path and intrinsic internal equilibrium is embedded into a potential field. The external equilibrium could be achieved by a further numerical relaxation. The RBSL magnetic field, $\bfB_\mathrm{FR}$, generated by a net current $I$ and net flux $F$ within a thin magnetic flux rope with a minor radius $a(l)$ is expressed as:
\begin{eqnarray}
  \label{eq:rbsl1}
  & & \bfB_\mathrm{FR} = \nabla \times \boldsymbol{A}_I + \nabla \times \boldsymbol{A}_F ,  \\
  \label{eq:rbsl2}
  & & \boldsymbol{A}_I(\boldsymbol{x}) = \frac{\mu_0 I}{4\pi}\int_{\mathcal{C} \cup \mathcal{C^*}} K_I(r) \boldsymbol{R}'(l) \frac{dl}{a(l)} , \\
  \label{eq:rbsl3}
  & & \boldsymbol{A}_F(\boldsymbol{x}) = \frac{F}{4\pi}\int_{\mathcal{C} \cup \mathcal{C^*}} K_F(r) \boldsymbol{R}'(l) \times \boldsymbol{r} \frac{dl}{a(l)^2} ,
\end{eqnarray}
where $\boldsymbol{A}_I(\boldsymbol{x})$ and $\boldsymbol{A}_F(\boldsymbol{x})$ are the vector potentials, $\mu_0$ the vacuum permeability, $\mathcal{C}$ and $\mathcal{C}^*$ the axis paths above and below the reference plane, $K_I(r)$ and $K_F(r)$ the integration kernels, $\boldsymbol{R}'=d\boldsymbol{R}/dl$ the unit tangential vector, $l$ the arc length along the axis paths, and $\boldsymbol{r} \equiv \boldsymbol{r}(l) = (\boldsymbol{x} - \boldsymbol{R}(l))/a(l)$. \citet{2018Titov} have provided the analytical forms of the integration kernels by assuming a straight force-free flux rope with a constant circular cross-section. The axial electric current is distributed in a parabolic profile along the minor radius of the flux rope. With such analytical integration kernels, a flux rope with arbitrary path could be derived via Eqs.~\eqref{eq:rbsl1}--\eqref{eq:rbsl3}.

\begin{figure}
    \centering
    \includegraphics[width=0.48\textwidth]{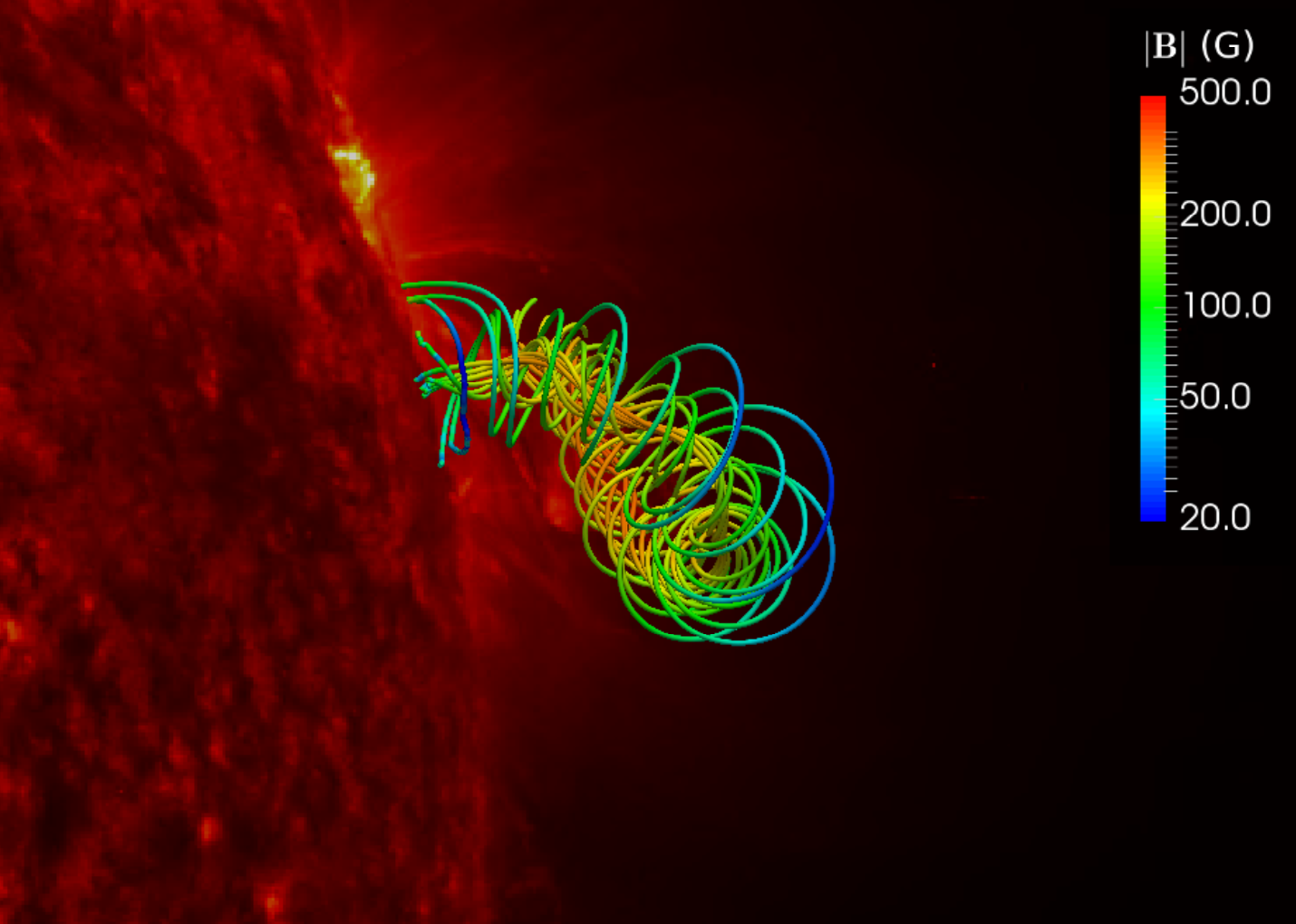}
    \caption{A magnetic flux rope constructed by the regularized Biot--Savart laws (RBSL) overlaid on the STEREO\_B/EUVI 304~\AA \ image observed at 01:11 UT on 2011 June 11.}
    \label{fig:rbsl}
\end{figure}

\citet{2019Guo2} implemented the RBSL method in \texttt{MPI-AMRVAC}, now available in the module {\tt src/physics/mod\_rbsl.t}. The module works for 3D Cartesian settings, allowing AMR. Fig.~\ref{fig:rbsl} shows the magnetic flux rope constructed by the RBSL method overlaid on the 304~\AA \ image observed by STEREO\_B/EUVI. In practice, one needs to determine four parameters to compute the RBSL flux rope, namely, the axis path $\mathcal{C}$, minor radius $a$, magnetic flux $F$, and electric current density $I$. The axis path is determined by triangulation of stereoscopic observations of STEREO\_A/EUVI, SDO/AIA, and STEREO\_B/EUVI at 304 \AA . The sub-photospheric counterpart $\mathcal{C}^*$ is the mirror image of $\mathcal{C}$ to keep the normal magnetic field outside the flux rope unchanged. The minor radius is determined by half the width of the filament observed in 304 \AA . The magnetic flux is then determined by the magnetic field observed by SDO/HMI at the footprints of the flux rope. Finally, the electric current $I=(\pm 5\sqrt{2}F)/(3\mu_0 a)$, where the sign is determined by the helicity sign of the flux rope. 

With all these four parameters, the RBSL flux rope is computed via Eqs.~\eqref{eq:rbsl1}--\eqref{eq:rbsl3}. It has to be embedded into a potential magnetic field to construct a magnetic configuration conforming with magnetic field observations. When $\mathcal{C}^*$ is the mirror image of $\mathcal{C}$, the RBSL flux rope has zero normal magnetic field outside the flux rope, while the magnetic flux inside the flux rope is $F$. Therefore, we subtract the magnetic flux $F$ inside the two footprints to compute the potential field. The RBSL flux rope field is embedded into this potential field. So, the normal magnetic field on the whole bottom boundary is left unchanged. The combined magnetic field might be out of equilibrium. We could relax this configuration by the magnetofrictional method \citep{2016Guo2,2016Guo1}. The final relaxed magnetic field is shown in Fig.~\ref{fig:rbsl}. This magnetic field could serve as the initial condition for further MHD simulations.

The RBSL method can also be applied to construct the analytical Titov--D\'emoulin modified model. An example is presented in \citet{2021Guo}, where the flux rope axis has to be a semicircular shape, and it is closed by a mirror semicircle under the photosphere. The major radius $R_c$ is a free parameter. The flux rope axis is placed on the $y=0$ plane and its center is located at $(x,y)=(0,0)$. The minor radius $a$ is also a free parameter. To guarantee the internal force-free condition, $a$ has to be much smaller than $R_c$. Then, the flux rope has to be embedded into a potential field to guarantee the external force-free condition. The potential field is constructed by two fictional magnetic charges of
strength $q$ at a depth of $d_q$ under the photosphere, which are along the $y$-axis at $y=\pm L_q$. The electric current and magnetic flux are determined by Eqs.~(7) and (10) in \citet{2014Titov}.

\subsection{Synthetic observations}\label{sec:euv}

For solar applications, it is customary to produce synthetic views on 3D simulation data, and various community tools have been developed specifically for post-processing 3D MHD data cubes. E.g., the 
{\tt FoMo} code \citep{2016Fomo} was originally designed to produce optically thin coronal EUV images based on density, temperature and velocity data, using CHIANTI (see \citet{2021DelZanna} and references therein) to quantify emissivities. {\tt FoMo} includes a more complete coverage of radio emission of up to optically thick regimes, and \citet{2020Pant} provides a recent
example of combined \texttt{MPI-AMRVAC}-{\tt FoMo} usage addressing non-thermal linewidths related to MHD waves. Another toolkit called {\tt FORWARD} \citep{2016Forward} includes the possibility to synthesize coronal magnetometry, but is {\tt Idl}-based (requiring software licenses) and an implementation for AMR grid structures is as yet lacking.

Since synthetic data for especially optically thin coronal emission is key for validation purposes, we now provide modules that directly perform the needed ray-tracing on AMR data cubes, in 
{\tt src/physics/mod\_thermal\_emission.t}. The module contains temperature tables specific for AIA, IRIS and EIS instruments, with coverage for various wavebands. One can synthesize both EUV and soft X-ray (SXR) emission, and the module can be used for synthetic images or for spectral quantifications. Images in {\tt vti} format are readily produced either during runtime, or in a post-processing step, where the user controls the relative orientation of the line-of-sight (LOS) to the data cube. We use this for 3D Cartesian data sets from MHD, allowing AMR.

\begin{figure}
    \centering
    \includegraphics[width=0.48\textwidth]{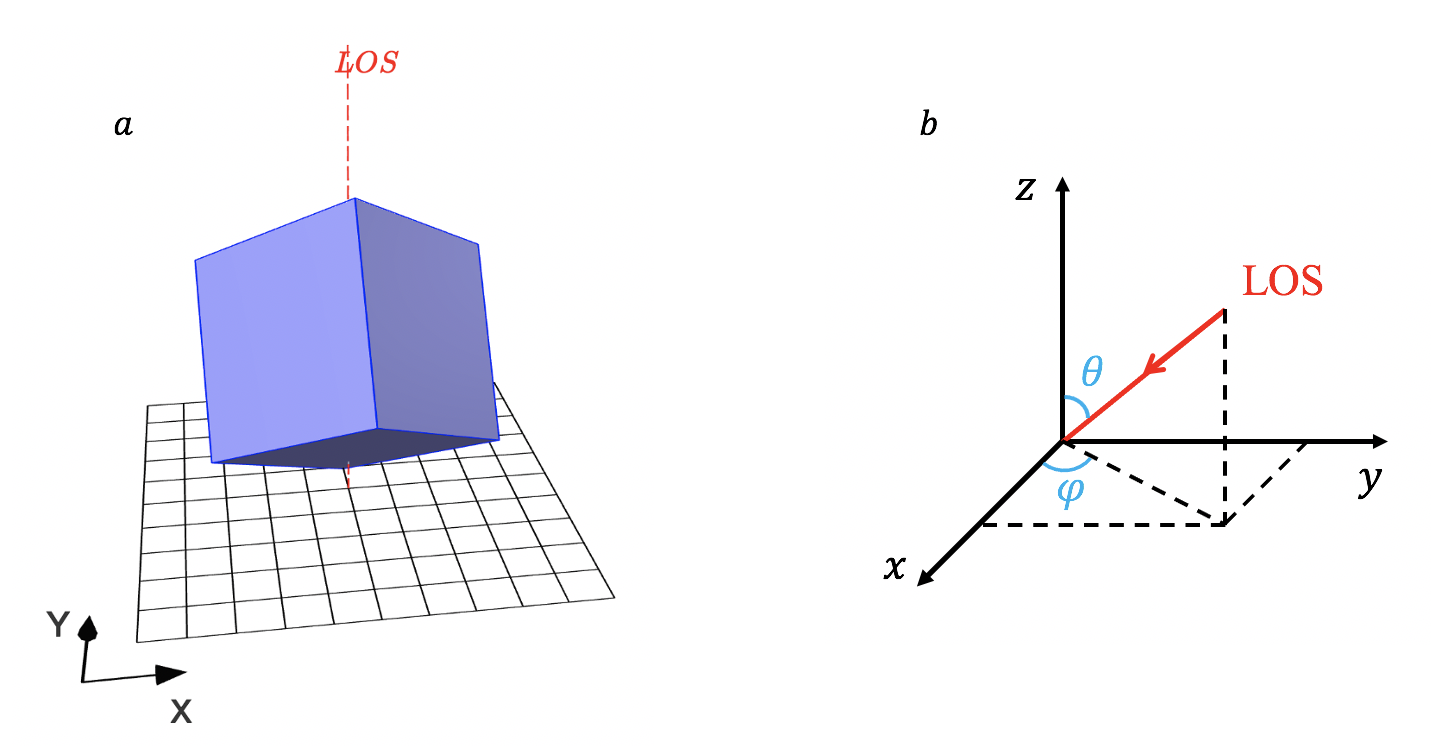}
    \caption{(a) Simulation box (blue cube), LOS (red dashed line) and EUV image plane (black mesh). The EUV image plane is perpendicular to LOS. (b) LOS depends on $\theta$ and $\varphi$ at simulation box coordinates. }
    \label{fig:syn_euv}
\end{figure}

As the first step in synthesizing an EUV image, a 2D table is created to record the luminosity of each image pixel. We assume that an EUV image has uniform pixel size, e.g.\ the pixel size of SDO/AIA images is 0.6 arcsec \citep{2012Lemen}. The spatial link between the EUV image plane and the 3D simulation box refers to Fig.~\ref{fig:syn_euv}a, where the mapping between simulation box coordinates $(x,y,z)$ and EUV image coordinates $(X,Y)$ is accomplished using the unit direction vectors $\vec{X_I}$ and $\vec{Y_I}$ of the image plane at simulation coordinates:
\begin{eqnarray}
X&=&(x,y,z) \cdot \vec{X_I} - (x_0,y_0,z_0) \cdot \vec{X_I}, \\
Y&=&(x,y,z) \cdot \vec{Y_I} - (x_0,y_0,z_0) \cdot \vec{Y_I}.
\end{eqnarray}
The vectors $\vec{X_I}$ and $\vec{Y_I}$ are both perpendicular to the line of sight and to each other, and are given by
\begin{eqnarray}
\vec{X_I} &=& \frac{ \vec{V}_{LOS} \times (0,0,1) }{ |\vec{V}_{LOS} \times (0,0,1)| } , \\
\vec{Y_I} &=& \frac{ \vec{X_I} \times \vec{V}_{LOS} }{ |\vec{X_I} \times \vec{V}_{LOS}| } ,
\end{eqnarray}
where $\vec{V}_{LOS}$ is the line of sight vector in simulation box coordinates. $\vec{V}_{LOS}$ can be written as
\begin{equation}
\vec{V}_{LOS}=(-\cos\varphi \sin\theta,-\sin\varphi \sin\theta,-\cos\theta),
\end{equation}
with the user-given parameters $\theta$ and $\varphi$ (Fig.~\ref{fig:syn_euv}b). 
The user-defined parameter $(x_0,y_0,z_0)$, which has a default value of $(0,0,0)$, can be any point in the simulation box and can then be mapped to the EUV image coordinate origin $(X=0,Y=0)$.

The integral EUV flux from each cell in the simulation box is computed and then distributed over the table, where the cell flux is given by
\begin{equation}
I_c = N_{ec}^2 G(T_{ec}) \Delta x \Delta y \Delta z ,
\label{eq:cell_euv_flux}
\end{equation}
where $N_{ec}$ is the cell electron number density, $T_{ec}$ is the cell electron temperature, $G$ is the contribution function of the corresponding EUV line given by the {\tt CHIANTI} atomic database, and $\Delta x$, $\Delta y$, and $\Delta z$ are the cell sizes in three directions \citep{2021DelZanna}. Due to instrument scattering, a single point source will appear as a blob in EUV observations. This effect is taken into consideration when distributing cell flux to image pixels by multiplying a Gaussian-type point spread function (PSF) \citep{2013Grigis}. The resulting pixel flux is given by
\begin{eqnarray}
I_p &=& \sum_i I_{c,i} \int_{X_{min}}^{X_{max}} \int_{Y_{min}}^{Y_{max}} \frac{1}{2\pi \sigma^2} \\
& & \times\exp\left[\frac{-(X-X_{c,i})^2 - (Y-Y_{c,i})^2}{2 \sigma^2}\right] dX dY \nonumber \\
&=& \sum_i \frac{I_{c,i}}{4} \left[\mathrm{erfc}\left(\frac{X_{min}-X_{c,i}}{\sqrt{2} \sigma}\right) - \mathrm{erfc}\left(\frac{X_{max}-X_{c,i}}{\sqrt{2} \sigma}\right)\right] \nonumber \\    
& & \times\left[\mathrm{erfc}\left(\frac{Y_{min}-Y_{c,i}}{\sqrt{2} \sigma}\right) - \mathrm{erfc}\left(\frac{Y_{max}-Y_{c,i}}{\sqrt{2} \sigma}\right)\right],
\label{eq:flux_int}
\end{eqnarray}
where $i$ is the cell index, $I_{c,i}$ is the integral EUV flux from the $i$th cell, $(X_{c,i},Y_{c,i})$ is the mapping result of the cell center at the image plane, and $X_{min}$, $X_{max}$, $Y_{min}$ and $Y_{max}$ give the borders of the pixel. The standard deviation $\sigma$ of the PSF is taken from the related papers of the corresponding spacecraft, and is usually close to the pixel size. When the cell size is not small enough compared to the EUV image pixel size (for example, the projection of any cell edge at the image plane is larger than half the pixel size), a cell is split into multiple parts before the calculation of Eqs.~\eqref{eq:cell_euv_flux}--\eqref{eq:flux_int} in order to improve the integral accuracy.

\begin{figure}
    \centering
    \includegraphics[width=0.48\textwidth]{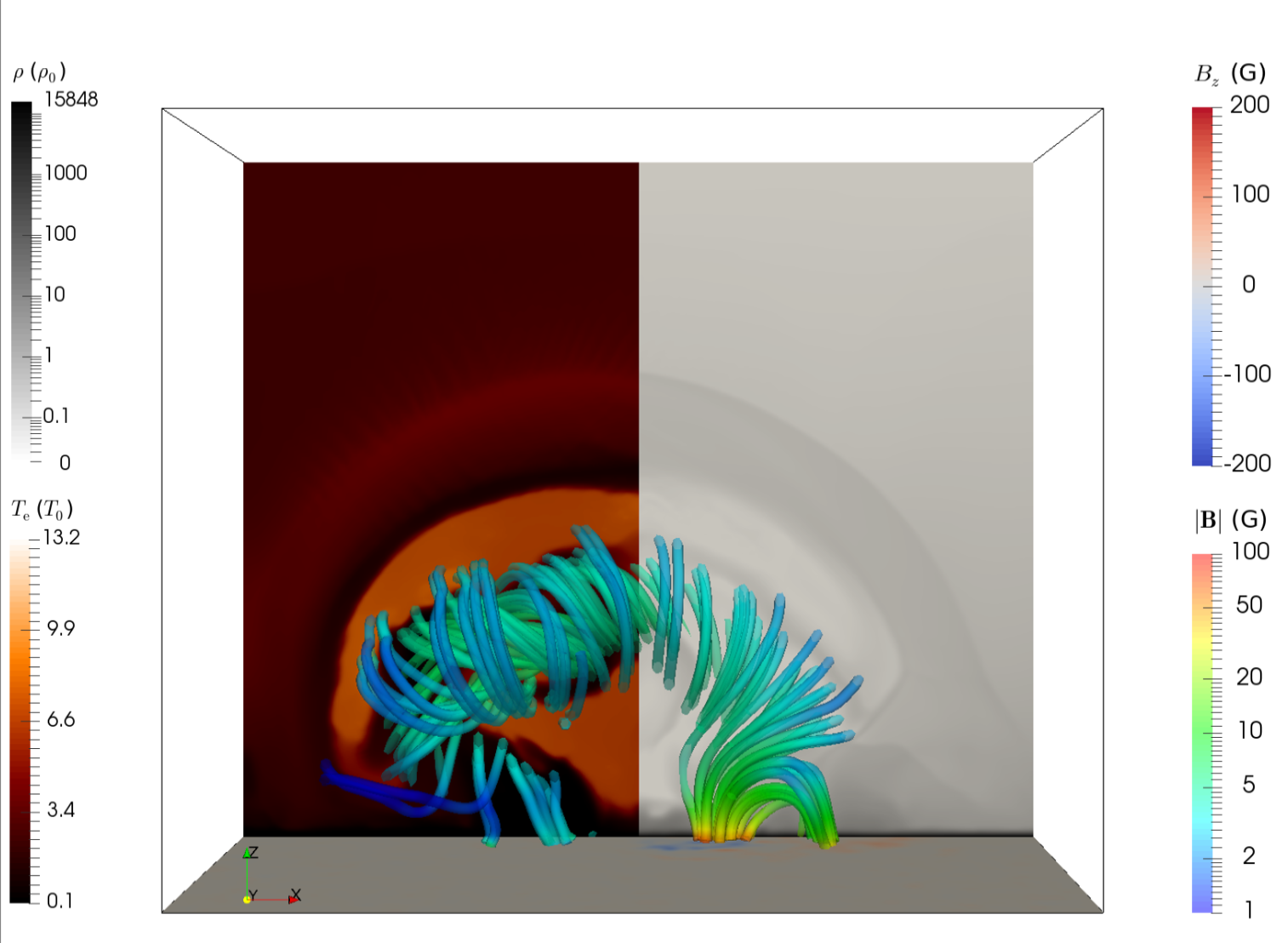}
    \caption{A data-driven MHD model, with all energetics incorporated. The vertical slices show the temperature on the left and density on the right, respectively. The magnetic field lines are colored by the field strength. The bottom slice displays the normal magnetic field, $B_z$.}
    \label{fig:mhd}
\end{figure}
\begin{figure*}
  \FIG{
    \centering
    \includegraphics[width=0.8\textwidth]{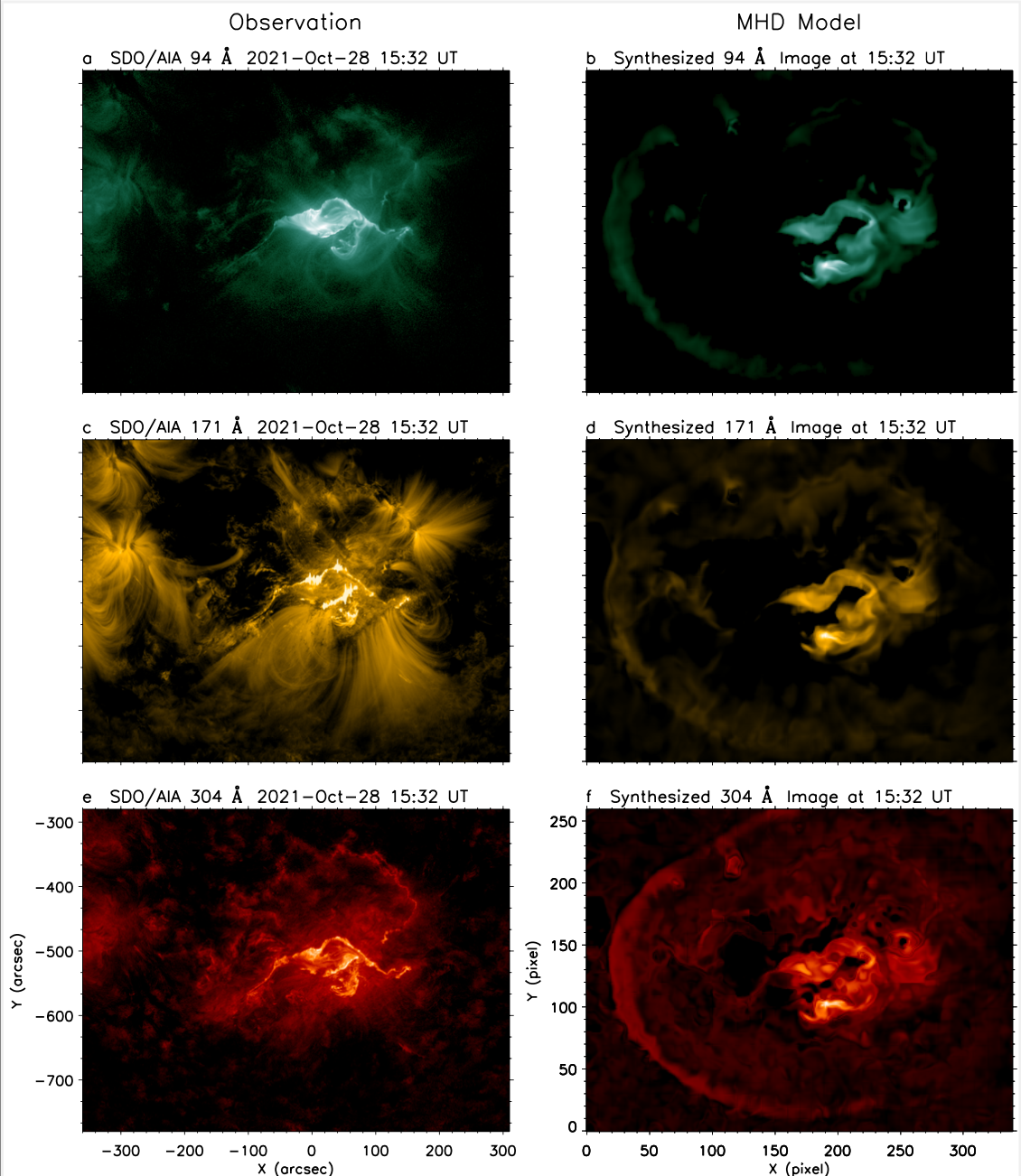}
    \caption{Comparison between SDO/AIA observations and synthesized EUV images from a data-driven MHD model (as in Fig.~\ref{fig:mhd}), including a full energy treatment. The left column shows the SDO/AIA observations at wavebands of (a) 94 \AA , (c) 171 \AA , and (e) 304 \AA . The right column shows the emission at the same waveband as the left synthesized from the MHD model at the same time.}
    \label{fig:euv}
  }
\end{figure*}

Fig.~\ref{fig:mhd} shows a snapshot of a data-driven MHD model for the X1.0 flare at 15:35 UT on 2021 October 28. The 3D MHD model considers a full energy equation with background heating, thermal conduction, and optically thin radiation losses. A detailed analysis of this simulation will be presented in a future paper. Here, we use a single snapshot from the evolution to synthesize EUV images, to demonstrate this new capability of \texttt{MPI-AMRVAC}. Fig.~\ref{fig:euv} shows comparisons of SDO/AIA observations and the synthesized EUV images from the data-driven MHD model. We select three different wavebands at 94 \AA , 171 \AA , and 304 \AA . It is found that the simulation and its synthesized EUV images reproduces qualitatively various aspects seen in the flare ribbons. In contrast to the actual observed images, coronal loops are not reproduced very accurately (as shown in Fig.~\ref{fig:euv}c and \ref{fig:euv}d), and the simulation displays a relatively strong spherical shock front, seen in all three wavebands. These aspects rather call for further improvement of the (now approximate) radiative aspects incorporated in the data-driven MHD model, but these can still be improved by adjusting the heating-cooling prescriptions and the magnetic field strength. Here, we only intend to show the synthesizing ability of \texttt{MPI-AMRVAC}, which has been clearly demonstrated.

\section{Handling particles and fieldlines}

\subsection{Sampling, tracking or Lorentz dynamics}\label{sec:part}

\begin{figure*}
    \centering
    \includegraphics[width=0.91\textwidth]{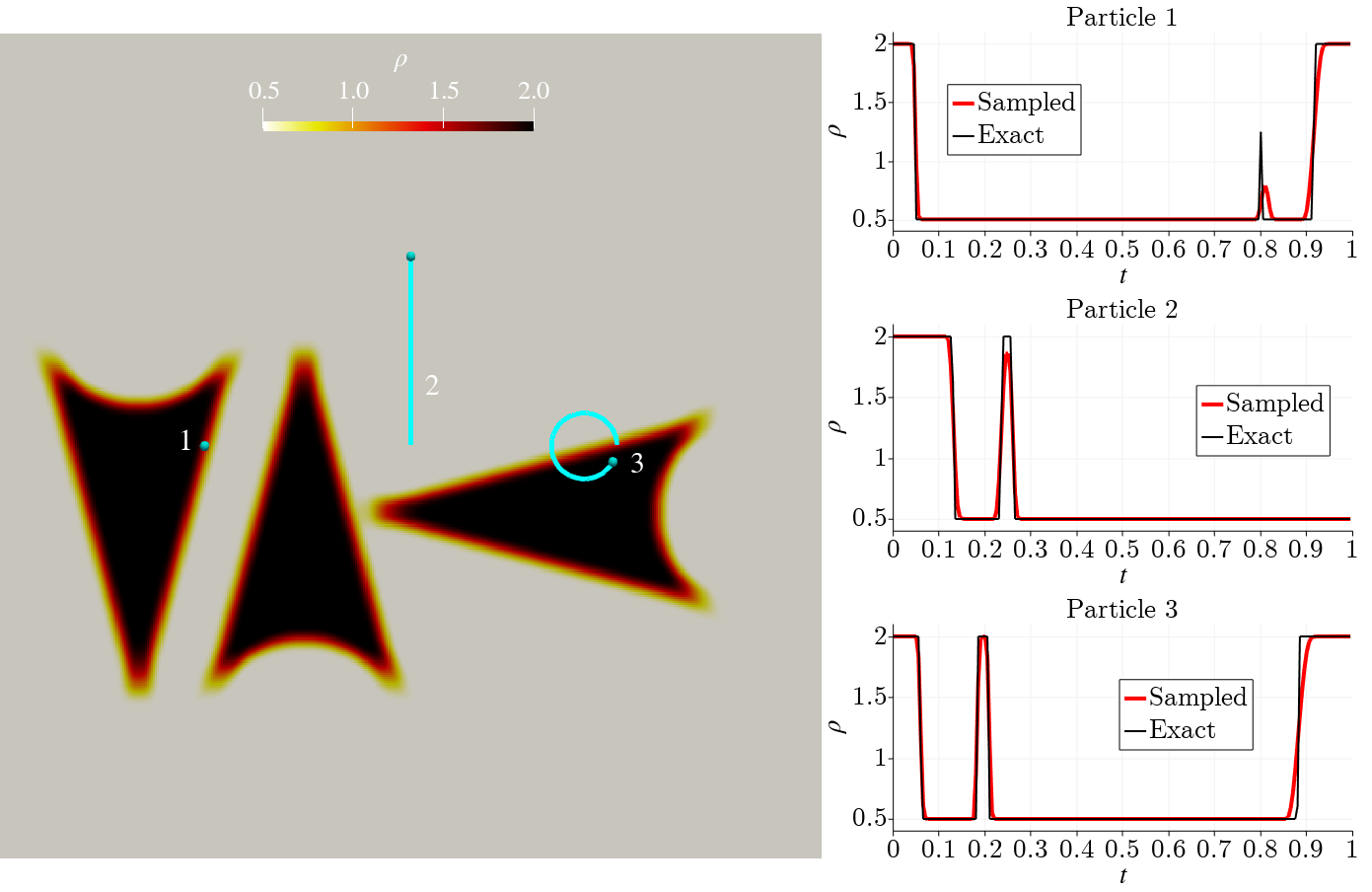}
    \caption{A demonstration of the sampling possibilities, where a 2D scalar linear advection problem is augmented with sampled solutions at three locations that follow user-specified orbits. The left panel shows the solution at $t=0.92$, along with the trajectories of sampling particles (blue spheres and lines). The right panels show the numerical (in red) and the analytic solution (black) as function of time for the three locations. An animation is provided.}
    \label{fig:vacparticles}
\end{figure*}

The {\tt src/particle} folder contains all options for handling `particle' dynamics that one may relate to the PDE system at hand. For any of the PDE modules listed in Table~\ref{tab:eqs}, computational particles can be employed to meaningfully sample all solution values at pre-chosen fixed or dynamically evolving locations. We illustrate this with a 2D linear scalar advection problem, where the particle module samples the solution at three distinct locations: one fixed in space and time, another one moving with constant speed on a vertical straight line, and a third location that follows a uniform circular motion. On a square domain, the diagonally advected profile corresponds to our {\tt VAC} logo, with constant values $\rho=0.5$ exterior, and $\rho=2$ interior to the letters, and Fig.~\ref{fig:vacparticles} shows the $\rho(x,y,t=1)$ distribution for a 4-level AMR run using a three-step TVDLF run with `koren' limiter \citep{1993Koren}. The sampling `particles' and their trajectories are shown in blue. The plots on the right show the sampled data for the three particles as function of time. The complete setup is provided in {\tt tests/demo/Advect\_ParticleSampling\_2D}, and demonstrates how the user can add and define additional variables (here corresponding to the exact solution $\rho_{\mathrm{exact}}(x,y,t)$ at given time $t$ and the error with respect to the exact solution) and how to add a payload (namely the sampled exact solution) to the particle sampler. This kind of sampling on prescribed user-defined trajectories could be relevant for 3D space weather related MHD simulations as done by {\tt ICARUS} \citep{2022Verbeke}, for comparison with actual spacecraft data.
The interpolations from (AMR) grid cell centers to locally sampled data are done by linear (bilinear/trilinear in 2D/3D) interpolation, where also linear interpolation in time is performed when dynamic evolutions occur. The actual subroutine for interpolating any field variable to a specific grid location can be modified at will, and is called \texttt{interpolate\_var} in the {\tt particle/mod\_particle\_base.t} module. Other interpolation schemes (e.g.\ quadratic or higher-order) can be implemented there.

\begin{figure}
    \centering
    \includegraphics[width=\columnwidth]{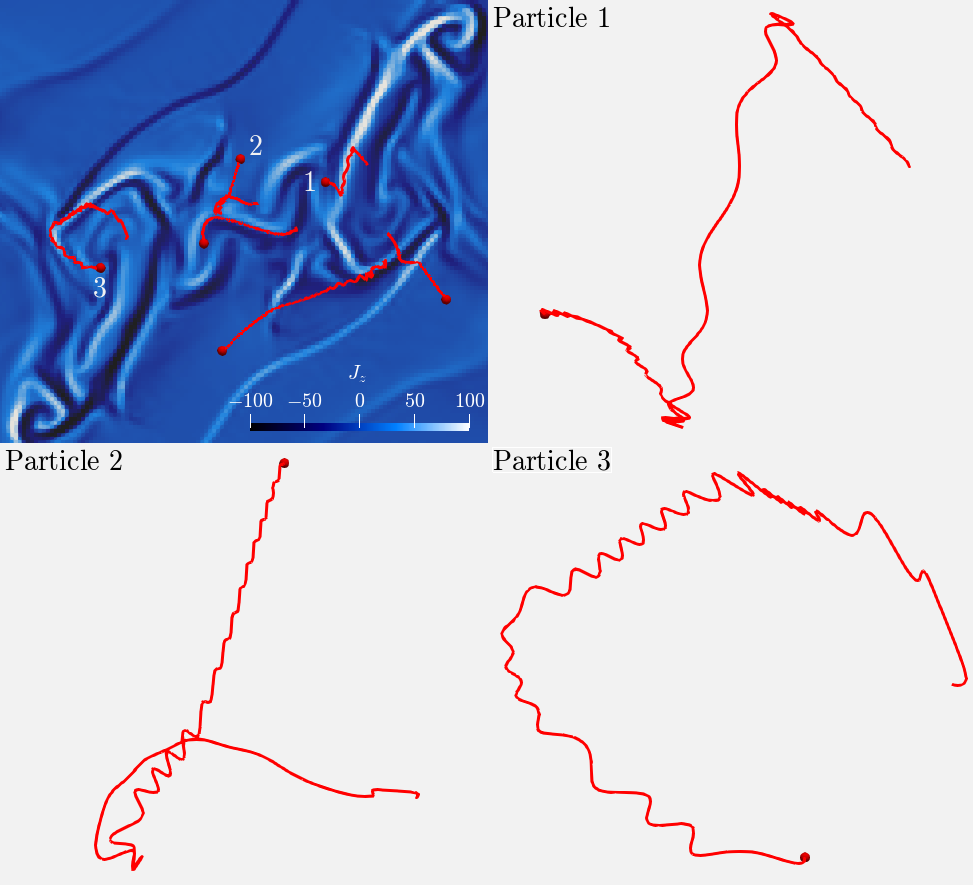}
    \caption{Demonstration of charged-particle tracing in an MHD simulation. The background fluid state is taken from Section \ref{sec:tilt} and kept fixed while several charged particles are traced (red lines in the top-left panel) by numerically solving the equations of motion. Selected zoomed-in trajectories (top-right, bottom-left, and bottom-right panels) show the typical oscillatory and drift motion of charged particles around and across magnetic field lines. An animation is provided.}
    \label{fig:tiltparticles}
\end{figure}

A second use-case for the {\tt src/particle} folder is specific to any PDE system featuring a vector velocity field, such as the HD or MHD systems. In that case, we may be interested in gas or plasma properties at locations that follow the flow, hence positions that are merely advected in a Lagrangian fashion. This is possible with the {\tt mod\_particle\_advect.t} module. 

Whenever plasma is involved, such as in an MHD (or two-fluid plasma-neutral) setting, we can also quantify instantaneous magnetic and electric field data from the MHD variables. This can then be used to compute the trajectories of charged test particles with given mass $m$ and charge $q$, according to the standard Lorentz equation, where acceleration $\mathbf{a}$ follows from $m{\mathbf{a}}=q\left(\bfE+\bfv\times\bfB\right)$, or its fully relativistic variant (see e.g.\ \citealt{2018Ripperda}). The latter has been used in \citet{2021Zhao} to analyse relativistic particle acceleration processes in 2D resistive MHD simulations of chaotic island formation (also occuring in the tilt evolution from Section~\ref{sec:tilt}). We demonstrate here the capability of tracing charged particles in MHD simulations by using the electromagnetic field data from Section \ref{sec:tilt} at $t=8.5$ (roughly corresponding to Fig.~\ref{f:tilt}) and evolving particles in this fixed MHD background. Fig.~\ref{fig:tiltparticles} (top left) shows the trajectories of selected particles (evolved for a time $t=0.5$) in the central $xy$ region $[-0.25,0.25]\times[-0.125,0.125]$ of the aforementioned MHD run, where island interaction creates chaotic magnetic structures and strong out-of-plane currents $J_z$. The same Figure (top right, bottom left, bottom right) presents a zoom-in onto three selected particle trajectories, showing explicitly the oscillatory and drift motion of charged particles around and across magnetic field lines. Several integration methods are available to numerically solve charged-particle motion in \texttt{MPI-AMRVAC}, either in the Newtonian or in the relativistic limit.

In many astrophysically relevant situations, one may be faced with unusually large differences in the (small) lengthscale set by the charged particle local gyroradius, and the one associated with (gradients in) the background electromagnetic field quantities. In that case, it can be advantageous to work with the guiding center approximation (GCA) where one solves a set of ODEs where the fast gyromotion is averaged out. The use of GCA equations in \texttt{MPI-AMRVAC} was, e.g., demonstrated in 3D MHD setups of Kelvin-Helmholtz unstable magnetopause setups featuring particle trapping in \citet{2019Leroy}. The various options for the relativistic Lorentz equation integrators implemented in \texttt{MPI-AMRVAC}, as well as the governing GCA equations with references to their original sources can be found in \citet{2018Ripperda}. A summary in the online documentation is the markup-document in {\tt doc/particle.md}.

\subsection{Fieldline tracing}\label{sec:trace}

We introduce a generic fieldline tracing module (in {\tt src/module/mod\_trace\_field.t}, which is able to trace various types of fieldlines that start at user-specified points, either during runtime, or in post-processing fashion. At the moment, this functionality works for 3D Cartesian AMR grids (but does not account for possibly stretched grids discussed in Section~\ref{sec:stretch}). The flow chart for tracing a single magnetic fieldline through a memory-distributed block-AMR setting has been introduced in \citet{2020Ruan} (their Appendix~B, Fig.~B1). We now add a functionality to trace multiple field lines in parallel, where now multiple starting points can be provided for a set of field lines to be traced through the AMR grid hierarchy. We also generalize the implementation to handle any type of vector field, such that we can plot or trace (1) magnetic fieldlines in MHD (or multi-fluid) settings, but also (2) flow streamlines for the velocity field, and (3) any user-defined vector field (e.g.\ useful for visualizing or quantifying electric fields and currents). This functionality is  demonstrated in Fig.~\ref{fig:KH} where it is used to compute and visualize velocity streamlines, and in Fig.~\ref{f:tilt} where the magnetic fieldlines shown are also calculated with this module. For these 2D cases (the tracing implementation works in 2D and 3D), we employ the method presented in \citet{1997Jobard} to select seedpoints to get evenly spaced streamlines/fieldlines. 

During this tracing, users can ask to interpolate any set of self-defined derived quantities to the fieldlines. This ability is, e.g., crucial to the {\tt TRAC} method for handling sharp transitions in temperature and density fields, where along the trajectories of magnetic field lines, the temperature gradients along the line tangent direction are required \citep{2021Zhou}.
In \citet{2020Ruan}, magnetic field lines in a 2D reconnection setup inspired by the `standard solar flare model' were computed during runtime, and model equations along the fieldlines were solved to quantify how the energy release by reconnection gets dynamically deposited in remote locations (by energetic electron beams that collide with lower-lying, denser chromosphere material). This involves back and forth interpolations between grid and fieldlines. Together with the general functionality provided through the particle module (Section~\ref{sec:part}), it then becomes possible to provide dynamically evolving seedpoints such that one traces the exact same fieldlines, simply by using Eulerian advection on the seeds.

\section{Data processing and technical aspects}

\subsection{Customized user-interfaces}

For any new application, the minimal requirement is to code up an application-specific {\tt mod\_usr.t} file where at least the initial condition for all involved variables must be provided. The input parameter file {\tt amrvac.par} makes use of {\tt Fortran} namelists to then select the time stepping scheme, the spatial discretization, I/O aspects, boundary conditions, and parameters that control the evolving AMR grid structure. For every PDE system from Table~\ref{tab:eqs}, equation-specific parameters may need to be set as well. 

The actual code versatility follows from the many ways to allow user-customized adaptations. These include as most commonly used ones:
\begin{itemize}
    \item the addition of specific source terms in the equations, and their corresponding effect on time step constraints for explicit time stepping strategies;
    \item the definition of user- or application-specific (de)refinement strategies, based on combinations of either purely geometric or solution-derived quantities;
    \item the design of specific boundary conditions on user-selected variables;
    \item the way to add derived variables to the I/O routines or to carry out post-processing during conversion or runtime on data files;
    \item the possibility to process data at every timestep; 
    \item the way to handle particle data and payloads.
\end{itemize}
The complete list of all customized subroutine interfaces is given in {\tt src/mod\_usr\_methods.t}. Many examples of their intended usage can be found in the {\tt tests} folder.

\subsection{Initializing from datacubes or driving boundaries}

\begin{figure}
    \centering
    \includegraphics[width=\columnwidth]{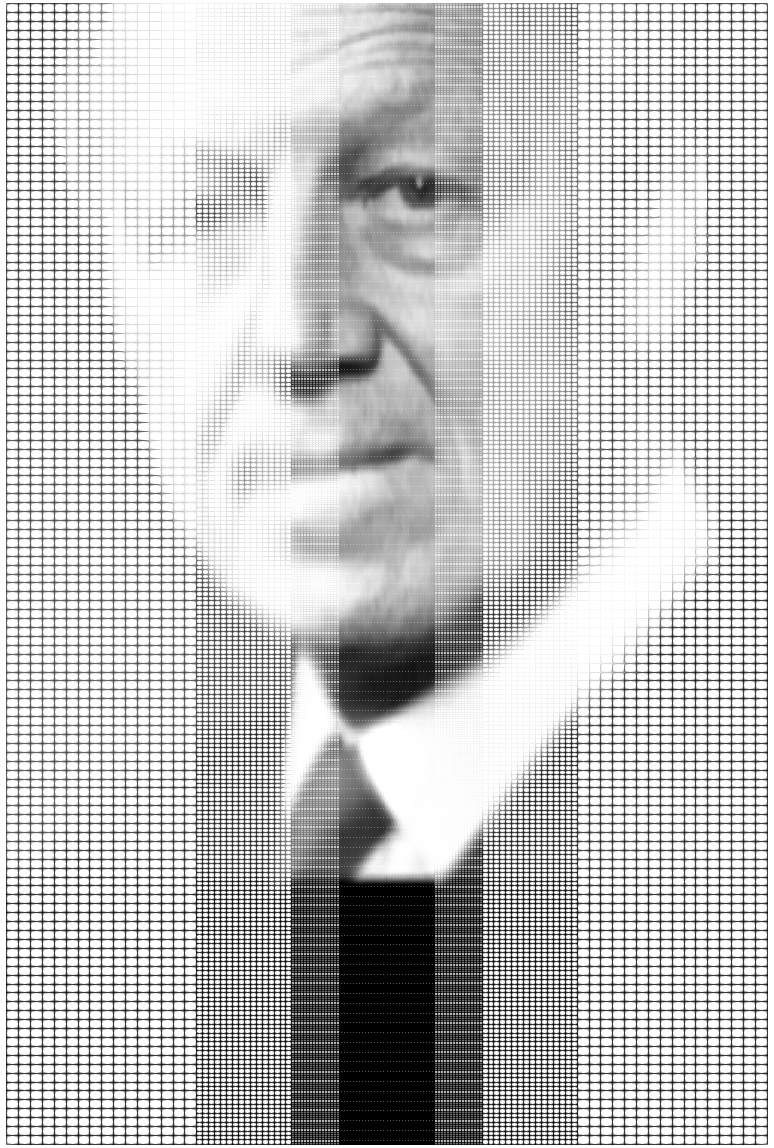}
    \caption{The density from a 2D, boundary-driven advection problem: the {\tt alfven.vtk} image is used to set time-dependent boundary values at the top edge, advected into the domain by a constant advection speed $\bfv=-\mathbf{e}_y$. We show the solution at $t=1.15$, on a $[0,1]\times[0,1.5]$ domain. The 4-level AMR hierarchy is purely geometrically controlled in this example, forcing the central band of the image to appear sharpest.  
    }
    \label{f:alfvenbc}
\end{figure}

The code offers various possibilities to 
read in structured data sets, which can be used for initialization or boundary driving purposes. E.g., the Alfv\'en shock test in Section~\ref{sec:alfven} showed how a {\tt vtk} image can be used to set the density at time $t=0$ in some part of the domain, while 2D and 3D advection tests can use the same image as a boundary condition. An impression of such 2D boundary-driven advection using the {\tt alfven.vtk} image (provided in {\tt tests/demo/Advect\_BC\_from\_file\_2D}), while enforcing user-controlled, purely geometric AMR (only a central region is at the highest AMR level) is provided in Fig.~\ref{f:alfvenbc}. Such 2D advection of the image mimics a faxing process where the image gets advected into the domain as time proceeds (this is a time-dependent 1D boundary driving realized by a single 2D image), and 3D variants would realize time-dependent 2D boundary driving. Initializing data uses the {\tt src/mod\_init\_datafromfile.t} module, which currently expects a {\tt vtk} formatted dataset (2D or 3D). This can easily be adjusted to read in structured data from other simulation codes, for detailed intercomparisons, or for revisiting structured grid data evolutions with AMR runs that may zoom in on details. All functionality discussed above stores the read-in data in a lookup-table, which allows for easy and efficient interpolations to the AMR grid.
As usual, when reading in external data, the user must beware of a possible need to introduce floor values (e.g.\ ensuring positive densities/pressures), or handling data outside the region covered by the given datacube. The time-dependent magnetofrictional module discussed in Section~\ref{sec:mf} is yet another example of how one can handle time-dependent boundary data, as provided from external files.

\subsection{AMR and multi-dimensional stretching in orthogonal coordinates}\label{sec:stretch}
\texttt{MPI-AMRVAC} allows for anisotropic grid stretching, that is, independent grid stretching prescriptions for each spatial coordinate, combined with AMR, for any geometry. A `stretched' grid does not imply enlarging the domain, but rather the use of non-uniform grid cells to cover a given domain, with cells that become gradually larger/smaller in a `stretched' direction. This generalizes our previously documented purely radial grid stretching for spherical (or polar/cylindrical) applications as documented in \citet{2018Xia}. Currently two stretching configurations are supported:
\begin{itemize}
  \item Unidirectional stretching, where every cell is stretched by a constant factor (the ``stretching factor'') in the specified spatial direction from cell to cell.
  \item Symmetric stretching, where one can specify the amount of blocks that should remain uniform, with the remaining blocks on either side stretched. This must have an even number of grid blocks on the base AMR level, and adopts a central symmetry.
\end{itemize}

\begin{figure*}
    \begin{minipage}{0.48\textwidth}
        \centering 
        \includegraphics[width=\textwidth]{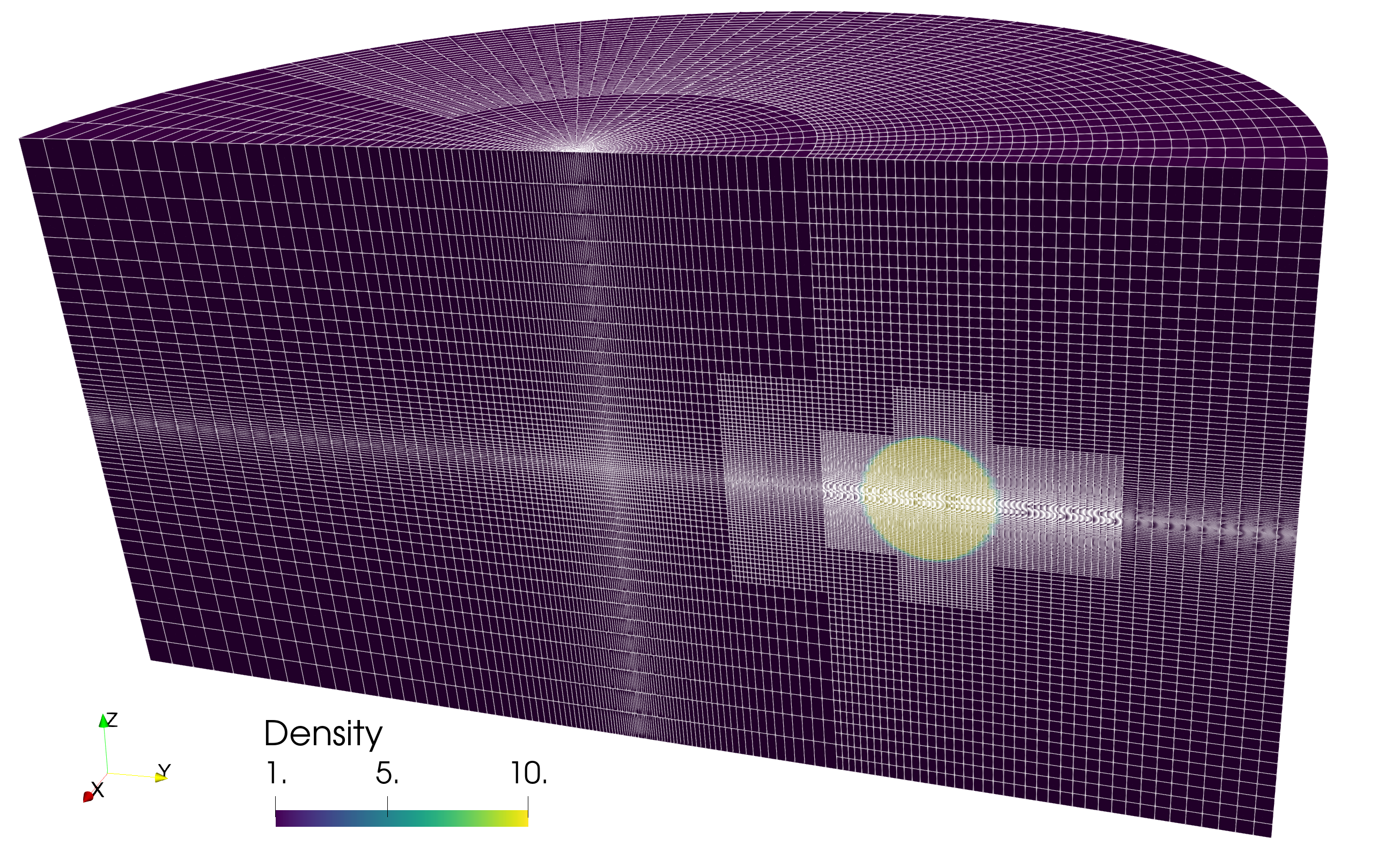}
    \end{minipage}
    \begin{minipage}{0.48\textwidth}
        \centering 
        \includegraphics[width=\textwidth]{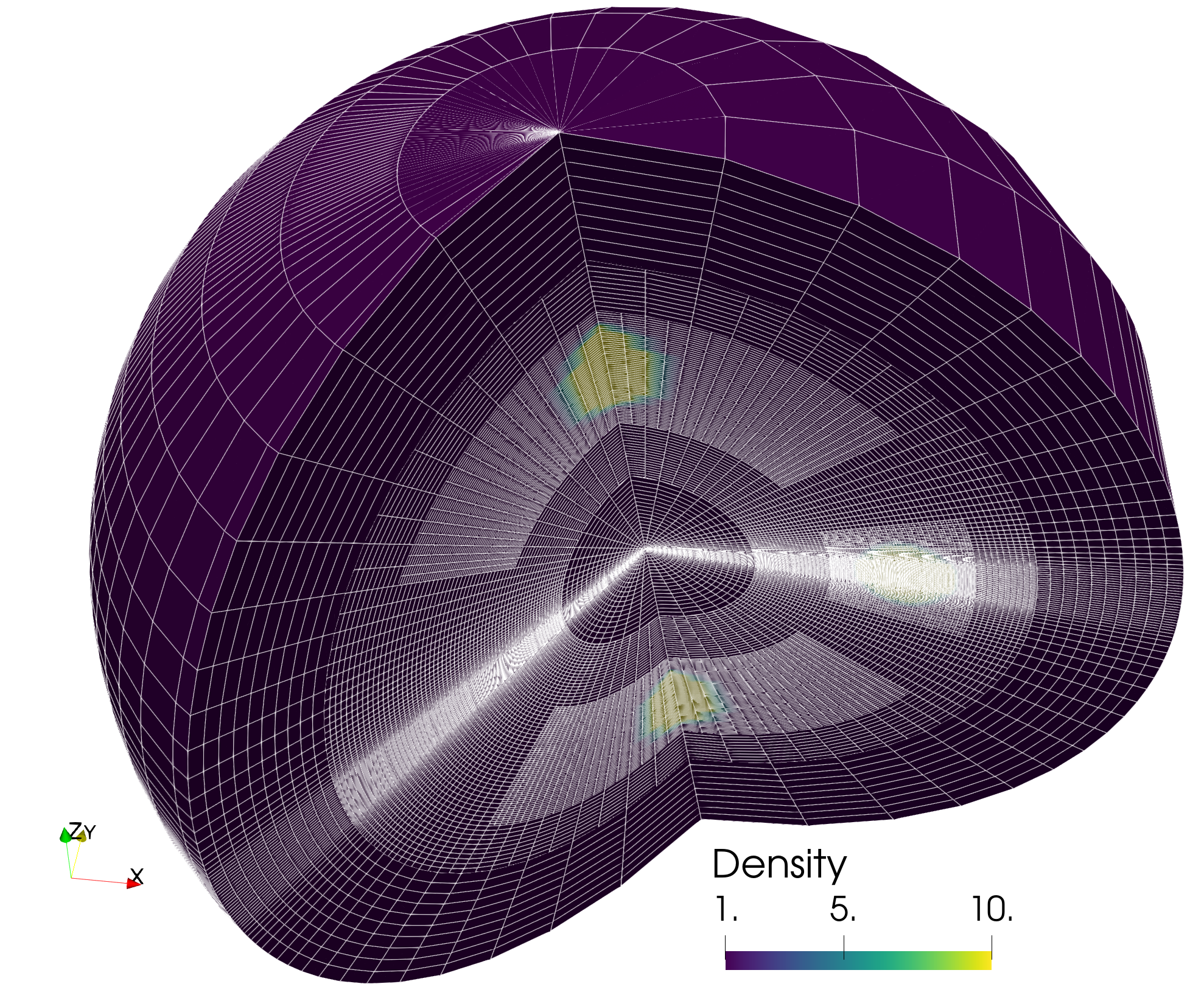}
    \end{minipage}
    \caption{
        A 3D cylindrical (left) and spherical (right) AMR grid with anisotropic stretching.
        Both cases are unidirectionally stretched in the radial direction, with additional symmetric $z$-stretching for the cylindrical one and both $\theta$ and $\phi$ symmetric stretching for the spherical one.
    }
    \label{fig: stretched_grids_3d}
\end{figure*}

Fig.~\ref{fig: stretched_grids_3d} showcases this capability for 3D cylindrical (left panel) and spherical (right panel) geometries. In both examples a uniform medium has been set up in which denser, uniform spheres have been added to the domain to trigger AMR. The domain itself consists of a $60^3$ grid with $6$ blocks in each direction. The cylindrical case employs unidirectional radial stretching with a stretching factor of $1.05$, combined with symmetric stretching for all $6$ blocks in the $z$-direction (so no uniform central blocks along the $z$-axis). The spherical case on the other hand employs anisotropic stretching in all three directions, $r$ is unidirectionally stretched with stretching factor $1.02$. Both the $\theta$- and $\phi$ directions are symmetrically stretched in four blocks with stretching factors $1.15$ and $1.2$, respectively. Note that these factors have been exaggerated for visual effects, in typical use cases a stretching factor between $1.01$ and $1.05$ is reasonable.

We caution that not all reconstruction procedures to compute cell edge from cell center values are fully compatible with this stretching, but e.g.\ WENO flavors listed in Table~\ref{t:limiters} with the `nm' extension all can be used reliably. These have been introduced as the Non-uniform Modified WENO variants in \citet{2018Huang}.

\subsection{Bootstrapping: handling small thermodynamic quantities}\label{sec:bootstrap}

Whenever extreme contrasts occur in thermodynamic (densities, pressures, temperatures) or magneto-kinetic quantities (high/low plasma beta and/or Mach numbers) within the simulated domain, it may become crucial to impose bootstrapping strategies to prevent unphysical states (negative pressure, or too low densities) from ever occuring. In the typical finite-volume methods as used here (and in many other software frameworks), this can be enforced numerically in a variety of ways. \texttt{MPI-AMRVAC} imposes such strategies by the global logical parameters {\tt check\_small\_values} and {\tt fix\_small\_values}, which by default are {\tt .true.} and {\tt .false.}. This default strategy means that a simulation will stop and simply report the error. This may signal user (or developer) implementation errors, which must be dealt with first. It may also signal problems in boundary treatments, or relate to as yet unresolved initial or instantaneous variations that may need higher resolutions. The advantage of having many choices of schemes/limiters then come in handy, since when problems persist for the most robust or diffusive combinations, a bug is almost surely at play.

However, there may be (rare) situations where round-off errors, or the various intricate non-linear prescriptions (such as involved in limited reconstructions) themselves create unphysical conditions, e.g., when recently updated conserved variables no longer translate to physical primitive variables. Therefore, we allow for activating fixes controlled by {\tt small\_pressure} and {\tt small\_density} user inputs, and offer a number of ways to recover from unphysical states. In practice, these bootstrapping procedures are then in effect on entry for the step translating conservative to primitive variables, can be used when switching total to partial energy density contributions, and could enforce a thermal pressure calculation to always return values equal to or above {\tt small\_pressure}. Fixes could also be enforced at the end of specific source term additions, but the bootstrapping will never modify magnetic fields, and will obviously no longer strictly obey energy conservation (or mass conservation when having positive {\tt small\_density}). It is then advised to monitor this possible accumulation of errors. Such bootstrapping was also found necessary to handle the occuring near-vacuum regions in thermal runaway simulations (as in Section~\ref{sec:otc}) in an MHD setting \citep{2021Hermans}. Bootstrapping measures may simply replace faulty values by the user-set {\tt small\_pressure} and {\tt small\_density} combinations, or work with a user-controlled area (line, square or cube in 1D, 2D or 3D) around the problem cell, to perform averaging from surrounding non-problematic cells on the primitive entries. In combination with specific checks enforced on the reconstructions from cell center to cell edges, this bootstrapping should then avoid fatal code crashes, without sacrificing physical reality. It is to be noted that various codes handle this in undocumented fashion, and may exploit (excessive) hyper-diffusion or parameter-rich clipping strategies whenever strong shocks, unresolved gradients, or actual discontinuities are at play.

\subsection{Data analysis routines and visualizations}\label{sec:vtu}
Natively, \texttt{MPI-AMRVAC} outputs its data in a custom dataformat (hereafter referred to as ``datfiles'' after its use of the typical \texttt{.dat} extension), which relies on standard I/O routines. This allows for optimized read/write I/O operations, thereby minimizing the amount of time needed to write data to disk. The custom (and compact) datfile format also implies efficient reading of the data, which the code can use to restart and/or continue the simulation from a previous snapshot. One major drawback of storing data this way is that a custom dataformat as employed here is unrecognizable by third-party software used for analysis or visualization. To mitigate this issue the code has various conversion routines implemented, which can convert existing datfiles to more standardized formats such as \texttt{.vtu} (VTK unstructured data) that are directly accessible by visualization packages such as {\tt ParaView} or {\tt VisIt}. While this approach is reasonable for smaller simulations it raises various issues for large-scale runs that output huge datfiles. The need for \texttt{.vtk} files for example results in additional usage of diskspace and more I/O-time due to converting. Furthermore, datfiles from large simulations are usually too big to load directly in memory.

For large datasets users can make use of \texttt{yt}, an open-source Python package for data analysis and visualization tailored to astrophysical codes \citep{2011Turk}. \texttt{Yt}'s core functionality is written in Cython and heavily relies on NumPy (and its C-API) for fast and efficient array operations. We developed a dedicated frontend that ensures \texttt{yt} can directly load native \texttt{MPI-AMRVAC} datfiles, eliminating the need for conversion to other data formats. Additionally, data loading through \texttt{yt} makes use of memory-mapping, where a first pass over the data maps the on-disk structure to physical space prior to actually loading. Data is then selectively retrieved based on queries by the user, which avoids caching all data directly into memory. All of \texttt{MPI-AMRVAC}'s possible geometries are supported as well. Some features such as staggered/stretched grids and magnetic field splitting, amongst others, are not yet supported, but will be in due time.

Historically, \texttt{MPI-AMRVAC} only saves the conserved variables to the datfile, but the corresponding \texttt{.vtu} files may switch to primitive variables, or even add any number of user-added additional fields (like temperature, divergence of the velocity, etc.).  Similarly, used-defined fields can be saved in datfiles, that can be later loaded in \texttt{yt}. This is demonstrated in {\tt tests/demo/AlfvenShock\_MHD2D} and
 {\tt tests/demo/Tilt\_Instability\_MHD2D}. Of course, adding more variables implies even more diskspace. To mitigate this, the user has full control on how many (and which) derived variables are stored to the \texttt{.vtu}  and \texttt{.dat} files, and can generate these post-process. A major advantage \texttt{yt} has over third-party visualization tools is the use of so-called ``derived fields''. These are quantities that are derived from one or more primitive variables and can be defined by the user. These fields are then composed using the queried data and allows visualization of (most) quantities that can be derived from others without having to save them to disk.

It should be noted that the use of {\tt ParaView} or {\tt VisIt} versus {\tt yt} is application-dependent and should be regarded as complementary. For smaller datasets users may prefer the graphical user interface of third-party software for visualisation rather than Python code. For large datasets {\tt yt} is (or should be) preferred.

\section{Future directions}

We provide an updated account of the open source \texttt{MPI-AMRVAC 3.0} framework, one among many AMR-supporting frameworks in active use for computational solar and astrophysically motivated research. We close this paper with an outlook to further promising developments.

Most of the currently implemented physics modules (HD, MHD, plasma-neutral 2-fluid) assume no or purely local (adequate for optically thin conditions) interactions between the gas/plasma and the radiation field. The radiative-hydro module \citep{2022Moens} using flux-limited-diffusion (FLD)  now included in the {\tt 3.0} version realizes only the first step towards more consistent radiative (M)HD treatments. First intrinsically coupled radiative-MHD modeling with \texttt{MPI-AMRVAC} include 2D axisymmetric magnetized wind studies for hot stars \citep{2021Driessen}, where line driving is essential both for the wind generation, and for realizing conditions prone to the line-deshadowing instability, causing clumpy, magnetized transmagnetosonic winds. These models use an isothermal closure relation, but handle the line-driven aspect of hot, massive star winds by means of a cumulative force contribution\footnote{This module is provided by F. Driessen as {\tt src/physics/mod\_cak\_force.t} that can be combined with HD and MHD modules, documented as {\tt doc/cakforce.md}.} as introduced by \citet{1975cak}. 

The FLD formulation can easily be carried over to an MHD setting, while both hydro and MHD formulations could also exploit first moment M1 closure formulations, as e.g.\ realized in {\tt RAMSES-RT} \citep{2015Rosdahl}. Both FLD and M1 radiative-(M)HD formulations have the distinct advantage that they in essence translate to fully hyperbolic PDE systems (with source term couplings), and this is readily adjusted to any framework like \texttt{MPI-AMRVAC}. Of course, in solar physics contexts, current state-of-the-art (non-AMR) codes like {\tt Stagger} (as used in \citealt{2006Stein}), {\tt Bifrost} \citep{2011Gudiksen,2020Nobrega}, {\tt MURaM} \citep{2005Vogler}, {\tt MANCHA3D} \citep{2018Khomenko,2022Navarro}, {\tt RAMENS} \citep{2017Iijima}, or {\tt CO5BOLD} \citep{2012Freytag} focus on 3D radiative-MHD simulations which include magnetoconvection in optically thick sub-photospheric layers, use optically thin prescriptions for the corona, and have a more sophisticated treatment of radiative effects known to be important in solar chromospheric layers. This includes handling partial ionization effects through MHD with ambipolar diffusion or two-fluid plasma-neutral modeling as implemented in \texttt{MPI-AMRVAC 3.0} \citep{2021Popescu,2022Braileanu}, but is especially concerned with more advanced radiative transfer aspects (going beyond LTE, approximating the complex chromospheric radiative aspects, as e.g.\ realized recently within {\tt MURaM} by \citealt{2022muram}). Future efforts towards such truly realistic radiative-MHD or radiative-multi-fluid models on evolving, block-AMR settings are highly desirable.

\texttt{MPI-AMRVAC 3.0} still uses standard {\tt Fortran} with {\tt MPI} for parallelization purposes (described in \citealt{2012Keppens}), and our suite of automated tests includes compiling the code in 1D to 3D setups with various versions of {\tt gfortran} or {\tt Intel} compilers. The code can use hybrid {\tt OpenMP-MPI} parallelism, where the {\tt OpenMP} pragma's ensure thread-based parallelism over the blocks available within a shared memory. The related {\tt BHAC} code considerably improved on this latter hybrid parallelization aspect \citep{2022Cielo}. For truly extreme resolution simulations, one needs further code restructuring (especially the internal boundary exchange involved with AMR meshes) towards 
task-based parallelism. For optimally exploiting the modern mixed CPU-GPU platforms, we plan to explore {\tt OpenACC}\footnote{\url{www.openacc.org}} or the possibilities for GPU offloading provided by {\tt OpenMP}. Similar modern restructuring efforts of {\tt Athena++} to {\tt K-Athena} to make efficient usage of tens of thousands of GPUs are documented in \citet{2021Philipp}.

Even when none or only approximate radiative effects are incorporated in our 2D or 3D (M)HD settings, one can use dedicated radiative transfer codes to assess the model appearance in synthetic views. For infrared views on colliding stellar winds with dust production zones in \citet{2016Hendrix}, this was handled by post-processing where the native \texttt{MPI-AMRVAC} datfiles were inputs to {\tt SKIRT} \citep{2015Camps}. A similar coupling of \texttt{MPI-AMRVAC} to the recent {\tt MAGRITTE} radiative transfer solver is presented in \citet{2020DeCeuster}. For solar physics applications, currently ongoing research uses the {\tt Lightweaver} \citep{2021Osborne} framework for synthesizing spectral info from the latest prominence formation models~\citep{2022Jenkins}.

Ultimately, doing justice to  plasma physics processes that are intrinsically multi-scale may need to go beyond pure fluid treatments. This is the main reason for developing visionary frameworks like {\tt DISPATCH} \citep{2018Nordlund} and  {\tt PATCHWORK} \citep{2018Shiokawa}. These envision hierarchically coupled, physics and grid-adaptive aspects, which still pose various technical algorithmic challenges. First steps towards such coupled Particle-In-Cell (PIC)-MHD efforts in \texttt{MPI-AMRVAC} contexts have been explored in \citet{2017Makwana,2018Makwana}. Complementary hybrid particle-MHD modeling has been realized in \citet{2018vanMarle}, in a 2D3V setting applied to particle acceleration processes at MHD shocks, following earlier approaches from \citet{Bai2015}. This functionality is currently not included in the \texttt{MPI-AMRVAC 3.0} release, since we prioritized some basic restructuring of the particle and field tracing modules as documented here. Note that the development part of the code can be inspected at 
\href{http://dev.amrvac.org}{http://dev.amrvac.org}, and the corresponding new branches followed on \texttt{github}.
Test particle studies may in the future benefit from the need for hybrid particle pushers \citep{Bacchini2020}, where automated switching between GCA and Lorentz treatments was demonstrated. The open-source strategy followed implies that many more research directions can be pursued, and we welcome any addition to the framework.

\begin{acknowledgements} We thank the referee for constructive comments and suggestions.
      YZ acknowledges funding from Research Foundation – Flanders FWO under the project number 1256423N.
      BP acknowledges funding from Research Foundation – Flanders FWO under the project number 1232122N.
      FB acknowledges support from the FED-tWIN programme (profile Prf-2020-004, project ``ENERGY'') issued by BELSPO.
      CX acknowledges funding from the Basic Research Program of Yunnan Province (202001AW070011), the National Natural Science Foundation of China (12073022).
      YG is supported by the National Key Research and Development Program of China (2020YFC2201201) and NSFC (11773016 and 11961131002).
      This project received funding from the European Research Council (ERC) under the European Union’s Horizon 2020 research and innovation program (grant agreement No. 833251 PROMINENT ERC-ADG 2018). Further, this research is supported by Internal funds KU Leuven, project C14/19/089 TRACESpace and FWO project G0B4521N. Visualisations used the open source software \href{https://www.paraview.org/}{ParaView}, \href{https://www.python.org/}{Python} and \href{https://yt-project.org/}{yt}. Resources and services used in this work were provided by the VSC (Flemish Supercomputer Center), funded by the Research Foundation - Flanders (FWO) and the Flemish Government.
      We acknowledge code contributions by master students Robbe D'Hondt and Brecht Geutjens, by Dr. Jannis Teunissen, Dr. Florian Driessen, and Dr. Cl\'ement Robert, and by PhD candidates Veronika Jer\v{c}i\'{c}, Joris Hermans, and Nicolas Moens. 
\end{acknowledgements}

\end{document}